\documentclass[a4paper,twoside,11pt]{article}

\usepackage{amsmath}
\usepackage{amstext}
\usepackage{amsopn}
\usepackage{amsthm}
\usepackage[plain]{algorithm}
\usepackage{algorithmic}
\usepackage{graphicx}
\usepackage{subfigure}
\usepackage{index}

\newcommand{\nb}[1]{\mathit{not}\,(#1)}
\newcommand{\Nb}[1]{\mathit{not}\bigl(#1\bigr)}
\newcommand{\nB}[1]{\mathit{not}\ #1}
\newcommand{\atleast}[1]{\mathit{Atleast}(#1)}
\newcommand{\atmost}[1]{\mathit{Atmost}(#1)}
\newcommand{\patoms}[1]{{#1}^+}
\newcommand{\natoms}[1]{{#1}^-}
\DeclareMathOperator{\Mid}{\big\vert}
\newcommand{\from}{\leftarrow}
\newcommand{\atoms}{\mathit{Atoms}}
\newcommand{\Atoms}[1]{\mathit{Atoms}(#1)}
\newcommand{\reduct}[2]{#1^{#2}}
\newcommand{\abs}[1]{\lvert#1\rvert}
\newcommand{\maps}{\rightarrow}
\newcommand{\lfp}[1]{\mathit{lfp}\left(#1\right)}
\newcommand{\gfp}[1]{\mathit{gfp}\left(#1\right)}

\newcommand{\expand}{\mathit{expand}}
\newcommand{\smodels}{\mathit{smodels}}
\newcommand{\conflict}{\mathit{conflict}}
\newcommand{\lookahead}{\mathit{lookahead}}
\newcommand{\heuristic}{\mathit{heuristic}}
\newcommand{\unacceptable}{\mathit{unacceptable}}
\newcommand{\stable}{\mathit{stable}}
\newcommand{\patleast}{\mathit{atleast}}
\newcommand{\patmost}{\mathit{atmost}}

\theoremstyle{plain}
\newtheorem{theorem}{Theorem}[section]
\newtheorem{lemma}[theorem]{Lemma}

\newtheorem{proposition}[theorem]{Proposition}
\newtheorem{corollary}[theorem]{Corollary}
\theoremstyle{definition}
\newtheorem{definition}[theorem]{Definition}
\newtheorem{example}[theorem]{Example}
\newtheorem*{examplecont}{Example~\ref{example}}
\theoremstyle{remark}
\newtheorem*{remark}{Remark}

\makeindex

\begin{document}

\pagestyle{empty}

\newcommand{\coverpage}{%
\noindent\textsf{Helsinki University of Technology
\hfill Laboratory for Theoretical Computer Science \\
Research Report 58 \\}
\textsf{\small Teknillisen korkeakoulun tietojenk\"asittelyteorian
laboratorion tutkimusraportti 58} \\
\textsf{\small Espoo 2000\hfill HUT-TCS-A58} \\
\vspace*{50mm}

\noindent\textsf{\LARGE\raggedright Extending and Implementing the
  Stable Model \\ Semantics \\ \mbox{}\\}
\noindent\textsf{\large Patrik Simons}
}

\coverpage

\cleardoublepage
\coverpage
\bigskip

\noindent\textsf{Dissertation for the degree of Doctor of Technology
  to be presented with due permission for public examination and
  debate in Auditorium T2 at Helsinki University of Technology (Espoo,
  Finland) on the 28th of April, 2000, at 12 o'clock noon.}
\vfill

\noindent\textsf{\small\raggedright
Helsinki University of Technology \\
Department of Computer Science and Engineering \\
Laboratory for Theoretical Computer Science \\
\mbox{}\\
Teknillinen korkeakoulu \\
Tietotekniikan osasto \\
Tietojenk\"asittelyteorian laboratorio}

\cleardoublepage
\begin{abstract}
  An algorithm for computing the stable model semantics of logic
  programs is developed. It is shown that one can extend the semantics
  and the algorithm to handle new and more expressive types of rules.
  Emphasis is placed on the use of efficient implementation
  techniques. In particular, an implementation of lookahead that
  safely avoids testing every literal for failure and that makes the
  use of lookahead feasible is presented. In addition, a good
  heuristic is derived from the principle that the search space should
  be minimized.
  
  Due to the lack of competitive algorithms and implementations for
  the computation of stable models, the system is compared with three
  satisfiability solvers. This shows that the heuristic can be
  improved by breaking ties, but leaves open the question of how to
  break them. It also demonstrates that the more expressive rules of
  the stable model semantics make the semantics clearly preferable
  over propositional logic when a problem has a more compact logic
  program representation. Conjunctive normal form representations are
  never more compact than logic program ones.
\end{abstract}

\cleardoublepage

\pagestyle{plain}

\pagenumbering{roman}

\section*{Preface}

I began working at the Laboratory for Theoretical Computer Science in
1995. This thesis is the result. I would like to thank Professor Leo
Ojala and Docent Ilkka Niemel\"a for giving me this opportunity, and I 
would also like to thank my colleagues in the laboratory for creating
a pleasant working atmosphere. I am especially grateful for the advice 
and comments that I have got from Ilkka during these years.

The research has been founded by the Academy of Finland (project
43963) and the Helsinki Graduate School in Computer Science and
Engineering. The financial support from the Jenny and Antti Wihuri
Foundation is acknowledged with gratitude.

I dedicate this work to my love Eeva, to my parents, and to my
sister.

\newlength{\tmp}
\settowidth{\tmp}{Otaniemi, January 2000}
\bigskip
\hfill%
\begin{minipage}{\tmp}
Otaniemi, January 2000

\vspace*{1cm}
Patrik Simons
\end{minipage}

\cleardoublepage
\tableofcontents
\newpage
\listoffigures
\listofalgorithms

\cleardoublepage
\pagenumbering{arabic}

\section{Introduction}
\label{sec:intro}

Logic programming with the stable model semantics has emerged as a
viable constraint programming
paradigm~\cite{MT:cs.LO/9809032,Niemela98:cnmr}. In this paradigm
problems are expressed as logic programs, and the stable models of the 
programs give the solutions to the problems. Since it is in general
hard to compute stable models, the typical algorithm employs an
exhaustive search when it tries to find a stable model. Different
algorithms vary in how much of the structure of a program they exploit 
when they limit their search.

We extend the stable model semantics to include three new types of
rules:
\begin{itemize}
\item choice rules for encoding subsets of a set,
\item cardinality rules for enforcing cardinality limits on the
subsets, and
\item weight rules for writing inequalities over weighted linear sums.
\end{itemize}

In addition, we define optimize statements that can be used to find
the largest or smallest stable models. The new rules can be translated
into normal rules, but not without introducing extra atoms and rules.
The motivation for including the new rules comes from reflecting on
how to solve problems by encoding them as logic programs. More
expressive rules lead to smaller programs that are easier to solve.
The reason for these three particular types is found in their
usefulness and in the ease and efficiency with which they can be
implemented.

We only consider variable-free programs. Ground programs can be
produced from non ground ones by, for example, the tool
\textit{lparse}~\cite{TS98}. Accordingly, we see the propositional
rules as primitives that can be employed as building blocks for a high
level language.

A procedure for computing the stable models of logic programs
containing the new rules has been implemented. The procedure and its
implementation bear the name $\smodels$. The purpose of this work is 
to explain the workings of $\smodels$.

\subsection{Related Work}

The stable model semantics is a form of nonmonotonic reasoning that is
closely related to the default logic of Reiter~\cite{Reiter80,MS92},
circumscription of
McCarthy~\cite{McCarthy80,McCarthy86,Lifschitz86:Minker}, and the
autoepistemic logic of Moore~\cite{Moore85,Gelfond87}. Since the
stable model semantics has become a standard method for supplying
semantics to nonmonotonic logic programs, there is a considerable
interest in automating its computation. The earliest methods for
finding stable models were based on the truth maintenance system of
Doyle~\cite{Doyle79} and the assumption-based TMS of
deKleer~\cite{deKleer86}, see~\cite{Elkan90,Eshghi90,Rodi91}. One of
the first algorithms that took advantage of the well-founded
semantics~\cite{VGRS91} was the one presented in~\cite{LRRS93} and
generalized in~\cite{LRR97}. Lately, more specialized algorithms have
been
developed~\cite{BNNS94,Subrahmanian95,CW96,Dimopoulos96,CMMT99,WNG:lpnmr99}.
However, they all suffer from exponential space complexity or weak
pruning techniques.

If we look in a broader context, then finding a stable model is a
combinatorial search problem. Other forms of combinatorial search
problems are propositional satisfiability, constraint satisfaction,
constraint logic programming and integer linear programming problems,
and some other logic programming problems such as those expressible in
\textsc{np-spec}~\cite{Cadoli:padl99}. The difference between these
problem formalisms and the stable model semantics is that they do not
include default negation. In addition, all but the last one are not
nonmonotonic.

From an algorithmic standpoint the progenitor of the $\smodels$
algorithm is the Davis-Putnam (-Logemann-Loveland)
procedure~\cite{DLL62} for determining the satisfiability of
propositional formulas. This procedure can be seen as a backtracking
search procedure that makes assumptions about the truth values of the
propositional atoms in a formula and that then derives new truth
values from these assumptions in order to prune the search space.

While the extended rules of this work are novel, there are some
analogous constructions in the literature. The choice rule can be seen
as a generalization of the disjunctive rule of the possible model
semantics~\cite{SI94}. The disjunctive rule of disjunctive logic
programs~\cite{P91} also resembles the choice rule, but the semantics
is in this case different. The stable models of a disjunctive program
are subset minimal while the stable models of a logic program are
grounded, i.e., atoms can not justify their own inclusion. If a
program contains choice rules, then a grounded model is not
necessarily subset minimal.

Since the optimize statements lexicographically order the stable
models, they can be used for prioritized reasoning. Priorities have
previously been used to lexicographically order
rules~\cite{Rintanen96:thesis,BLR:lpnmr96} and to order
atoms~\cite{SI:jicslp96}.

\subsection{Applications}

The program {\em smodels} has been used in several contexts. In the
field of verification of distributed systems, it has successfully been
applied to deadlock and reachability
problems~\cite{Heljanko:CSP98,Heljanko:Tacas99,Heljanko:csp99}. In
addition, it has been used to lessen the impact of the state space
explosion inherent in the reachability analysis of place/transition
nets~\cite{kva:csp99}. It has also been used for model checking in a
system for computing alternating fixed points~\cite{LRS98:tacas}.
In the area of product configuration it has provided a base for a
rule-based language with favorable computational
properties~\cite{SN:padl99}. An application of {\em smodels} in the
planning domain has resulted in performance comparable to and
sometimes better than that of other efficient general purpose
planners~\cite{DNK97}. Lastly, it has served as an implementation base
for dynamic constraint satisfaction problems~\cite{SGN99} and for
logic programs with weight constraint rules~\cite{NSS99:lpnmr}.

\subsection{A Brief History}

My work on {\em smodels} began in February 1995. For my Master's
Thesis~\cite{S95} I implemented a decision procedure for the
stable model semantics. The procedure had arisen during the work of my 
instructor, Ilkka Niemel\"a, on autoepistemic
logic~\cite{Niemela93:thesis}, and it was directly applicable to both
default logic~\cite{Niemela95:ijcai} and the stable model
semantics~\cite{NS95}. The algorithm was needlessly complex
and I was able to simplify it while at the same time making it prune
the search space more. In response to this, Ilkka devised a way to
strengthen the algorithm~\cite{NS:FB96NR7,NS96:jicslp} by employing
the Fitting semantics~\cite{Fitting85}. I further improved the
procedure~\cite{S97} by making systematic use of backward
chaining~\cite{CW96} and lookahead. \nocite{NS97:lpnmr}
Different types of rules were then introduced~\cite{S99:lpnmr}
and more optimizations were done.

\subsection{Contributions}

The key contributions of this work are: the new rule types and the
extension of the stable model semantics, an algorithm for computing
the stable models of sets of extended rules, and its generalization to 
compute specific stable models. An important contribution is the
derivation of a heuristic. The heuristic was not found by
experimentation, instead it was derived using the principle that the
search space should be minimized.

There are also some contributions that concern the efficient
implementation of the $\smodels$ algorithm. Naturally, one must be
familiar with the algorithm to understand the concepts involved.
The first contribution is an implementation of lookahead that safely
avoids testing every literal for failure. The implementation makes the
use of lookahead feasible. The second contribution decreases the
amount of work needed when pruning the search space and it consists of
the use of source pointers and strongly connected components in the
computation of the upper closure. The last contribution is an
improvement of the way $\smodels$ backtracks and is a type of
backjumping.

We also note that the central pruning function, the $\expand$
function, is an efficient implementation of the well-founded
semantics~\cite{VGRS91,S95}.

\subsection{Outline of the Work}

The stable model semantics and its extension to choice, cardinality,
and weight rules are presented in Section~\ref{sec:stable}. In
Section~\ref{sec:algorithm} we present an implementation of the
semantics, the $\smodels$ procedure, that computes stable models of
logic programs. We also show how the procedure can be extended to find
specific stable models such as the lexicographically smallest one. The
implementation is described in greater detail in
Section~\ref{sec:implementation}, and the complexity of the reasoning
tasks are discussed in Section~\ref{sec:complexity}. In
Section~\ref{sec:comparison} we compare $\smodels$ with the
Davis-Putnam procedure for testing satisfiability of propositional
formulas and with some algorithms for computing stable models, and in
Section~\ref{sec:experiments} we compare it with some propositional
satisfiability checkers by running experiments on random satisfiability
problems, on pigeon-hole problems, and on Hamiltonian cycle problems.
The conclusions follow in Section~\ref{sec:conclusions}. Finally, we
state some useful properties of monotone functions in
Appendix~\ref{appendix:monotone}.

\cleardoublepage
\section{The Stable Model Semantics}
\label{sec:stable}

In this section we introduce the stable model semantics for normal
logic programs. We extend the semantics to cover three new
types of rules: choice rules that encode subsets of a set, cardinality
rules that enforce cardinality limits on the subsets, and
weight rules that express inequalities over weighted linear sums.
In addition, we present the compute statement, which is used when one
searches for models that contain certain atoms, and the optimize
statements, which are used when one searches for models of optimal
weight. We begin by defining the stable model semantics.

\subsection{The Stable Model Semantics}
\label{subsec:stable}

Let $\atoms$ be a set of primitive propositions, or
atoms\index{atom}. A logic program is a set of rules of the form
\[h \from a_1,\dotsc,a_n,\nB{b_1}, \dotsc ,\nB{b_m,}\]
where $h,a_1,\dotsc,a_n,b_1,\dotsc,b_m$ are members of $\atoms$. The
atom $h$ is the head of the rule and the other atoms in the rule
make up the body. We call the expression $\nB{b}$ a
not-atom\index{not-atom} --- atoms and not-atoms are referred to as
literals\index{literal}.

\begin{example}\label{example}
Think of a stable model as the set of atoms that are true, any other
atoms are false. The program
\begin{align*}
a &\from b \\
b &\from c,\nB{d} \\
d &\from \nB{b} \\
c &\from a
\end{align*}
has, perhaps surprisingly, only one stable model: the set $\{d\}$.
\end{example}

The stable model semantics\index{stable model semantics} for a logic
program $P$ is defined as follows~\cite{GL88}. The reduct\index{reduct}
$\reduct{P}{A}$\index{$P$@$\reduct{P}{A}$} of $P$ with respect to the set
of atoms $A$ is obtained by
\begin{enumerate}
\item deleting each rule in $P$ that has a not-atom $\nB{x}$ in its
body such that $x\in A$, and by
\item deleting all not-atoms in the remaining rules.
\end{enumerate}
The deductive closure\index{deductive closure} of $\reduct{P}{A}$ is
the smallest set of atoms that is closed under $\reduct{P}{A}$ when
the rules in $\reduct{P}{A}$ are seen as inference rules.

\begin{definition}\index{stable model}
A set of atoms $S$ is a stable model of $P$ if and only if $S$ is the
deductive closure of $\reduct{P}{S}$.
\end{definition}

\begin{examplecont}[continued]
The set $\{a,b,c\}$ is not a stable model of the program $P$
\begin{align*}
a &\from b \\
b &\from c,\nB{d} \\
d &\from \nB{b} \\
c &\from a
\end{align*}
since the deductive closure of
\[\reduct{P}{\{a,b,c\}} = \{a\from b,\; b\from c,\; c\from a\}\]
is the empty set.
\end{examplecont}

In order to facilitate the definition of more general forms of rules,
we introduce an equivalent characterization of the stable model
semantics.

\begin{proposition}\index{closure}\index{stable model}
We say that $f_P : 2^\atoms\maps 2^\atoms$ is a closure of the program 
$P$ if
\begin{multline*}
f_P(S) = \{ h \mid h \from
a_1,\dotsc,a_n,\nB{b_1},\dotsc,\nB{b_m} \in P, \\
 a_1,\dotsc,a_n\in f_P(S),\ b_1,\dotsc,b_m\not\in S \}.
\end{multline*}
Let
\[g_P(S)\index{$gP$@$g_P(S)$} = \bigcap\{f_P(S) \mid
\text{$f_P : 2^\atoms\maps 2^\atoms$ is a closure} \}.\]
Then, $S$ is a stable model of $P$ if and only if
\[S = g_P(S).\]
\end{proposition}

\begin{proof}
  Note that the deductive closure of the reduct $\reduct{P}{S}$ is
  a closure, and note that for every $f_P$ that is a closure, the
  deductive closure of $\reduct{P}{S}$ is a subset of $f_P(S)$.
\end{proof}

A stable model is therefore a model that follows from the complement
of itself by means of the smallest possible closure. One speaks of
stable models as grounded\index{grounded} models, as atoms in the
models do not imply themselves. Atoms are not true without grounds.

\begin{remark}
  The alternative definition of the stable model semantics is
  basically a variation of the definition of the semantics of default
  logic given by Reiter~\cite{Reiter80}. The function $g_P$ just
  computes the least fixed point of the monotonic operator
  \begin{multline*}\index{$fP$@$f_P^S(A)$}
    f_P^S(A) = \{ h \mid h \from
    a_1,\dotsc,a_n,\nB{b_1},\dotsc,\nB{b_m} \in P, \\
    a_1,\dotsc,a_n\in A,\ b_1,\dotsc,b_m\not\in S \}.
  \end{multline*}
\end{remark}

\begin{example}
The problem of deciding whether a program has a stable model is
NP-complete~\cite{MT91:jacm}. Hence, one can encode satisfiability
problems as logic programs. The satisfying assignments of the formula
\[(a\lor b\lor\neg c)\land (\neg a\lor b\lor\neg d)\land (\neg b\lor
c\lor d)\]
correspond to the stable models of the program
\begin{align*}
a &\from \nB{a'} & a' &\from \nB{a} \\
b &\from \nB{b'} & b' &\from \nB{b} \\
c &\from \nB{c'} & c' &\from \nB{c} \\
d &\from \nB{d'} & d' &\from \nB{d} \\
\mathit{false} &\from \nB{a},\nB{b},c &
\mathit{false} &\from a,\nB{b},d \\
\mathit{false} &\from b,\nB{c},\nB{d} &
\mathit{contradiction} &\from \nB{\mathit{contradiction}},\mathit{false}
\end{align*}
In this case, there are ten satisfying assignments. The encoding works 
by rejecting models that contain $\mathit{false}$ and by stating that
an unsatisfied clause implies $\mathit{false}$. The intersection of a
stable model and the atoms in the formula is a satisfying assignment.
\end{example}

\subsection{More Expressive Rules}
\label{subsec:morerules}

Consider the problem of ensuring that at most $k$ of the atoms
$a_1,\dotsc,a_n$ are included in every stable model of a program. A
naive programmatic solution would consist of the rules 
\[\{\mathit{false} \from a_{i_1},\dotsc,a_{i_{k+1}} \mid
1\leq i_1 < \dotsb < i_{k+1} \leq n\}\]
of which there are $\binom{n}{k+1}$ and of the stable models that do
not include $\mathit{false}$. There is also a quadratic solution, or
to be more precise, a solution that needs on the order of $nk$ rules.

Let the atom $l(a_i,j)$ represent the fact that at least $j$ of the
atoms in $\{a_i,\dotsc,a_n\}$ are in a particular stable model. Then,
the demand that at most $k$ of $a_1,\dotsc,a_n$ are in a model can be
handled by the constraint $\mathit{false}\from l(a_1,k+1)$. The
definition of $l(a_i,j)$ is given by the program
\begin{align*}
  l(a_i,j) &\from l(a_{i+1},j) \\
  l(a_i,j+1) &\from a_i, l(a_{i+1},j) \\
  l(a_i,1) &\from a_i
\end{align*}
and we need about $nk$ instances of it to cover all values of $i$ and
$j$.

There is a need for a representation that is more compact. We propose
to introduce three types of rules that have sufficient expressiveness
to compactly describe problems such as the one above. The first type
is the cardinality rule\index{cardinality rule}, which is of the form
\[h \from k\,\{ a_1,\dotsc,a_n, \nB{b_1}, \dotsc ,\nB{b_m}\}.\]
It is interpreted as follows: if at least $k$ literals in the set
\[\{a_1,\dotsc,a_n, \nB{b_1}, \dotsc ,\nB{b_m}\}\]
are satisfied by a stable model, then the atom $h$ must be in the stable
model. An atom $a$ is satisfied by a stable model $S$ if $a\in S$ and a
not-atom $\nB{b}$ is satisfied by $S$ if $b\not\in S$.

\begin{example}
  Once is a mistake, twice is a habit. Or,
  \[\mathit{habit} \from 2\,\{\mathit{mistake}_1,\dotsc,\mathit{mistake}_n\}.\]
\end{example}

The second type is the choice rule\index{choice rule},
\[\{h_1,\dotsc,h_k\} \from a_1,\dotsc,a_n, \nB{b_1}, \dotsc ,\nB{b_m},\]
which implements a nondeterministic choice over the atoms in
$\{h_1,\dotsc,h_k\}$ whenever the literals in the body are satisfied
by a stable model. That is, the rule gives ground for the inclusion of
any number of atoms in its head. For example, the rule
\[\{h_1,\dotsc,h_k\}\from\]
can be encoded by the program
\begin{align*}
  h_i &\from \nB{h_i'} \\
  h_i' &\from \nB{h_i}, \qquad i=1,\dotsc,k.
\end{align*}

\begin{example}
  Each stable model of the program
  \begin{align*}
    \{a_1,a_2,a_3,a_4\} &\from \\
    \mathit{false} &\from \nB{a_1}, \nB{a_2}, \nB{a_3}, \nB{a_4}
  \end{align*}
  that does not contain $\mathit{false}$ includes at least one atom.
\end{example}

Finally, the third type of rule, the weight rule\index{weight rule},
is of the form
\[h \from \{ a_1 = w_{a_1},\dotsc,a_n = w_{a_n},
\nB{b_1} = w_{b_1},\dotsc,\nB{b_m} = w_{b_m}\} \geq w,\]
and its head $h$ will be in a stable model $S$ if
\[\sum_{a_i\in S} w_{a_i}+\sum_{b_i\not\in S} w_{b_i} \geq w.\]
Here, the weights are real numbers.

We can restrict ourselves to positive weights, since a rule with
negative weights can be translated into a rule with only positive
weights. Namely,
\[h \from \{ a = w_a, b = -w_b,
\nB{c} = w_c, \nB{d} = -w_d\} \geq w, \quad w_a,w_b,w_c,w_d>0\]
is equivalent to
\begin{multline*}
  h \from \{ a = w_a, b = -w_b, b = w_b, \nB{b} = w_b, \\
  \nB{c} = w_c, \nB{d} = -w_d, \nB{d} = w_d, d = w_d\} \geq w+w_b+w_d
\end{multline*}
which is equivalent to
\[h \from \{ a = w_a, \nB{b} = w_b,
\nB{c} = w_c, d = w_d\} \geq w+w_b+w_d.\]
Furthermore, the rule
\[h \from \{ a_1 = w_{a_1},\dotsc,a_n = w_{a_n},
\nB{b_1} = w_{b_1},\dotsc,\nB{b_m} = w_{b_m}\} \leq w\]
can be transformed into
\begin{multline*}
  h \from \{ a_1 = -w_{a_1},\dotsc,a_n = -w_{a_n}, \\
  \nB{b_1} = -w_{b_1},\dotsc,\nB{b_m} = -w_{b_m}\} \geq -w.
\end{multline*}

\begin{example}
  The stable models of the program
  \begin{align*}
    \{a_1,\dotsc,a_n\} &\from \\
    \mathit{false} &\from \{a_1 = w_1,\dotsc,a_n = w_n\} \geq w \\
    \mathit{true} &\from \{a_1 = v_1,\dotsc,a_n = v_n\} \geq v
  \end{align*}
  containing the atom $\mathit{true}$ but not the atom $\mathit{false}$
  correspond to the ways one can pack a subset of $a_1,\dotsc,a_n$ in a
  bin such that the total weight is less than $w$ and the total value is
  at least $v$. The weights and values of the items are given
  by respectively $w_1,\dotsc,w_n$ and $v_1,\dotsc,v_n$.
\end{example}

Often only stable models including or excluding certain atoms are of
interest. We can guarantee that every stable model of a program
includes a specific atom $\mathit{true}$ by adding a rule
\[\mathit{true}\from\nB{\mathit{true}}\]
to the program. Similarly, we can exclude an atom $\mathit{false}$
from all stable models of a program with the help of the rule
\[\mathit{contradiction} \from \nB{\mathit{contradiction}},\mathit{false}\]
provided that the atom $\mathit{contradiction}$
does not appear anywhere else in the program. Despite the simplicity
of the two previous rules, we will use compute
statements\index{compute statement} of the form
\[\mathit{compute}\,\{ a_1,\dotsc,a_n, \nB{b_1}, \dotsc ,\nB{b_m}\}\]
to state that we only accept stable models that contain
$a_1,\dotsc,a_n$ but not $b_1,\dotsc,b_m$.

\begin{example}
  The satisfying assignments of the formula
  \[(a\lor b\lor\neg c)\land (\neg a\lor b\lor\neg d)\land (\neg b\lor
  c\lor d)\]
  correspond to the stable models of the program
  \begin{gather*}
    \begin{split}
      \{a,b,c,d\} &\from \\
      \mathit{false} &\from \nB{a},\nB{b},c \\
      \mathit{false} &\from a,\nB{b},d \\
      \mathit{false} &\from b,\nB{c},\nB{d}
    \end{split} \\
    \mathit{compute}\,\{\nB{\mathit{false}}\}
  \end{gather*}
\end{example}

Sometimes one wants to find the stable model with the least number of
atoms. Other times the atoms have been given priorities and one wants
to find the stable model with the highest priority. In order to be
able to express such preferences, we introduce two optimize
statements:\index{optimize statement} the minimize statement and its
dual the maximize statement. The minimize statement\index{minimize
  statement}
\[\mathit{minimize}\,
\{a_1 = w_{a_1},\dotsc,a_n = w_{a_n},
\nB{b_1} = w_{b_1},\dotsc,\nB{b_m} = w_{b_m}\}\]
declares that we want to find a stable model $S$ with the smallest
weight \[\sum_{a_i\in S} w_{a_i}+\sum_{b_i\not\in S} w_{b_i}.\]
If there are several minimize statements, then we order the stable
models lexicographically according to the weights of the
statements. In this case, the first statement is the most
significant. The maximize statement\index{maximize statement}
\[\mathit{maximize}\,\{a = w_a, \nB{b} = w_b\}\]
is just another way to write
\[\mathit{minimize}\,\{a = -w_a, \nB{b} = -w_b\}\]
or
\[\mathit{minimize}\,\{a = k-w_a, \nB{b} = k-w_b\},\]
if one wants to avoid negative weights.

\begin{example}
  The lexicographically first stable model of the program
  \begin{gather*}
    \begin{split}
      a &\from \nB{b} \\
      b &\from \nB{a} \\
    \end{split} \\
    \begin{split}
      \mathit{minimize}\, &\{a=1\} \\
      \mathit{minimize}\, &\{b=1\}
    \end{split}
  \end{gather*}
  is $\{b\}$.
\end{example}

\begin{example}
  Multiple minimize statements can be written as one:
  \begin{align*}
    \mathit{minimize}\, &\{a_1=w_{a_1},\dotsc,a_n=w_{a_n}\} \\
    \mathit{minimize}\, &\{b_1=w_{b_1},\dotsc,b_m=w_{b_m}\}
  \end{align*}
  is equivalent to
  \[\mathit{minimize}\, \{a_1=2^kw_{a_1},\dotsc,a_n=2^kw_{a_n},
  b_1=w_{b_1},\dotsc,b_m=w_{b_m}\}\]
  if $2^k>\sum w_{b_i}$. 
\end{example}

We now turn to the formal definition of the stable model semantics for 
the extended syntax.

\begin{definition}
  A basic rule\index{basic rule} $r$ is of the form
  \[h \from a_1,\dotsc,a_n,\nB{b_1}, \dotsc ,\nB{b_m}\]
  and is interpreted by the function
  $f_r:2^\atoms\times 2^\atoms\maps 2^\atoms$
  as follows.
  \[f_r(S,C)\index{$fr$@$f_r(S,C)$}
  = \{h \mid a_1,\dotsc,a_n\in C,\ b_1,\dotsc,b_m\not\in S\}.\]
\end{definition}

The intuition behind the function $f_r$ is that it produces the result
of a deductive step when applied to a candidate stable model $S$ and
its consequences $C$. In other words, if $C$ is a deductive closure,
then $f_r(S,C)\subseteq C$. By this reasoning, the definition of $f_r$
for the new rules follows straightforwardly.

\begin{definition}
  A cardinality rule\index{cardinality rule} $r$ is of the form
  \[h \from k\,\{ a_1,\dotsc,a_n, \nB{b_1}, \dotsc ,\nB{b_m}\}\]
  and is interpreted by
  \[
  f_r (S,C) = \bigl\{h \Mid
  \abs{\{a_1,\dotsc,a_n\}\cap C} +
  \abs{\{b_1,\dotsc,b_m\} - S} \geq k\bigr\}.
  \]
  A choice rule\index{choice rule} $r$ is of the form
  \[\{h_1,\dotsc,h_k\} \from a_1,\dotsc,a_n, \nB{b_1}, \dotsc ,\nB{b_m}\]
  and is interpreted by
  \[
  f_r (S,C) = \bigl\{h \Mid h\in\{h_1,\dotsc,h_k\}\cap S,
  a_1,\dotsc,a_n\in C,\ b_1,\dotsc,b_m\not\in S\bigr\}.
  \]
  Finally, a weight rule\index{weight rule} $r$ is of the form
  \[
  h \from \{ a_1 = w_{a_1},\dotsc,a_n = w_{a_n},
  \nB{b_1} = w_{b_1},\dotsc,\nB{b_m} = w_{b_m}\} \geq w,
  \]
  for $w_{a_i},w_{b_i} \geq 0$,
  and is interpreted by
  \[f_r (S,C) = \{h \mid \sum_{a_i\in C} w_{a_i} +
  \sum_{b_i\not\in S} w_{b_i} \geq w\}.\]
\end{definition}

\begin{example}
  Let $S = \{a\}$ be a model of a logic program, one of whose rules is
  $r$: \[h \from \{a = 1,b = 2,\nB{c} = 3\} \geq 4.\]
  Since we expect a stable model to be a deductive closure, we set
  $C = \{a\}$ and compute $f_r(S,C) = \{h\}$. As $h\not\in C$, we note
  that $S$ can not be a stable model.
\end{example}

\begin{definition}
  Let $P$ be a set of rules. As before we say that
  $f_P:2^\atoms\maps 2^\atoms$ is a closure if
  \[f_P(S) = \bigcup_{r\in P} f_r\bigl(S,f_P(S)\bigr),\]
  and we define
  \[g_P(S)\index{$gP$@$g_P(S)$} = \bigcap\{f_P(S) \mid
  \text{$f_P : 2^\atoms\maps 2^\atoms$ is a closure}\}.\]
  Then, $S$ is a stable model\index{stable model} of the program $P$ if
  and only if
  \[S = g_P(S).\]
\end{definition}

\newcommand{\Lfp}[1]{\mathit{lfp}(#1)}
\newcommand{\Gfp}[1]{\mathit{gfp}(#1)}

\begin{lemma}\label{lemma:stable}
  Let $P$ be a logic program and let $S$ be a set of atoms.
  Define
  \[f_P^S(A) = \bigcup_{r\in P} f_r(S,A)\index{$fP$@$f_P^S(A)$}.\]
  Then, the least fixed point $\Lfp{f_P^S}$ of $f_P^S$ is equal to
  $g_P(S)$.
\end{lemma}

\begin{proof}
  Note that the operator $f_P^S(A)$ is monotonic and that it has a
  fixed point $f_P(S)$ for any closure $f_P$. Hence,
  $\Lfp{f_P^S}\subseteq f_P(S)$ for all $f_P$ and therefore
  $\Lfp{f_P^S}\subseteq g_P(S)$. Since $\Lfp{f_P^S}$, taken as a
  function of $S$, defines a closure, $g_P(S)\subseteq\Lfp{f_P^S}$.
  Thus, $g_P(S)=\Lfp{f_P^S}$ follows.
\end{proof}

\begin{example}
  Let $P$ be the logic program
  \begin{align*}
      \{a, b, c\} &\from \\
      \mathit{true} &\from 2\,\{a, b, c\} \\
      \mathit{compute}\, &\{\mathit{true}\} \\
      \mathit{minimize}\, &\{a=1,b=2\}.
  \end{align*}
  The stable models of the two rules in $P$ are $\emptyset$, $\{a\}$,
  $\{b\}$, $\{c\}$, $\{a,b,\mathit{true}\}$, $\{a,c,\mathit{true}\}$,
  $\{b,c,\mathit{true}\}$, and $\{a,b,c,\mathit{true}\}$, as we can
  easily check using Lemma~\ref{lemma:stable}. For instance,
  $f_P^{\{a\}}(\emptyset) = \{a\} = f_P^{\{a\}}(\{a\})$. Because of
  the compute statement we only want stable models that contain the
  atom $\mathit{true}$, namely the models $\{a,b,\mathit{true}\}$,
  $\{a,c,\mathit{true}\}$, $\{b,c,\mathit{true}\}$, and
  $\{a,b,c,\mathit{true}\}$. But since there is also a minimize
  statement we are really only interested in the smallest of these:
  $\{a,c,\mathit{true}\}$.
\end{example}

The stable models of a logic program that do not contain choice rules
are subset minimal, whereas the stable models of a program with choice
rules can be subsets of each other. It is therefore not possible to
translate a program containing choice rules into one with no choice
rules without introducing new atoms. In this sense, the introduction
of choice rules makes the stable model semantics more expressive.

\begin{proposition}
Let $P$ be a logic program that does not contain any choice rules. If
$S$ and $S'$ are stable models of $P$, then $S\subseteq S'$ implies
$S=S'$. 
\end{proposition}

\begin{proof}
Let $P$ be a logic program that does not contain any choice rules and
let $S\subseteq S'$ be two stable models of $P$. As
\[f_P^{S'}(S)=\bigcup_{r\in P} f_r(S',S)
\subseteq \bigcup_{r\in P} f_r(S,S)=f_P^{S}(S)=S\]
since $f_r(S',S)$ is anti-monotonic in its first argument,
$S'=\Lfp{f_P^{S'}}\subseteq S$.
\end{proof}

\begin{corollary}
  Let $P$ be the set of all logic programs and let $P'$ be the set of
  all logic programs that do not contain choice rules. Then, there is
  no mapping from $P$ to $P'$ that preserves stable models.
\end{corollary}

\begin{example}
We can translate the disjunctive rules of the possible
model\index{possible model semantics} semantics~\cite{SI94} into
choice rules and basic rules such that possible models correspond to
stable models. Namely, change every disjunctive rule
\[a_1\lor\dotsb\lor a_k\from b_1\land\dotsb\land b_n\land
\nB{c_1}\land\dotsb\land\nB{c_m}\]
into a choice rule
\begin{align*}
\{a_1,\dotsc,a_k\}&\from b_1,\dotsc,b_n,\nB{c_1},\dotsc,\nB{c_m} \\
\intertext{and a basic rule}  
\mathit{false}&\from b_1,\dotsc,b_n,\nB{c_1},\dotsc,\nB{c_m},
\nB{a_1},\dotsc,\nB{a_k},
\end{align*}
and add the compute statement
\[\mathit{compute}\,\{\nB{\mathit{false}}\}.\]
\end{example}

\cleardoublepage
\section{The Algorithm}
\label{sec:algorithm}

Having defined the stable model semantics, we want to find a
method that can be used to compute stable models. We take the obvious
approach and enumerate all subsets of the atoms in a program and test
each subset for stability. During the exhaustive search we make use
of the properties of the stable model semantics to prune away large
numbers of subsets. The procedure is put forth as a backtracking
search algorithm. Since we want to be able to handle large programs,
we avoid constructs that require more than a linear amount of space.

\begin{figure}
\centering
\includegraphics{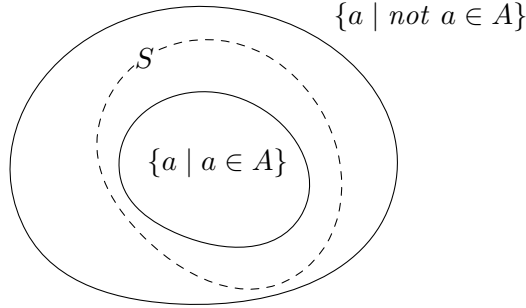}
\caption{The relation between the set $A$ and one of its stable models
  $S$}
\label{fig:A}
\end{figure}

At the heart of the algorithm is a set of literals, which we name $A$,
that represents a set of stable models. The atoms in the set $A$ are
members of these models and the not-atoms in the set are atoms that
are not in the stable models. It follows that if there is an atom in
$A$ that also appears as a not-atom in $A$, then the set of models that
$A$ represents is empty. Hence, our algorithm begins with $A$ as the
empty set. It adds atoms and not-atoms to $A$ and checks whether the
resulting set corresponds to at least one stable model. If it does
not, then it backtracks by removing atoms and not-atoms from $A$ and
by changing atoms into not-atoms and vice versa. The relationship
between the partial model $A$ and one of its stable models is
exemplified in Figure~\ref{fig:A}.

The search space consisting of all possible configurations of $A$ is
pruned by deducing additions to $A$ from the program using the
properties of the stable model semantics. For example, if the rule
\[a\from b,\nB{c}\]
is in a program and $b,\nB{c}\in A$, then we deduce that every
stable model of the program that contains $b$ but not $c$ must contain 
$a$. Consequently, we add $a$ to $A$. Expanding $A$ can lead to
situations in which an atom in $A$ is also a not-atom in $A$. If such
a conflict\index{conflict} takes place, then the algorithm backtracks.

We can prune the search space some more by wisely choosing what
literals we add to $A$. A good heuristic helps, but choosing literals
that immediately give rise to conflicts avoids a lot of
backtracking. We find these literals by looking ahead: for each
literal not in $A$ we temporarily add the literal to $A$, expand it,
and check for conflicts.

We explain the algorithm in greater detail in the rest of the section.

\subsection{The Decision Procedure}
\label{subsec:decision}

For an atom $a$, let $\nb{a} = \nB{a}$, and for a
not-atom $\nB{a}$, let \[\nb{\nB{a}} = a.\] For a set of literals $A$,
define \[\nb{A}\index{$notA$@$\nb{A}$} = \{ \nb{a} \mid a\in A\}.\]
Let $\patoms{A}=\{a\in\atoms\mid a\in A\}$\index{$A+$@$\patoms{A}$} and
let $\natoms{A}=\{a\in\atoms\mid \nB{a}\in A\}$.\index{$A-$@$\natoms{A}$}
Define
$\Atoms{A}=\patoms{A}\cup\natoms{A}$\index{$AtomsA$@$\Atoms{A}$}, and 
for a program $P$, define
$\Atoms{P}=\Atoms{L}$\index{$AtomsP$@$\Atoms{P}$}, where $L$ is the
set of literals that appear in the program.

A set of literals $A$ is said to cover\index{cover} a set of atoms $B$
if $B\subseteq\Atoms{A}$, and $B$ is said to agree\index{agree} with
$A$ if \[\patoms{A}\subseteq B \qquad\text{and}\qquad
\natoms{A}\subseteq\atoms-B.\]

Algorithm~\ref{alg:smodels}\index{$smodels$@$\smodels(P,A)$} displays
a decision procedure for the stable model semantics. The function
$\smodels(P,A)$ returns true whenever there is a stable model of $P$
agreeing with the set of literals $A$. In fact, the function computes
a stable model and, as will become apparent, it can be modified to
compute all stable models of $P$ that agree with $A$.

\begin{algorithm}
\caption{A decision procedure for the stable model semantics}
\label{alg:smodels}
\begin{algorithmic}
\item[\textbf{function} $\smodels(P,A)$]
  \index{$smodels$@$\smodels(P,A)$}
 \STATE $A:=\expand(P,A)$
 \STATE $A:=\lookahead(P,A)$
 \IF{$\conflict(P,A)$}
  \STATE return false
 \ELSIF{$A$ covers $\Atoms{P}$}
  \STATE return true \COMMENT{$\patoms{A}$ is a stable model}
 \ELSE
  \STATE $x:=\heuristic(P,A)$
  \IF{$\smodels(P,A\cup \{x\})$}
   \STATE return true
  \ELSE
   \STATE return $\smodels\bigl(P,A\cup\{\nb{x}\}\bigr)$
  \ENDIF
 \ENDIF.
\medskip
\item[\textbf{function} $\expand(P,A)$]\index{$expand$@$\expand(P,A)$}
 \REPEAT
  \STATE $A':=A$
  \STATE $A:=\atleast{P,A}$
  \STATE $A:=A\cup\{\nB{x}\mid x\in\Atoms{P}$ and
$x\not\in\atmost{P,A}\}$
 \UNTIL{$A = A'$}
 \STATE return $A$.
\medskip
\item[\textbf{function} $\conflict(P,A)$]
  \index{$conflict$@$\conflict(P,A)$}
 \STATE \COMMENT{Precondition: $A = \expand(P,A)$}
 \IF{$\patoms{A}\cap\natoms{A} \neq \emptyset$}
  \STATE return true
 \ELSE
  \STATE return false
 \ENDIF.
\medskip
\item[\textbf{function} $\lookahead(P,A)$]
 \STATE return $A$.
\end{algorithmic}
\end{algorithm}

The decision procedure calls four functions: $\expand(P,A)$,
$\conflict(P,A)$, $\lookahead(P,A)$, and $\heuristic(P,A)$. The
function $\expand(P,A)$ expands the set $A$ using the functions
$\atleast{P,A}$ and $\atmost{P,A}$, the function $\conflict(P,A)$
discovers conflicts, and the function $\heuristic(P,A)$ computes
heuristically good literals that can be included in $A$. For now we
let the function $\lookahead(P,A)$ return $A$.

Let $A' = \expand(P,A)$\index{$expand$@$\expand(P,A)$}. We assume that
\begin{description}
\item[E1] $A\subseteq A'$ and that
\item[E2] every stable model of $P$ that agrees with $A$
also agrees with $A'$.
\end{description}
Moreover, we assume that the function $\conflict(P,A)$
\index{$conflict$@$\conflict(P,A)$} satisfies the two conditions
\begin{description}
\item[C1] if $A$ covers $\Atoms{P}$ and there is no stable model that
agrees with $A$, then $\conflict(P,A)$ returns true, and
\item[C2] if $\conflict(P,A)$ returns true, then there is no
stable model of $P$ that agrees with $A$.
\end{description}
In addition, we expect $\heuristic(P,A)$ to return a literal
not covered by $A$.

\begin{theorem}
Let $P$ be a set of rules and let $A$ be a set of literals. Then,
there is a stable model of $P$ agreeing with $A$ if and only if
$\smodels(P,A)$ returns true.
\end{theorem}

\begin{proof}
Let $\mathit{nc}(P,A) = \Atoms{P}-\Atoms{A}$ be the atoms not covered
by $A$. We prove the claim by induction on the size of
$\mathit{nc}(P,A)$.

Assume that the set $\mathit{nc}(P,A) = \emptyset$. Then,
$A'=\expand(P,A)$ covers $\Atoms{P}$ by E1 and $\smodels(P,A)$ returns
true if and only if $\conflict(P,A')$ returns false. By E2, C1, and C2,
this happens precisely when there is a stable model of $P$ agreeing
with $A$.

Assume $\mathit{nc}(P,A)\neq\emptyset$. If $\conflict(P,A')$
returns true, then $\smodels(P,A)$ returns false and by E2 and 
C2 there is no stable model agreeing with $A$. On the other hand, if
$\conflict(P,A')$ returns false and $A'$ covers $\Atoms{P}$,
then $\smodels(P,A)$ returns true and by E2 and C1 there is a
stable model that agrees with $A$. Otherwise, induction together with
E1 and E2 show that $\smodels(P,A'\cup\{x\})$ or
$\smodels\bigl(P,A'\cup\{\nb{x}\}\bigr)$ returns true if and
only if there is a stable model agreeing with $A$.
\end{proof}

Let $S$ be a stable model of $P$ agreeing with the set of literals $A$.
Then, $f_r(S,S)\subseteq S$ for $r\in P$, and we make the following
observations. Let
\[\mathit{min}_r(A)\index{$minr$@$\mathit{min}_r(A)$}
= \negthickspace \bigcap_{\substack{\patoms{A}\subseteq C \\
\natoms{A}\cap C=\emptyset}} \negthickspace f_r(C,C)\]
be the inevitable consequences of $A$, and let
\[\mathit{max}_r(A)\index{$maxr$@$\mathit{max}_r(A)$}
= \negthickspace \bigcup_{\substack{\patoms{A}\subseteq C \\
\natoms{A}\cap C=\emptyset}} \negthickspace f_r(C,C)\]
be the possible consequences of $A$. Then,
\begin{enumerate}
\item if $r\in P$, then $S$ agrees with $\mathit{min}_r(A)$,
\item if there is an atom $a$ such that for all $r\in P$,
  $a\not\in\mathit{max}_r(A)$, then $S$ agrees with $\{\nB{a}\}$,
\item if the atom $a\in A$, if there is only one $r\in P$ for which
  $a\in\mathit{max}_r(A)$, and if there exists a literal $x$ such that
  $a\not\in\mathit{max}_r(A\cup\{x\})$, then $S$ agrees with
  $\{\nb{x}\}$, and
\item if $\nB{a}\in A$ and if there exists a literal $x$ such that
  for some $r\in P$, $a\in\mathit{min}_r(A\cup\{x\})$, then $S$ agrees
  with $\{\nb{x}\}$.
\end{enumerate}
Note that if $a,\nB{a}\in A$, then $\mathit{min}_r(A) = \atoms$ and
$\mathit{max}_r(A) = \emptyset$.

\begin{proposition}
  The claims 1--4 hold.
\end{proposition}

\begin{proof}
  Recall from Lemma~\ref{lemma:stable} that
  \[f_P^S(A) = \bigcup_{r\in P} f_r(S,A)\]
  is monotonic and that its least fixed point is equal to $g_p(S)$.
  Hence, if $S$ is a stable model of $P$, then for any $r\in P$,
  $f_r(S,S)\subseteq S$.

  Let the stable model $S$ agree with the set $A$. As
  $\mathit{min}_r(A)\subseteq f_r(S,S)$, the first claim follows.
  Notice that \[S=\bigcup_{r\in P} f_r(S,S).\]
  If for all $r\in P$, $a\not\in\mathit{max}_r(A)$, then
  for all $r\in P$, $a\not\in f_r(S,S)$ and consequently
  $a\not\in S$. This proves the second claim. 

  If $a\in A$ and there is only one $r\in P$ for which
  $a\in\mathit{max}_r(A)$, then $a\in f_r(S,S)\subseteq S$. If
  $a\not\in\mathit{max}_r(A\cup\{x\})$ for some literal $x$ and if $S$
  agrees with $\{x\}$, then $a\not\in f_r(S,S)$ which is a
  contradiction. Thus, the third claim holds.

  Finally, if $\nB{a}\in A$ and there is a literal $x$ such that for
  some $r\in P$, $a\in\mathit{min}_r(A\cup\{x\})$, then by the first
  claim $a\in S$ which is a contradiction. Therefore, also the fourth
  claim is true.
\end{proof}

The four statements help us deduce additional literals that are in
agreement with $S$. Define $\atleast{P,A}$ as the smallest set of literals
containing $A$ that can not be enlarged using 1--4 above, i.e., let
$\atleast{P,A}$\index{$Atleast$@$\atleast{P,A}$} be the least fixed
point of the operator 
\begin{align*}\index{$fB$@$f(B)$}
f(B) = A&\cup B \\
&\cup\{a\in \mathit{min}_r(B) \mid \text{$a\in\Atoms{P}$ and
$r\in P$}\} \\
&\cup\{\nB{a}\mid \text{$a\in\Atoms{P}$ and for all $r\in P$,
$a\not\in\mathit{max}_r(B)$}\} \\
&\cup\bigl\{\nb{x}\Mid \text{there exists $a\in B$ such that
$a\in\mathit{max}_r(B)$} \\
&\qquad\qquad\qquad\text{for only one $r\in P$ and
$a\not\in\mathit{max}_r(B\cup\{x\})$}\bigr\} \\
&\cup\bigl\{\nb{x}\Mid \text{there exists $\nB{a}\in B$ and $r\in P$
such that} \\
&\qquad\qquad\qquad a\in\mathit{min}_r(B\cup\{x\})\bigr\}.
\end{align*}

\begin{example}
Let $P$ be the program
\begin{align*}
  a &\from b,\nB{c} \\
  d &\from \nB{a} \\
  e &\from \nB{b}
\end{align*}
We will compute $A=\atleast{P,\{d\}}$. Since $c$ does not appear in
the head of any rule in $P$, $\nB{c}\in A$ by claim 2. As
$d\in A$, $\nB{a}\in A$ by claim 3. It follows that $\nB{b}\in
A$ by 4. Finally, $e\in A$ by 1. Hence,
$\atleast{P,\{d\}}=\{\nB{a},\nB{b},\nB{c},d,e\}$.
\end{example}

\begin{lemma}\label{lemma:atleast}
The function $\atleast{P,A}$ is monotonic in its second argument.
\end{lemma}

\begin{proof}
Observe that the function $\mathit{min}_r(B)$ is monotonic and that
the function $\mathit{max}_r(B)$ is anti-monotonic. Hence, 
\begin{gather*}
\{a\in \mathit{min}_r(B) \mid r\in P\}, \\
\{\nB{a}\mid \text{$a\in\Atoms{P}$ and for all $r\in P$,
$a\not\in\mathit{max}_r(B)$}\}, \\
\intertext{and}
\bigl\{\nb{x}\Mid \text{there exists $\nB{a}\in B$ and $r\in P$
such that $a\in\mathit{min}_r(B\cup\{x\})$}\bigr\}
\end{gather*}
are monotonic with respect to $B$. Assume that there exists $a\in B$
such that $a\in\mathit{max}_r(B)$ for only one $r\in P$ and
$a\not\in\mathit{max}_r(B\cup\{x\})$. If $B\subseteq B'$ and 
$a\not\in\mathit{max}_r(B')$, then
\begin{multline*}
  \nB{a}\in\{\nB{a'}\mid \text{$a'\in\Atoms{P}$ and for all $r\in P$,} \\
  a'\not\in\mathit{max}_r(B')\} \subseteq f(B').
\end{multline*}
Consequently, both $a,\nB{a}\in f(B')$ and therefore
\begin{gather*}
\mathit{min}_r\bigl(f(B')\bigr) = \atoms, \\
\intertext{and}
\mathit{max}_r\bigl(f(B')\bigr) = \emptyset.
\end{gather*}
It follows that
$f\bigl(f(B')\bigr) = \Atoms{P}\cup\Nb{\Atoms{P}}$. Thus, $f^2$
is monotonic and has a least fixed point. Finally, notice that $f$ has
the same fixed points as $f^2$. By the definition of $f$,
$B\subseteq f(B)$. Thus, $f^2(B)=B$ implies
\[B\subseteq f(B)\subseteq f\bigl(f(B)\bigr)=B.\]
\end{proof}

We conclude,

\begin{proposition}
If the stable model $S$ of $P$ agrees with $A$, then $S$ agrees with
$\atleast{P,A}$.
\end{proposition}

Furthermore, we can bound the stable models from above.

\begin{definition}
For a choice rule $r$ of the form
\[\{h_1,\dotsc,h_k\} \from a_1,\dotsc,a_n, \nB{b_1}, \dotsc ,\nB{b_m},\]
let
\[f_r'(S,C)\index{$fr'$@$f_r'(S,C)$}
= \bigl\{h\in\{h_1,\dotsc,h_k\} \Mid
a_1,\dotsc,a_n\in C,\ b_1,\dotsc,b_m\not\in S\bigr\},\]
and for any other type of rule, let $f_r' (S,C) = f_r (S,C)$.
Let $P$ be a set of rules and let $A$ be a set of literals. Define
$\atmost{P,A}$\index{$Atmost$@$\atmost{P,A}$} as the least fixed point
of
\[f'(B)\index{$fB'$@$f'(B)$} =
\bigcup_{r\in P} f_r'(\patoms{A},B-\natoms{A})-\natoms{A}.\]
\end{definition}

\begin{proposition}
Let $S$ be a stable model of $P$ that agrees with $A$. Then,
$S\subseteq \atmost{P,A}$.
\end{proposition}

\begin{proof}
Note that $f_r'(S,C)$ is anti-monotonic in its first argument, i.e.,
$S\subseteq S'$ implies $f_r'(S',C)\subseteq f_r'(S,C)$, and monotonic
in its second argument. Fix a program $P$, a stable model $S$ of $P$,
and a set of literals $A$ such that $S$ agrees with $A$. Define
\[f_P^S(B) = \bigcup_{r\in P} f_r(S,B)\]
and
\[f'(B) = \bigcup_{r\in P} f_r'(\patoms{A},B-\natoms{A})-\natoms{A}.\]
Let $L$ be the least fixed point of $f'$. Since $S$ agrees with $A$,
\[f_r(S,S\cap L)\subseteq f_r'(\patoms{A},S\cap L -
\natoms{A})-\natoms{A},\]
and $f_P^S(S\cap L)\subseteq f'(S\cap L) \subseteq L$. Hence, for varying
$B$, the least fixed point of $f_P^S(S\cap B)$, which is equal to the
least fixed point of $f_P^S$, is a subset of $L$ by
Lemma~\ref{lemma:monotone}. In other words, $S\subseteq L$.
\end{proof}

It follows that $\expand(P,A)$ satisfies the conditions E1 and 
E2. The function $\conflict(P,A)$ obviously fulfills C2, and 
the next proposition shows that also C1 holds.

\begin{proposition}\label{prop:stable}
If $A = \expand(P,A)$ covers the set $\Atoms{P}$ and
$\patoms{A}\cap\natoms{A} = \emptyset$, then $\patoms{A}$ is a stable
model of $P$. 
\end{proposition}

\begin{proof}
Assume that $A = \expand(P,A)$ covers $\Atoms{P}$ and that
$\patoms{A}\cap\natoms{A} = \emptyset$. Then,
$\patoms{A} = \atmost{P,A}$ and
\[f_r(\patoms{A},B) = f_r'(\patoms{A},B)-\natoms{A} =
f_r'(\patoms{A},B-\natoms{A})-\natoms{A}\]
for every $B\subseteq\patoms{A}$, since $f_r(\patoms{A},B)\subseteq
f_r(\patoms{A},\patoms{A})=\mathit{min}_r(A)\subseteq\patoms{A}$.
Thus, $\patoms{A}$ is the least fixed point of
\[f_P^{\patoms{A}}(B) = \bigcup_{r\in P} f_r(\patoms{A},B),\] from 
which we infer, by Lemma~\ref{lemma:stable}, that $\patoms{A}$ is a
stable model of $P$.
\end{proof}

\begin{example}
  Let $P$ be the program
  \begin{align*}
    a &\from \nB{b} \\
    c &\from a
  \end{align*}
  Then, $\atmost{P,\emptyset} = \{a,c\}$ as
  $f'_{a\from\nB{b}}(\emptyset,\emptyset) = \{a\}$ and
  $f'_{c\from a}(\emptyset,\{a\}) = \{c\}$. But,
  $\atmost{P,\{\nB{a}\}} = \emptyset$.
\end{example}

\begin{example}
  Let $P$ be the program
  \begin{align*}
    a &\from \nB{b} \\
    b &\from \nB{a},c \\
    c &\from c
  \end{align*}
  By computing $A=\expand(P,\emptyset)$ we see that this program has
  only one stable model. Namely, $\atleast{P,\emptyset} = \emptyset$
  and $\atmost{P,\emptyset} = \{a\}$. Hence,
  $\nB{b},\nB{c}\in A$. Since
  $\atleast{P,\{\nB{b},\nB{c}\}} = \{a,\nB{b},\nB{c}\}$ and since
  $\atmost{P,\{a,\nB{b},\nB{c}\}} = \{a\}$, $A=\{a,\nB{b},\nB{c}\}$.
  Therefore, $A$ covers $\Atoms{P}$ and $\{a\}$ is the only stable
  model of $P$.
\end{example}

\begin{figure}
\centering
\includegraphics{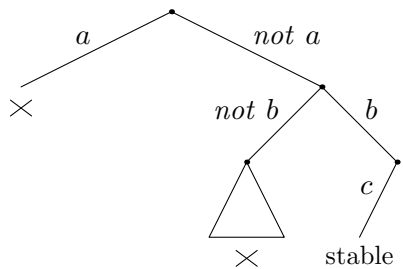}
\caption{A visualization of the search process}
\label{fig:tree}
\end{figure}

\begin{example}
  The $\smodels$ algorithm traverses a search space consisting of sets
  of literals. One can visualize the path the algorithm takes as a
  tree whose nodes are the sets and whose edges are labeled with the
  heuristic choices that have been made during the computation. By
  convention we assume that the algorithm goes down the left branch
  first. Hence, the whole computation corresponds to an in-order
  traversal of the tree. For example, in Figure~\ref{fig:tree} the
  algorithm first tries the atom $a$ and experiences a conflict. It
  then changes to $\nB{a}$ and tries $\nB{b}$ from which it continues
  through some unspecified choices that all lead to a conflict. After
  this $b$ is asserted and the choice of the atom $c$ ends in a stable
  model.
\end{example}

\subsection{Looking Ahead}
\label{subsec:lookahead}

The $\expand$ function does a good job of reducing the search space. It
makes use of some simple properties of the stable model semantics to
refine a partially computed model. Even if the function only has to
satisfy the two general conditions E1 and E2, it is in practice
severely constrained by the small amount of time it can take to do its 
work. Since the function is called so often, it must be fast.
Otherwise, the algorithm will not achieve acceptable performance on
programs with many stable models.

By E2 the $\expand$ function must not loose any stable models. It can
therefore not enlarge the partial model $A$ by much if there are many
stable models agreeing with $A$. A slower but slightly better $\expand$
function does not help. On the other hand, if a partial model
does not agree with any stable models, then $\expand$ should
return as large a set as possible. A slower function is not such an
objection then. This dichotomy is addressed here.

\begin{figure}
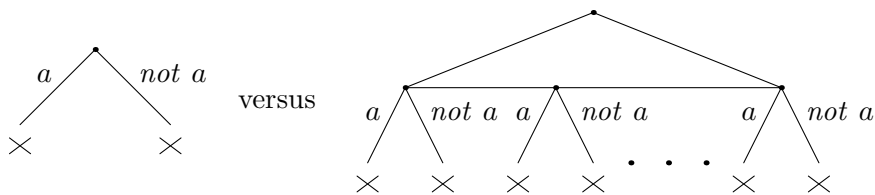

\centering
\valign{\vfil#\vfil\cr\hbox{\includegraphics{figlookahead.1}}\cr
\hbox{\quad versus \quad}\cr
\hbox{\includegraphics{figlookahead.2}}\cr}
\caption{Looking ahead yields a potentially smaller search space}
\label{fig:lookahead}
\end{figure}

The $\expand$ function as presented is a compromise that sacrifices
optimality for performance. We want to strengthen it to better handle
partial models that can not be enlarged to stable models. Consider a
program $P$, a partial model $A$, and an atom $a$ such that both
$\expand(P,A\cup\{a\})$ and $\expand(P,A\cup\{\nB{a}\})$ contain
conflicts, i.e., $\conflict(P,A')$ returns true for
$A' = \expand(P,A\cup\{x\})$, where $x=a,\nB{a}$. If $\smodels(P,A)$
chooses $a$ or $\nB{a}$ immediately, then it will return after only
two $\expand$ calls. If it does not, then the number of $\expand$ calls
can be potentially very large, as is illustrated in
Figure~\ref{fig:lookahead}.

The literals that instantly give rise to conflicts can be found by
testing. We call the testing procedure lookahead, as it corresponds
to looking ahead and seeing how $\smodels$ behaves when it has chosen
a literal. Observe that if the stable model $S$ agrees with the
partial model $A$ but not with $A\cup\{x\}$ for some literal $x$, then
$S$ agrees with $A\cup\{\nb{x}\}$. Hence, we make progress as soon as
we find a literal $x$ that causes a conflict. That is to say, we can
then enlarge $A$ by $\nb{x}$. In addition, since
$x'\in\expand(P,A\cup\{x\})$ implies
\begin{align*}
\expand(P,A\cup\{x'\})
&\subseteq\expand\bigl(P,\expand(P,A\cup\{x\})\bigr) \\
&=\expand(P,A\cup\{x\})
\end{align*}
due to the monotonicity of $\expand$, it is not even necessary to
examine all literals in $\Atoms{P}$ not covered by $A$. As we test a
literal $x$ we can directly rule out all atoms in
$\expand(P,A\cup\{x\})$. We implement lookahead according to these
observations by rewriting the function
$\lookahead$\index{$lookahead$@$\lookahead(P,A)$} as in
Algorithm~\ref{alg:lookahead}. Possibly calling $\expand$ on the order
of $\abs{\Atoms{P}-\Atoms{A}}$ times for each new literal might not
seem like a good idea, but in practice it has proven amazingly
effective.

\begin{algorithm}[t]
\caption{Looking ahead}
\label{alg:lookahead}
\begin{algorithmic}
\item[\textbf{function} $\lookahead(P,A)$]
  \index{$lookahead$@$\lookahead(P,A)$}
 \REPEAT
  \STATE $A' := A$
  \STATE $A := \mathit{lookahead\_once}(P,A)$
 \UNTIL{$A = A'$}
 \STATE return $A$.
\medskip
\item[\textbf{function} $\mathit{lookahead\_once}(P,A)$]
 \STATE $B := \Atoms{P}-\Atoms{A}$
 \STATE $B := B\cup\nb{B}$
 \WHILE{$B \neq \emptyset$}
  \STATE Take any literal $x\in B$
  \STATE $A' := \expand(P,A\cup\{x\})$
  \STATE $B:=B-A'$
  \IF{$\conflict(P,A')$}
   \STATE return $\expand(P,A\cup\{\nb{x}\})$
  \ENDIF
 \ENDWHILE
 \STATE return $A$.
\end{algorithmic}
\end{algorithm}

The idea that one can use the detection of conflicts to prune the
search space has been presented in the context of propositional
satisfiability checkers by Zabih and McAllester~\cite{ZM88}. It is
interesting to note that they concluded that the pruning method seems
promising but causes too much overhead. Modern satisfiability checkers
avoid the overhead by only employing lookahead on a small
heuristically chosen subset of all atoms.

\subsection{Heuristics}
\label{subsec:heuristics}

The heuristic choices that are made in a backtracking algorithm can
drastically affect the time the algorithm has to spend searching for a
solution. Since a correct choice brings the algorithm closer to a
solution while a wrong choice leads the algorithm astray, great effort
is often expended on creating heuristics that find the correct
choices. This seems to be a bad approach. A heuristic invariably fails
at some point, otherwise it would not be a heuristic, and then it
tries to find nonexistent solutions when it should be minimizing the
duration of the search in that part of the search space. We will
therefore optimize our heuristic\index{heuristic} for the case of no
stable models. That is, we will try to minimize the size of the
remaining search space.

For a literal $x$, let
\[A_p = \expand(P,A\cup\{x\})\]
and
\[A_n = \expand\bigl(P,A\cup\{\nb{x}\}\bigr).\]
Assume that the search space is a full binary tree of height $H$. A
full binary tree is a binary tree whose paths from the root to the
leaves are all of equal length. Let $p = \abs{A_p-A}$ and
$n = \abs{A_n-A}$. Then,
\[2^{H-p}+2^{H-n} = 2^H\frac{2^n+2^p}{2^{p+n}}\]
is an upper bound on the size of the remaining search space.
Minimizing this number is equal to minimizing
\[\log \frac{2^n+2^p}{2^{p+n}} = \log(2^n+2^p) - (p+n).\]
Since \[2^{\max(n,p)} < 2^n+2^p \leq 2^{\max(n,p)+1}\]
is equivalent to
\[\max(n,p) < \log(2^n+2^p) \leq \max(n,p)+1\]
and
\[-\min(n,p) < \log(2^n+2^p) - (p+n)\leq 1-\min(n,p),\]
it suffices to maximize $\min(n,p)$. If two different literals have
equal minimums, then one chooses the one with the greater maximum,
as this minimizes $2^{-\max(n,p)}$.

When the best literal $x$ has been found, we have to return one of $x$
and $\nb{x}$. If there are no stable models agreeing with $A$, then it 
does not matter which one. But if there are stable models agreeing
with both $A\cup\{x\}$ and $A\cup\{\nb{x}\}$, then we should again try to
minimize the remaining search space. Hence, we return the one that
shrinks the search space the most.

The function $\heuristic$\index{$heuristic$@$\heuristic(P,A)$} is
shown in Algorithm~\ref{alg:heuristic}. In an implementation of
$\smodels$ one naturally integrates $\lookahead$ and $\heuristic$ to
avoid unnecessary work. At the same time one can take advantage of the
fact that $x'\in\expand(P,A\cup\{x\})$ implies
\[\abs{\expand(P,A\cup\{x'\})-A} \leq \abs{\expand(P,A\cup\{x\})-A}\]
to evade some of the $\expand$ computations.

Freeman noted in~\cite{Freeman95} that a good satisfiability heuristic 
should choose the literal that minimizes the quantity
$2^{H-p}+2^{H-n}$, where $n$ and $p$ are computed using unit
propagation instead of $\expand$. It is again interesting to note that
he then concluded that such a heuristic is too slow in practice.

\begin{algorithm}
\caption{The heuristic}
\label{alg:heuristic}
\begin{algorithmic}
\item[\textbf{function} $\heuristic(P,A)$]
  \index{$heuristic$@$\heuristic(P,A)$}
 \STATE $B := \Atoms{P}-\Atoms{A}$
 \STATE $\mathit{min} := 0$
 \STATE $\mathit{max} := 0$
 \WHILE{$B \neq \emptyset$}
  \STATE Take any atom $a\in B$
  \STATE $B := B-\{a\}$
  \STATE $p := \abs{\expand(P,A\cup\{a\})-A}$
  \IF{$p \geq \mathit{min}$}
   \STATE $n := \abs{\expand\bigl(P,A\cup\{\nB{a}\}\bigr)-A}$
   \IF{$\min(n,p) > \mathit{min}$ or $\bigl(\min(n,p) = \mathit{min}$
        and $\max(n,p) > \mathit{max}\bigr)$}
    \STATE $\mathit{min} := \min(n,p)$
    \STATE $\mathit{max} := \max(n,p)$
    \IF{$p = \max(n,p)$}
     \STATE $x := a$
    \ELSE
     \STATE $x := \nB{a}$
    \ENDIF
   \ENDIF
  \ENDIF
 \ENDWHILE
 \STATE return $x$.
\end{algorithmic}
\end{algorithm}

\subsubsection*{Lookahead order}

The $\lookahead$ function does not test the literals in any particular
order. Since the function returns as soon as it finds a conflict, it
is desirable that it should find them as soon as possible. We notice
that in a large program $P$ and for a small set of literals $A$,
$\expand(P,A\cup\{x,y\})$ probably does not contain a conflict if 
$\expand(P,A\cup\{x\})$ does not. Therefore, if we first test a literal $x$
and do not find a conflict and then test another literal $y$ and find
a conflict, then expanding $A\cup\{x,\nb{y}\}$ will quite likely not
lead to a conflict. Hence, keeping the literals in a least recently
used order is reasonable. Indeed, it has been verified that it often
removes many needless $\expand$ calls.

\subsection{Searching for Specific Stable Models}
\label{subsec:minmax}

Besides searching for one stable model, one may also search for many,
all, or just certain stable models. All of these variants can easily
be incorporated into the $\smodels$ algorithm. In particular, we will
deal with the case of optimize statements, as we must at least
implicitly examine all stable models to find the smallest or largest
one.

The simplest way to handle the search for specific stable models is to 
modify the $\conflict(P,A)$ function such that it returns true
whenever one can be sure that we are not searching for any stable
models agreeing with $A$. This can be thought of as changing what it
means for a stable model to agree with a set of literals. Only the
stable models that we accept can agree with a set of literals. Hence,
we need not redo the proof of soundness and completeness for the
modified algorithm. The new $\conflict$ function is presented in
Algorithm~\ref{alg:conflict2}. The function
$\unacceptable(P,A)$\index{$unacceptable$@$\unacceptable(P,A)$}
implements the test for the acceptable stable models. If it returns
true, then there must not be any acceptable stable models agreeing
with $A$. Moreover, if $\patoms{A}$ is a stable model that is not
acceptable, then $\unacceptable(P,A)$ must return true.

\begin{algorithm}
\caption{Removing unacceptable stable models}
\label{alg:conflict2}
\begin{algorithmic}
\item[\textbf{function} $\conflict(P,A)$]
  \index{$conflict$@$\conflict(P,A)$}
 \STATE \COMMENT{Precondition: $A = \expand(P,A)$}
 \IF{$\patoms{A}\cap\natoms{A} \neq \emptyset$}
  \STATE return true
 \ELSE
  \STATE return $\unacceptable(P,A)$
 \ENDIF.
\end{algorithmic}
\end{algorithm}

\begin{algorithm}
\caption{A generalization of the decision procedure}
\label{alg:smodels2}
\begin{algorithmic}
\item[\textbf{function} $\smodels(P,A)$]
  \index{$smodels$@$\smodels(P,A)$}
 \STATE $A:=\expand(P,A)$
 \STATE $A:=\lookahead(P,A)$
 \IF{$\conflict(P,A)$}
  \STATE return false
 \ELSIF{$A$ covers $\Atoms{P}$}
  \STATE return $\stable(P,A)$ \COMMENT{$\patoms{A}$ is a stable model}
 \ELSE
  \STATE $x:=\heuristic(P,A)$
  \IF{$\smodels(P,A\cup \{x\})$}
   \STATE return true
  \ELSE
   \STATE return $\smodels\bigl(P,A\cup\{\nb{x}\}\bigr)$
  \ENDIF
 \ENDIF.
\end{algorithmic}
\end{algorithm}

What about computing all stable models of a program? We can control
how many stable models are computed by calling a function
$\stable(P,A)$\index{$stable$@$\stable(P,A)$} instead of returning
true when a stable model has been found, cf
Algorithm~\ref{alg:smodels2}. If $\stable(P,A)$ returns false, then
the algorithm will search for the next model. Otherwise, it stops.
Since returning false from $\stable$ is indistinguishable from
returning false from $\conflict$, it is guaranteed that $\stable$ is
only called for acceptable stable models.  If $\stable$ always returns
false, then it will be called once for every stable model of $P$.
Therefore, the function can be used to, e.g., display or count the
models.

We now turn to the optimize statements. Without loss of generality we 
consider only programs containing one minimize statement and only
positive weights:
\[\mathit{minimize}\,
\{a_1 = w_{a_1},\dotsc,a_n = w_{a_n},
\nB{b_1} = w_{b_1},\dotsc,\nB{b_m} = w_{b_m}\}.\]
We let the literals not in the minimize statement have zero weight.
Let $B$ be a global variable that is initially empty. We define the
acceptable models as models whose weight is smaller than the weight of 
$B$ if $B$ is not empty. When an acceptable stable model $\patoms{A}$
is found we assign $A$ to $B$. In addition, we let $\stable$ always
return false. Thus, when $\smodels$ returns, $B$ contains a stable
model of minimal weight if one exists. Otherwise, it is empty.
The corresponding $\unacceptable$ and $\stable$ functions are
portrayed in Algorithm~\ref{alg:minimal}.

\begin{algorithm}
\caption{Finding a minimal stable model}
\label{alg:minimal}
\begin{algorithmic}
\item[\textbf{function} $\unacceptable(P,A)$]
\index{$unacceptable$@$\unacceptable(P,A)$}
 \IF{$B\neq\emptyset$ and $\sum_{a\in A}w_a+\sum_{\nB{b}\in A}w_b
      \geq\sum_{a\in B}w_a+\sum_{\nB{b}\in B} w_b$}
  \STATE return true
 \ELSE
  \STATE return false
 \ENDIF.
\medskip
\item[\textbf{function} $\stable(P,A)$]\index{$stable$@$\stable(P,A)$}
 \STATE $B := A$
 \STATE return false.
\end{algorithmic}
\end{algorithm}

\cleardoublepage
\section{Implementation}
\label{sec:implementation}

In this section we will present an efficient implementation of the
functions $\atleast{P,A}$ and $\atmost{P,A}$. Both implementations are 
variations of a linear time algorithm of Dowling and
Gallier~\cite{DG84}. We will also examine various optimizations that
can be used to improve the implementation and the $\smodels$
algorithm.

The basic Dowling-Gallier algorithm\index{Dowling-Gallier algorithm}
computes the deductive closure of a set of basic rules in time linear
in the size of the set of rules. The rules must not contain any
not-atoms, i.e., they are Horn clauses. A variant of the basic
algorithm is shown in Algorithm~\ref{alg:dowlinggallier}. It follows
naturally from one observation and one implementation trick. We
observe that a deductive step is monotone. Namely, if the body of a
rule is in a set of atoms, then it is also in any superset of the same
set. Hence, the order in which the rules are applied does not matter.
The trick is to use a counter for each rule to find out when a rule
can be used in the deduction. The counters should initially hold the
number of atoms in the bodies of the rules. Every time an atom is
added to the closure, the counters of the rules in whose bodies the
atom appears are decremented. If any counter reaches zero, then the
body of the corresponding rule is in the closure and the head of the
rule is put in the closure. The basic algorithm is obviously correct.
An atom is in the closure if and only if the atom is the head of a
rule whose body is in the closure.

\begin{algorithm}
\caption{The basic Dowling-Gallier algorithm}
\label{alg:dowlinggallier}
\begin{algorithmic}
\item[\textbf{procedure} Dowling-Gallier $(P)$]
 \STATE\COMMENT{Invariant:
   $\abs{\mathrm{body}-\mathrm{closure}}=\mathrm{counter}$ and
   $\mathrm{counter}=0$ is equivalent to 
   $\mathrm{head}\in\mathrm{closure}\cup\mathrm{queue}$}
 \STATE Initialize the counter of every rule to the number of atoms
in its body
 \STATE let the queue contain the heads of the rules in $P$ whose
bodies are empty 
 \WHILE{the queue is not empty}
  \STATE remove the first element from the queue and call it $a$
  \IF{$a$ is not in the closure}
   \STATE add $a$ to the closure
   \FOR{each rule $r\in P$ in whose body $a$ appears}
    \STATE decrement the counter of $r$ by one
    \IF{the counter of $r$ is equal to zero}
     \STATE append the head of $r$ to the end of the queue
    \ENDIF
   \ENDFOR
  \ENDIF
 \ENDWHILE.
\end{algorithmic}
\end{algorithm}

\subsection{At Least}
\label{subsec:atleast}

\newcommand{\inpos}{\patoms{\mathit{inA}}}
\newcommand{\inneg}{\natoms{\mathit{inA}}}
\newcommand{\poslist}{\mathit{plist}}
\newcommand{\neglist}{\mathit{nlist}}
\newcommand{\headlist}{\mathit{hlist}}
\newcommand{\headof}{\mathit{headof}}
\newcommand{\body}{\mathit{body}}
\newcommand{\head}{\mathit{head}}
\newcommand{\fire}{\mathit{fire}}
\newcommand{\inactivate}{\mathit{inactivate}}
\newcommand{\inactive}{\mathit{inactive}}
\newcommand{\literal}{\mathit{literal}}
\newcommand{\backchaintrue}{\mathit{backchaintrue}}
\newcommand{\backchainfalse}{\mathit{backchainfalse}}

We begin with basic rules of the form
\[h \from a_1,\dotsc,a_n,\nB{b_1}, \dotsc ,\nB{b_m}.\]
For every rule $r\in P$ we create a literal counter
$r.\literal$ that holds the number of literals in the body of
$r$ that are not members of the partially computed closure
$\atleast{P,A}$. In addition, an inactivity counter
$r.\inactive$ is also created. If the set $A$ is a partially computed
closure, then the inactivity counter records the number of literals in
the body of $r$ that are in $\nb{A}$. The counter
$r.\inactive$ is therefore positive, and the rule $r$ is
inactive, if the set $\mathit{max}_r(A)$ is empty and one can not then
use $r$ to deduce its head. For every atom $a$ we create a head
counter $a.\headof$ that holds the number of active rules with head
$a$.

Recall that a literal can be brought into $\atleast{P,A}$ in four
different ways. We handle the four cases with the help of the three
counters.
\begin{enumerate}
\item If $r.\literal$ reaches zero, then the head of $r$ is
added to the closure.
\item If $a.\headof$ reaches zero, then $\nB{a}$ is added to the
closure.
\item If $a.\headof$ is equal to one and $a$ is in the closure,
then every literal in the body of the only active rule with head $a$
is added to the closure. 
\item Finally, if $a$ is the head of $r$, if $\nB{a}$ is in the
closure, and if $r.\literal = 1$ and
$r.\inactive = 0$, then there is precisely one literal $x$ in
the body of $r$ that is not in the closure, and $\nb{x}$ is added to
the closure.
\end{enumerate}

Cardinality rules and choice rules are easily incorporated into the
same framework. Specifically, one does neither use the first nor the
fourth case together with choice rules, and one does not compare the
literal and inactivity counters of a cardinality rule
\[h \from k\,\{ a_1,\dotsc,a_n, \nB{b_1}, \dotsc ,\nB{b_m}\}\]
with zero but with $m+n-k$. A weight rule
\[h \from \{ a_1 = w_{a_1},\dotsc,a_n = w_{a_n},
\nB{b_1} = w_{b_1},\dotsc,\nB{b_m} = w_{b_m}\} \geq w,\]
is managed using the upper and lower bound of the sum of the weights
in its body. Given a set of literals $A$, the lower bound is
\[\sum_{a_i\in \patoms{A}}w_{a_i} + \sum_{b_i\in \natoms{A}}w_{b_i}\]
and the upper bound is
\[\sum_{a_i\not\in\natoms{A}}w_{a_i} + \sum_{b_i\not\in
\patoms{A}}w_{b_i}.\]
If the upper bound is less than $w$, then the rule is inactive, and if
the lower bound is at least $w$, then the head is in the closure.

Note that by Lemma~\ref{lemma:atleast} the order in which the
four ways of bringing atoms into the closure are used is of no
importance for the end result. The actual implementation will not
compute all of $\atleast{P,A}$ when it equals
$\Atoms{P}\cup\Nb{\Atoms{P}}$, but a subset $B$ that satisfies 
$\patoms{B}\cap\natoms{B}\neq\emptyset$. This is enough since no
conflicts are lost.

In preparation for the procedure that computes $\atleast{P,A}$, we
construct two Boolean flags for each atom $a$: $a.\inpos$ is set to
true when $a$ is included in the closure $\atleast{P,A}$ and 
$a.\inneg$ is set to true when $\nB{a}$ is included in the closure.
All atoms have three lists as well. The list $a.\poslist$
contains pointers to every rule $r$ for which $a\in r.\body$, the
list $a.\neglist$ contains pointers to every rule $r$ for which
$\nB{a}\in r.\body$, and the list $a.\headlist$ contains pointers to
every rule $r$ for which $a\in r.\head$. To summarize, we have
the following variables:
\begin{list}{}{\settowidth{\labelwidth}{$r.\inactive$}
\setlength{\leftmargin}{\labelwidth}
\addtolength{\leftmargin}{\labelsep}}
\item[$a.\headof$\hfill] The number of active rules whose head
  contains $a$.
\item[$a.\inpos$\hfill] A flag that is true if $a$ is in the current
  closure.
\item[$a.\inneg$\hfill] A flag that is true if $\nB{a}$ is in the
  current closure.
\item[$a.\poslist$\hfill] The rules in whose bodies $a$ appears.
\item[$a.\neglist$\hfill] The rules in whose bodies $\nB{a}$ appears.
\item[$a.\headlist$\hfill] The rules in whose heads $a$ appears.
\item[$r.\literal$\hfill] The number of literals in the body that are in
the current closure.
\item[$r.\inactive$\hfill] The number of literals in the body that
are in the negation of the current closure, i.e., in $\nb{A}$ if $A$
is the current closure. 
\item[$r.\body$\hfill] The body of the rule $r$.
\item[$r.\head$\hfill] The head or heads of the rule $r$.
\end{list}

\begin{algorithm}
\caption{The implementation of $\protect\atleast{P,A}$}
\label{alg:atleast}
\begin{algorithmic}
\item[\textbf{procedure} $\patleast()$]\index{$Atleast$@$\atleast{P,A}$}
 \WHILE{$\mathit{posq}$ or $\mathit{negq}$ are not empty}
  \IF{$\mathit{posq}$ is not empty}
   \STATE $a := \mathit{posq}.\mathit{pop}()$
   \STATE $a.\inpos := $ true
   \FOR{each rule $r\in a.\poslist$}
    \STATE $r.\fire()$
   \ENDFOR
   \FOR{each rule $r\in a.\neglist$}
    \STATE $r.\inactivate()$
   \ENDFOR
   \IF{$a.\headof = 1$}
    \STATE Let $r$ be the only active rule in $a.\headlist$
    \STATE $r.\backchaintrue()$
   \ENDIF
  \ENDIF  
  \IF{$\mathit{negq}$ is not empty}
   \STATE $a := \mathit{negq}.\mathit{pop}()$
   \STATE $a.\inneg := $ true
   \FOR{each rule $r\in a.\neglist$}
    \STATE $r.\fire()$
   \ENDFOR
   \FOR{each rule $r\in a.\poslist$}
    \STATE $r.\inactivate()$
   \ENDFOR
   \IF{$a.\headof > 0$}
    \FOR{each rule $r$ in $a.\headlist$}
     \STATE $r.\backchainfalse()$
    \ENDFOR
   \ENDIF
  \ENDIF  
 \ENDWHILE.
\end{algorithmic}
\end{algorithm}

\begin{algorithm}
\caption{Auxiliary functions for the basic rules}
\label{alg:basicaux}
\begin{algorithmic}
\item[\textbf{function} $r.\fire()$]
 \STATE $r.\literal := r.\literal - 1$
 \IF{$r.\literal = 0$}
  \STATE $\mathit{posq}.\mathit{push}(r.\head)$
 \ELSIF{$r.\head.\inneg$}
  \STATE $r.\backchainfalse()$
 \ENDIF.
\medskip
\item[\textbf{function} $r.\inactivate()$]
 \STATE $r.\inactive := r.\inactive+1$
 \IF{$r.\inactive = 1$}
  \STATE $a := r.\head$
  \STATE $a.\headof := a.\headof-1$
  \IF{$a.\headof = 0$}
   \STATE $\mathit{negq}.\mathit{push}(a)$
  \ELSIF{$a.\inpos$ and $a.\headof = 1$}
    \STATE Let $r'$ be the only active rule in $a.\headlist$
    \STATE $r'.\backchaintrue()$
  \ENDIF
 \ENDIF.
\end{algorithmic}
\end{algorithm}
\begin{algorithm}
\caption{More auxiliary functions for the basic rules}
\label{alg:basicaux2}
\begin{algorithmic}
\item[\textbf{function} $r.\backchaintrue()$]
 \FOR{every $a\in\patoms{r.\body}$}
  \STATE $\mathit{posq}.\mathit{push}(a)$
 \ENDFOR
 \FOR{every $a\in\natoms{r.\body}$}
  \STATE $\mathit{negq}.\mathit{push}(a)$
 \ENDFOR.
\medskip
\item[\textbf{function} $r.\backchainfalse()$]
 \IF{$r.\literal = 1$ and $r.\inactive = 0$}
  \FOR{every $a\in\patoms{r.\body}$}
   \IF{$a.\inpos = $ false}
    \STATE $\mathit{negq}.\mathit{push}(a)$
    \STATE return
   \ENDIF
  \ENDFOR
  \FOR{every $a\in\natoms{r.\body}$}
   \IF{$a.\inneg = $ false}
    \STATE $\mathit{posq}.\mathit{push}(a)$
    \STATE return
   \ENDIF
  \ENDFOR
 \ENDIF.
\end{algorithmic}
\end{algorithm}

The procedure $\patleast()$ that computes $\atleast{P,A}$ is described
in Algorithm~\ref{alg:atleast}. It follows the basic Dowling-Gallier
algorithm, but extends it to handle the four ways a literal can be
included in $\atleast{P,A}$. The procedure is written such that the
main work happens in the four functions $\fire()$, $\inactivate()$,
$\backchaintrue()$, and $\backchainfalse()$, corresponding to the
cases 1, 2, 3, and 4, respectively. We need specific instances of
these functions for every type of rule. Therefore, we write
$r.\fire()$ to denote the $\fire()$ function that corresponds to the
rule $r$, and we use the same notation for the other functions. The
instances used together with basic rules are given in
Algorithms~\ref{alg:basicaux}--\ref{alg:basicaux2}. The rest are in
Appendix~\ref{appendix:atleast}. Since the procedure $\patleast()$ and
the functions must be efficient they do not take any arguments but
work on global data.

Two queues are used in $\patleast()$: $\mathit{posq}$ and
$\mathit{negq}$. We assume that they are implemented such that pushing
an atom onto the end of a queue does nothing if the atom is already on
the queue. In addition, we assume that pushing an atom $a$ whose
$a.\inpos$ flag is true onto the queue $\mathit{posq}$ or pushing
an atom $a$ whose $a.\inneg$ flag is true onto the queue
$\mathit{negq}$ also have no effect. These assumptions have a simple
Boolean flag implementation and they make the descriptions a bit
shorter.

One can use stacks instead of queues, but tests have shown that queues
are faster. The reason seems to be that if there is a conflict, then
it is found quicker if one derives literals breadth-first using a
queue than if one derives them depth-first using a stack. A stack
implementation might do a lot of work before it notices a conflict
that is only a few rules from the start of the derivation.

Before computing $\atleast{P,A}$ we have to initialize the queue
$\mathit{posq}$ with $\patoms{A}$ and the heads of the rules whose
bodies are empty. The queue $\mathit{negq}$ must be initialized with
the atoms in $\natoms{A}$. Observe that the procedure can also be used 
incrementally. If $\atleast{P,A}$ has been computed and we are going
to compute $\atleast{P,A\cup\{a\}}$, then we just initialize
$\mathit{posq}$ with $a$ and call $\patleast()$.

\begin{proposition}
  The procedure $\patleast()$ computes $\atleast{P,A}$ in time linear
  in the size of the program $P$.   
\end{proposition}

\begin{proof}
  Note that for every atom $a$, the list $a.\headlist$ is traversed at
  most twice and the lists $a.\poslist$ and $a.\neglist$ are traversed
  at most once. For every rule $r$, the list $r.\body$ is traversed at
  most twice. To be precise, the list $a.\headlist$ is only traversed
  in the procedure $\patleast()$ when $a.\inpos$ or $a.\inneg$ is
  being set and this can only happen once. The same holds for the
  lists $a.\poslist$ and $a.\neglist$. Moreover, the only other place
  where the list $a.\headlist$ is examined is in the function
  $\inactivate()$ and there too only once, when $a.\headof$ is
  decremented to one.
  
  The list $r.\body$ is only traversed in $\backchaintrue()$ and
  $\backchainfalse()$. These functions are, for each rule, called at
  most once from the procedure $\patleast()$. The function
  $\backchainfalse()$ is also called from $\fire()$, but then
  $r.\head.\inneg$ is true and the list is gone through if $r.\literal
  = 1$, which only happens once. Finally, $\backchaintrue()$ is called
  from $\inactivate()$, but only once when $r.\head.\headof = 1$.
\end{proof}

In conclusion, notice how the amount of work performed by the
procedure $\patleast()$ corresponds to the number of literals in
$\atleast{P,A}-A$. Specifically, observe that if $\patleast()$ is
called in succession with an increasingly larger set $A$, then the
total amount of work done is still linear in the size of $P$.

\subsection{At Most}
\label{subsec:atmost}

The deductive closure $\atmost{P,A}$ can be computed in a style
similar to that of $\atleast{P,A}$. However, since $\atmost{P,A}$
diminishes as $A$ grows, one would then have to compute it from
scratch each time $A$ changes. If $\atmost{P,A}$ is large and if it
changes only a little, then we would be doing a lot of extra work
if we computed it anew.

We will try to localize the computation by using the basic
Dowling-Gallier algorithm in two stages. Assume that we have computed
the deductive closure $\atmost{P,A}$ and that we want to compute
$\atmost{P,A'}$ for a set $A'\supset A$. We begin by calculating a
set $B\subseteq\atmost{P,A'}$ with the help of a version of the basic
Dowling-Gallier algorithm. Instead of deriving new atoms, this version
removes them. After this stage we apply the basic algorithm to
incrementally compute $\atmost{P,A'}$ from $B$.

The first stage removes atoms from the closure $\atmost{P,A}$ in the
following way. If it notices that a rule can no longer be used to
derive the atom or atoms in the head of the rule, then it removes the
atom or atoms from the closure. The removal may lead to the
inactivation of more rules and subsequent removals of more atoms.
Since there can be rules that still imply the inclusion of an atom in
$\atmost{P,A'}$ after the atom has been removed, the first stage might
remove too many atoms. We therefore need the second stage to add them
back. Sufficiently many atoms are removed by the first stage, as an
atom that is not removed is the head of a rule that is not inactive.
The atoms in the body of this rule are in the same way heads of other
rules that are not inactive, and this succession continues until one
reaches rules that imply the inclusion of their heads even if the
closure were empty.

\newcommand{\upper}{\mathit{upper}}
\newcommand{\inupper}{\mathit{inUpper}}
\newcommand{\propagateFalse}{\mathit{propagateFalse}}
\newcommand{\propagateTrue}{\mathit{propagateTrue}}
\newcommand{\isUpperActive}{\mathit{isUpperActive}}

The two stages need a new counter for every rule. We call this counter
$\upper$, since it is used to compute the upper closure\index{upper
  closure} that approximates stable models from above. In addition,
each atom $a$ has a Boolean flag $a.\inupper$ that is true if $a$ is a
member of the upper closure. The procedure that computes
$\atmost{P,A}$ is described in Algorithm~\ref{alg:atmost}. The main
work happens in the three functions $\propagateFalse()$,
$\propagateTrue()$, and $\isUpperActive()$, corresponding to
decrementing the counter, incrementing the counter, and testing if the
rule can be used to derive its head or heads, respectively. The
functions for the basic rules are given in
Algorithm~\ref{alg:upperaux}. The functions for the other types are in
Appendix~\ref{appendix:atmost}.

\begin{algorithm}[t]
\caption{The implementation of $\protect\atmost{P,A}$}
\label{alg:atmost}
\begin{algorithmic}
\item[\textbf{procedure} $\patmost()$]\index{$Atmost$@$\atmost{P,A}$}
 \STATE $F := \emptyset$
 \WHILE{$\mathit{queue}$ is not empty}
  \STATE $a := \mathit{queue}.\mathit{pop}()$
  \IF{$a.\inupper =$ true}
   \FOR{each rule $r\in a.\poslist$}
    \STATE $r.\propagateFalse()$
   \ENDFOR
   \STATE $a.\inupper :=$ false
   \STATE $F := F\cup\{a\}$
  \ENDIF
 \ENDWHILE
 \FOR{each atom $a\in F$}
  \FOR{each rule $r\in a.\headlist$}
   \IF{$r.\isUpperActive()$}
    \STATE $\mathit{queue}.\mathit{push}(a)$
   \ENDIF
  \ENDFOR
 \ENDFOR
 \WHILE{$\mathit{queue}$ is not empty}
  \STATE $a := \mathit{queue}.\mathit{pop}()$
  \IF{$a.\inupper =$ false and $a.\inneg = $ false}
   \FOR{each rule $r\in a.\poslist$}
    \STATE $r.\propagateTrue()$
   \ENDFOR
   \STATE $a.\inupper :=$ true
  \ENDIF
 \ENDWHILE.
\end{algorithmic}
\end{algorithm}

\begin{algorithm}
\caption{Auxiliary functions for the basic rules}
\label{alg:upperaux}
\begin{algorithmic}
\item[\textbf{function} $r.\propagateFalse()$]
 \STATE $r.\upper := r.\upper + 1$
 \IF{$r.\upper = 1$ and $r.\inactive = 0$}
  \STATE $\mathit{queue}.\mathit{push(r.\head)}$
 \ENDIF.
\medskip
\item[\textbf{function} $r.\propagateTrue()$]
 \STATE $r.\upper := r.\upper - 1$
 \IF{$r.\upper = 0$ and $r.\inactive = 0$}
  \STATE $\mathit{queue}.\mathit{push(r.\head)}$
 \ENDIF.
\medskip
\item[\textbf{function} $r.\isUpperActive()$]
 \IF{$r.\upper = 0$ and $r.\inactive = 0$}
  \STATE return true
 \ELSE
  \STATE return false
 \ENDIF.
\end{algorithmic}
\end{algorithm}

We assume that the queue $\mathit{queue}$ is implemented such that
pushing an atom onto the end of it does nothing if the atom is already
on the queue. The $\inactive$ counters are used by the auxiliary
functions and the $\inneg$ flag is used by the main procedure. For a
rule $r$, $r.\inactive > 0$ if
\[\mathit{max}_r'(A) = \negthickspace
\bigcup_{\substack{\patoms{A}\subseteq C \\
\natoms{A}\cap C=\emptyset}} \negthickspace f_r'(C,C)\]
is the empty set, and for an atom $a$, $a.\inneg$ is true if
$a\in\natoms{A}$. In addition, we assume that the flag $a.\inupper$ is
set to false whenever $a.\inneg$ is true.

The first and largest upper closure, $\atmost{P,\emptyset}$, can be
computed by the second stage if the queue is initialized with the
atoms that follow immediately from some rules. These are the atoms in
\[\bigcup_{r\in P} f_r'(\emptyset,\emptyset).\]

The problem that remains is initializing the queue such that the first
stage removes sufficiently many atoms from $\atmost{P,A}$. We must
initialize the queue with the atoms in the heads of the rules that
potentially become inactive when $A$ changes to $A'$. Recall that 
$\atmost{P,A}$ is the least fixed point of the operator
\[f'(B) = \bigcup_{r\in P} f_r'(\patoms{A},B-\natoms{A})-\natoms{A}.\]
For $A\subseteq A'$, $C = \atmost{P,A}$ and for every basic or choice 
rule $r$ we initialize the queue with the atoms in
\[f_r'(\patoms{A},C-\natoms{A})-f_r'(\patoms{A'},C-\natoms{A'})-\natoms{A'}.\]
For every cardinality and weight rule $r$ we initialize the queue with
the atoms in
\[f_r'(\patoms{A},C-\natoms{A})-f_r'(\patoms{A'},\emptyset)-\natoms{A'}.\]
The reason behind the more lax condition for cardinality and weight
rules is best illustrated by an example.
\begin{example}
Let $P$ be the program consisting of the rule $r$
\[a \from 1\,\{ a, \nB{b} \}.\]
Then, $C = \atmost{P,\emptyset} = \{a\}$ and
$\atmost{P,\{b\}} = \emptyset$. However, 
\[f_r'(\patoms{\{b\}},C-\natoms{\{b\}})-\natoms{\{b\}} = 
f_r'(\{b\},\{a\}) = \{a\}.\]
Hence, for $A=\emptyset$ and for $A'=\{b\}$,
\[f_r'(\patoms{A},C-\natoms{A})-f_r'(\patoms{A'},C-\natoms{A'})-\natoms{A'} 
= \emptyset\] and we miss an inactive rule if we assume that the atoms
in $C$ remain in the closure.
\end{example}

\begin{figure}
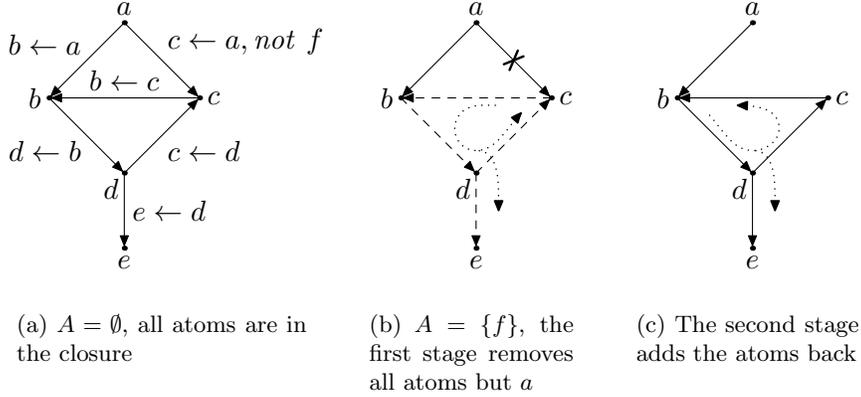

\centering
\subfigure[{\protect $A=\emptyset$, all atoms are in the closure}]{
\label{fig:atmost:a}
\begin{minipage}[b]{.35\textwidth}
\centering
\includegraphics{figatmost.1}
\end{minipage}}
\subfigure[{\protect $A=\{f\}$}, the first stage removes all atoms
but $a$]{
\label{fig:atmost:b}
\begin{minipage}[b]{.26\textwidth}
\centering
\includegraphics{figatmost.2}
\end{minipage}}
\subfigure[The second stage adds the atoms back]{
\label{fig:atmost:c}
\begin{minipage}[b]{.28\textwidth}
\centering
\includegraphics{figatmost.3}
\end{minipage}}
\caption{A visualization of the computation of {\protect $\atmost{P,A}$}}
\label{fig:atmost}
\end{figure}

\begin{example}
An illustration of the two stages of $\patmost()$ is shown in
Figure~\ref{fig:atmost}. The program $P$
\begin{align*}
b &\from a & c &\from a,\nB{f} \\
b &\from c & c &\from d \\
d &\from b & e &\from d \\
a &\from
\end{align*}
is displayed as a graph whose nodes are the atoms and whose edges are
the rules that partake in the computation. The initial state
is shown in~\ref{fig:atmost:a} and corresponds to
$\atmost{P,\emptyset} = \{a,b,c,d,e\}$. In~\ref{fig:atmost:b} the atom
$f$ is added to $A=\emptyset$ and the rule $c\from a,\nB{f}$ becomes
inactive. The first stage then removes the atoms $c$, $b$, $d$, and
$e$. The second stage notices that $b$ still follows from $b\from a$
and adds the atoms $b$, $d$, $e$, and $c$ back into the closure. The
end result $\atmost{P,\{f\}} = \{a,b,c,d,e\}$ is shown
in~\ref{fig:atmost:c}.
\end{example}

\begin{proposition}
  The procedure $\patmost()$ computes $\atmost{P,A}$ in time linear
  in the size of the program $P$.   
\end{proposition}

\begin{proof}
  Note that for every atom $a$, the list $a.\poslist$ is traversed at
  most twice and the lists $a.\headlist$ is traversed at most once.
\end{proof}

\subsection{Various Optimizations}

It is possible to optimize the $\smodels$ algorithm in various
ways. For example, if the algorithm has deduced that an atom can not
be part of any model, then one can remove the atom from the bodies of
the rules in which it appears. The bodies shorten and any following
traversal of them is faster. Similarly, an inactive rule can safely
be removed from the lists of the atoms. A reduction of this type leads 
only to a small performance improvement. Hence, reducing a program by
removing rules and atoms is best done just before the heuristic is
computed for the first time, as one then need not undo the changes
later.

The heuristic can in principle return any atom in $\Atoms{P}$ for a
program $P$. Since $\smodels(P,A)$ calls itself twice for every choice
point that the heuristic finds, the algorithm explores at worst a
search space of size $2^{\abs{\Atoms{P}}}$. Restricting the choice
points to a smaller set could substantially improve the algorithm.

The procedure $\patmost()$ will in the worst case remove all atoms
from the upper closure before adding them back again. If we could
localize the computation of $\patmost()$ to small parts of the
program, then we could avoid the worst case.

Below, we identify the choice points that can be ignored and present
two methods for localizing the computation of $\atmost{P,A}$.

\subsubsection*{Reducing the Search Space}
\addcontentsline{toc}{subsubsection}{\quad Reducing the Search Space}

Fix a program $P$. Let $B$ be the set of atoms that appear as
not-atoms in the bodies of the rules in $P$ or in the heads of the choice 
rules in $P$. Then, for any rule $r\in P$ and for any sets of atoms
$S, C\subseteq\Atoms{P}$,
\[f_r(S\cap B,C) = f_r(S,C)\]
by the definition of $f_r$. Hence, there is a one-to-one
correspondence between the stable models and their intersections with
$B$. In other words, we do not have to consider choice points that are
not in $B$.

The program $P$ defines a directed graph $G=(V,E)$ where the vertices
is the set $V=\Atoms{P}\cup P$ and where the edges is the set
\newcommand{\pair}[1]{\langle #1\rangle}
\[E = \{\pair{r,a} \mid a\in r.\head\}
\cup\{\pair{a,r} \mid \text{$a\in r.\patoms{body}$ or
$a\in r.\natoms{body}$}\}.\]
We say that the atom $a$ appears as a not-atom in a cycle in the graph 
if the edge $\pair{a,r}$ is part of the cycle and $a\in r.\natoms{body}$.

Let \[P_a = \{r \mid \text{there is a directed path from $r$ to $a$ in 
$G$}\}.\]
If $a$ does not appear as a not-atom in a cycle and if $a$ does not
appear in the head of a choice rule, then
$g_{P_a}(S) = g_{P_a}(S-\{a\})$ for all $S$. Since
\[g_{P_a}(S)\subseteq g_P(S)\]
and since $a\in g_P(S)$ implies $a\in g_{P_a}(S)$ for any $S$,
$a\in S=g_P(S)$ for a stable model $S$ if and only if $a\in
g_{P_a}(S)$. Thus, the choice points can be restricted to the set
$\Atoms{P}-\{a\}$. Furthermore, we can combine all restrictions as
$P_a\subseteq P_b$ if there is a path from $a$ to $b$. Therefore, the
set of choice points $B$ can be taken to be the atoms that appear as
not-atoms in some cycles or that appear in the heads of some choice
rules.

\begin{proposition}
  Let $P$ be a program, let $B$ be the set of atoms that appear as
  not-atoms in some cycles or that appear in the heads of some choice
  rules. Let $S$ and $S'$ be stable models of $P$. If
  $S\cap B=S'\cap B$, then $S=S'$.
\end{proposition}

\begin{proof}
  Let $B$ be the set of atoms that appear as not-atoms in some cycles
  or that appear in the heads of some choice rules of $P$. Define
  $B_0=B$ and
  \begin{multline*}
    B_{i+1}=B_i\cup\{a\in\Atoms{P}-B_i\mid
    \text{ for every $b\in\Atoms{P}-B_i$,} \\
    \text{if there is a path from $b$ to $a$, then there is a path
      from $a$ to $b$}\}.
  \end{multline*}
  If $a\in B_{i+1}-B_i$, then $a$ does not appear in the head of a
  choice rule nor does it appear as a not-atom in
  \[P_{i+1}=\bigcup_{b\in B_{i+1}-B_i}P_b.\]
  Hence, \[g_{P_{i+1}}(S\cap B_i)=g_{P_{i+1}}(S\cap B_{i+1})\]
  for any $S$. Now, for any $a\in B_{i+1}-B_i$ and for any $S$,
  $a\in g_P(S)$ if and only if $a\in g_{P_{i+1}}(S)$.

  Let $S$ and $S'$ be two stable models of $P$ for which
  $S\cap B_i=S'\cap B_i$ holds. If $a\in B_{i+1}-B_i$, then
  $a\in S=g_P(S)$ implies
  \begin{multline*}
    a\in g_{P_{i+1}}(S) = g_{P_{i+1}}(S\cap B_{i+1}) =
    g_{P_{i+1}}(S\cap B_i) = g_{P_{i+1}}(S'\cap B_i) \\
    = g_{P_{i+1}}(S'\cap B_{i+1}) = g_{P_{i+1}}(S')
    \subseteq g_P(S') = S'.
  \end{multline*}
  Thus, $S\cap B_{i+1}=S'\cap B_{i+1}$ by symmetry. Consequently,
  $S\cap B=S'\cap B$ implies $S=S'$.
\end{proof}

We need a stronger result than that the stable model semantics
makes the atoms in the set $B$ define the stable models. We
want $\smodels$ to stop searching as soon as $B$ is covered. This
holds if we use both $\atmost{P,A}$ and $\atleast{P,A}$ in $\expand$.

\begin{proposition}
  Let $P$ be a program, let $B$ be the set of atoms that appear as
  not-atoms in some cycles or that appear in the heads of some choice
  rules of $P$. Let $A$ be a set of literals such that
  $\Atoms{A}=B$. Then, $\expand(P,A)$ covers $\Atoms{P}$.
\end{proposition}

\begin{proof}
  Let $C=\expand(P,A)$. Assume that $\Atoms{C}\subset\Atoms{P}$.
  Then, there exists an atom $a\in\atmost{P,C}-\patoms{C}$
  and a rule $r\in P$ that is not a choice rule such that
  \[a\in f_r'(\patoms{C},\patoms{C}-\natoms{C})-\natoms{C}=
  f_r(\patoms{C},\patoms{C}).\]
  Since otherwise
  $f_r'(\patoms{C},\patoms{C}-\natoms{C})-\natoms{C}\subseteq\patoms{C}$,
  $f'(\patoms{C})\subseteq\patoms{C}$, and consequently
  $\atmost{P,C}\subseteq\patoms{C}$ by Lemma~\ref{lemma:monotone}.

  As the atoms in $\Atoms{P}-\Atoms{C}$ can not appear as not-atoms in
  the body of $r$, \[f_r(\patoms{C},\patoms{C}) = f_r(C',\patoms{C})\]
  for $\patoms{C}\subseteq C'$ and $\natoms{C}\cap C'=\emptyset$.
  Hence, $f_r(\patoms{C},\patoms{C}) = \mathit{min}_r(C)$
  since $f_r$ is monotonic in its second argument. But then,
  \[a\in\atleast{P,C}\subseteq C\] which is a contradiction. Thus, $C$ 
  covers $\Atoms{P}$.
\end{proof}

It is easy to make the $\smodels$ algorithm ignore choice points.
We just change the cost function of the heuristic. Recall that the
heuristic searches for the literal $x$ that minimizes $2^{-p}+2^{-n}$
for $p = \abs{A_p-A}$ and $n = \abs{A_n-A}$, where
\[A_p = \expand(P,A\cup\{x\}) \quad\text{and}\quad
A_n = \expand\bigl(P,A\cup\{\nb{x}\}\bigr).\]
To limit the relevant choice points to $B$ we simply redefine
$p$ and $n$ as
\[p = \abs{(A_p-A)\cap B} \quad\text{and}\quad
n = \abs{(A_n-A)\cap B}.\]

Notice that we do not really restrict the choice points to
$B$. Instead, we guarantee that if $\smodels$ branches on a choice
point, then at least one literal in $B$ will follow when we prune the
search space. Thus, we retain the greatest possible freedom in
choosing choice points and we can still be certain that the size of
the search space stays below $2^{\abs{B}}$.

\begin{example}
Let $P$ be the program
\begin{align*}
a &\from \nB{b} & b &\from c \\
c &\from \nB{a} & d &\from \nB{c}.
\end{align*}
Since $d$ does not appear as a not-atom in the program, it is not a
choice point. Since $c$ does not appear as a not-atom in a cycle, it is
not a choice point either. Thus, the set of choice points of $P$ are
$B=\{a,b\}$.
\end{example}

\subsubsection*{Strongly Connected Components}
\addcontentsline{toc}{subsubsection}{\quad Strongly Connected Components}

Consider the computation of $\atmost{P,A'}$ that starts from
$\atmost{P,A}$. The function $\patmost()$ operates on the graph
$G=(V,E)$ given by $V=\Atoms{P}$ and \[E = \{\pair{a,b} \mid
\text{there is a rule $r$ such that $a\in r.\patoms{body}$ and
$b\in r.\head$}\}.\] 
Take an atom $a\in\atmost{P,A}$. Observe that the only atoms that
determine whether $a$ is removed from the upper closure
during the first stage of $\patmost()$ are the atoms that have
a directed path to $a$ in $G$. Hence, we can localize $\patmost()$ by
computing both stages inside a strongly connected component of $G$
before moving on to the rest of the graph. A strongly connected
component\index{strongly connected component} of a directed graph is a
set of vertices such that there is a directed path from any vertex in
the component to any other.

In practice, this is done by removing rules from the list $a.\poslist$
for every atom $a$. Specifically, if $r\in a.\poslist$ and if $r.head$
is not in the same strongly connected component as $a$, then we remove 
$r$ from $a.\poslist$. Here we assume that $\patmost()$ and
$\patleast()$ use their own versions of $a.\poslist$.

One $\patmost()$ invocation will only compute a superset of the new
upper closure after this change. If the atom $a$ is removed, then we
know that $a.\inneg$ can be set to true. Setting $a.\inneg$ to true in
turn, influences the $\inactive$ counters of some rules. Therefore, we
can possibly initialize the queue with some atoms. If we call
$\patmost()$ every time we initialize the queue with new atoms, we
eventually remove everything not in $\atmost{P,A'}$.

Notice that we do not explicitly keep track of the strongly connected
components nor the acyclic graph they induce. Hence, the components
are not processed in any fixed order even if the order given by the
acyclic graph would be favorable. Changing the queue in $\patmost()$
into a priority queue would solve this problem. Also notice that if
there are no strongly connected components, then we need not compute
$\patmost()$.

\begin{example}
A partial program is shown in Figure~\ref{fig:scc:a}. The
corresponding graph is shown in Figure~\ref{fig:scc:b}. If we assume
that $a$ and $b$ are in the upper closure then the upper closure is
the set $\{a,b,c,d,e,f,g,h\}$. If we remove $a$, then the first stage
of $\patmost()$ removes all atoms except $b$ and the second stage adds
them back.

Removing the rules that leave a strongly connected component results
in the graph in Figure~\ref{fig:scc:c}. The first stage of $\patmost()$
now removes only the atoms $c$, $d$, and $e$.
\end{example}

\newsavebox{\progbox}
\savebox{\progbox}{%
\begin{minipage}[b]{.28\textwidth}
\centering
\begin{align*}
c &\from e \\
d &\from a \\
d &\from c \\
e &\from b \\
e &\from d \\
f &\from d \\
f &\from e \\
f &\from h \\
g &\from f \\
h &\from g
\end{align*}
\end{minipage}}

\begin{figure}
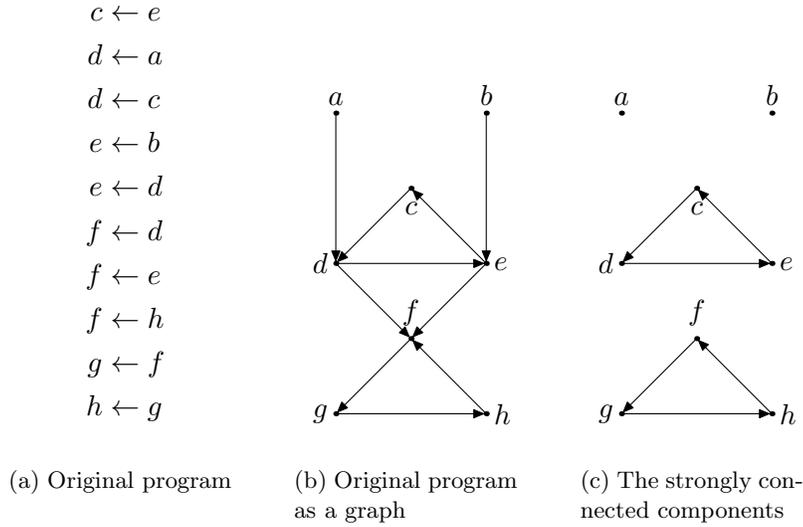

\centering
\subfigure[Original program]{
\label{fig:scc:a}
\usebox{\progbox}}
\subfigure[Original program as a graph]{
\label{fig:scc:b}
\begin{minipage}[b]{.28\textwidth}
\centering
\includegraphics{figscc.1}
\end{minipage}}
\subfigure[The strongly connected components]{
\label{fig:scc:c}
\begin{minipage}[b]{.28\textwidth}
\centering
\includegraphics{figscc.2}
\end{minipage}}
\caption{Localizing {\protect $\atmost{P,A}$} with strongly connected
components}
\label{fig:scc}
\end{figure}

\subsubsection*{The Source Pointer}
\addcontentsline{toc}{subsubsection}{\quad The Source Pointer}

Partitioning a program into strongly connected components localizes
the upper closure computation to a component if the component remains
in the upper closure. If the component is removed, then the
computation spreads to other components, perhaps in vain.

\newcommand{\source}{\mathit{source}}

We can partly overcome this problem with the help of the following
construct. For every atom $a$, we create a source pointer\index{source
  pointer} $a.\source$ whose mission is to point to the first rule
that causes $a$ to be included in the upper closure. During the first
stage of $\patmost()$, it suffices to only remove atoms which are to
be removed due to a rule in a source pointer. For if the rule in a
source pointer does not justify the removal of an atom, then the atom
is reentered into the closure in the second stage of the computation.
The source pointer is implemented by modifying the auxiliary functions
$\propagateFalse()$ and $\propagateTrue()$ as can be seen in
Algorithm~\ref{alg:sourceupperaux}.

\begin{algorithm}[t]
\caption{Auxiliary functions and the source pointer}
\label{alg:sourceupperaux}
\begin{algorithmic}
\item[\textbf{function} $r.\propagateFalse()$]
 \STATE $r.\upper := r.\upper + 1$
 \IF{$r.\upper = 1$ and $r.\inactive = 0$ and \\
     \quad $(r.\head.\source = 0$ or $r.\head.\source = r)$}
  \STATE $r.\head.\source := 0$
  \STATE $\mathit{queue}.\mathit{push(r.\head)}$
 \ENDIF.
\medskip
\item[\textbf{function} $r.\propagateTrue()$]
 \STATE $r.\upper := r.\upper - 1$
 \IF{$r.\upper = 0$ and $r.\inactive = 0$}
  \IF{$r.\head.\source = 0$}
   \STATE $r.\head.\source := r$
  \ENDIF
  \STATE $\mathit{queue}.\mathit{push(r.\head)}$
 \ENDIF.
\end{algorithmic}
\end{algorithm}

\begin{figure}
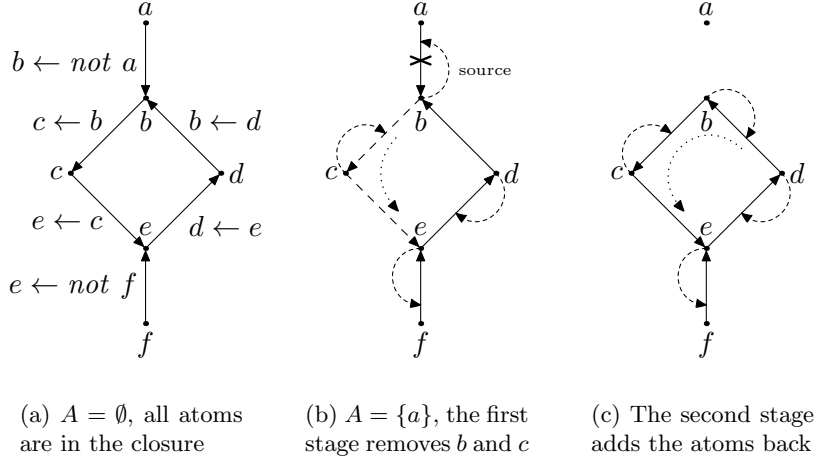

\centering
\subfigure[{\protect $A=\emptyset$}, all atoms are in the closure]{
\label{fig:source:a}
\begin{minipage}[b]{.28\textwidth}
\centering
\includegraphics{figsource.1}
\end{minipage}}
\subfigure[{\protect $A=\{a\}$, the first stage removes $b$ and $c$}]{
\label{fig:source:b}
\begin{minipage}[b]{.28\textwidth}
\centering
\includegraphics{figsource.2}
\end{minipage}}
\subfigure[The second stage adds the atoms back]{
\label{fig:source:c}
\begin{minipage}[b]{.28\textwidth}
\centering
\includegraphics{figsource.3}
\end{minipage}}
\caption{Localizing {\protect $\atmost{P,A}$} with source pointers}
\label{fig:source}
\end{figure}

\begin{example}
Let $P$ be the program
\begin{align*}
b &\from \nB{a} & b &\from d \\
c &\from b & d &\from e \\
e &\from c & e &\from \nB{f}.
\end{align*}
Three pictures illustrating the source pointers and the two stages of
$\patmost()$ are shown in Figure~\ref{fig:source}. The initial state
is shown in~\ref{fig:source:a} and corresponds to
$\atmost{P,\emptyset} = \{b,c,d,e\}$. In~\ref{fig:source:b} the atom
$a$ is added to $A=\emptyset$ and the rule $b\from \nB{a}$ becomes
inactive. The first stage then removes the atom $b$, as $b.\source$
points at the rule $b\from \nB{a}$, and the atom $c$, as $c.\source$
points at $c\from b$. The second stage notices that $b$ still
follows from $b\from d$ and adds the atoms $b$ and $c$ back into the
closure. The source pointer $b.\source$ is at the same time updated to 
point at $b\from d$. The end result $\atmost{P,\{a\}} = \{b,c,d,e\}$
is shown in~\ref{fig:source:c}.
\end{example}

\subsection{Backtracking}
\label{subsec:backtracking}

The $\smodels(P,A)$ procedure employs chronological backtracking. The
last literal that is included in a partial stable model is the first
that is removed when a conflict is detected. Since we want a
linear-space implementation of $\smodels(P,A)$, we can not store
snapshots of the states of the data structures. We must instead store
the changes that take place.

It is enough to keep track of the changes to the set $A$. For example,
if the atom $a$ is added to $A$, i.e., if the flag $a.\inpos$ is
set to true, then for every basic rule $r$ in $a.\poslist$ the counter 
$r.\literal$ is decremented by one. Hence, if $a$ is removed
from $A$, then $r.\literal$ should be incremented. Generally,
the value of every counter is completely determined from its previous
value, and the previous value is completely determined by its new
value. Hence, if we during backtracking undo the changes to $A$ in
the opposite order to which they took place, then we can directly
compute the old values of the counters.

Consequently, backtracking can be implemented with the help of a stack
of size $\Atoms{P}$. Every time a literal is added to $A$ it is also
pushed onto the stack. We backtrack by popping literals off the stack
and by computing the correct values for the relevant counters.

\subsubsection*{Backjumping}

Lookahead guarantees that conflicts are caused by the chronologically
next to last choice. Assume that the literal $x_1$ is the next to last
choice, as given by the heuristic. Furthermore, assume that
$\conflict(P,A')$ returns false for $A' = \expand(P,A\cup\{x_1\})$,
but that the choice of $x_1$ does not affect the conflict that arises
when the literal $x_2$ is added to $A'$. That is to say,
$\conflict(P,A'')$ returns true for $A''=\expand(P,A\cup\{x_2\})$.
Since $\lookahead$ examines all possible ways of adding a literal to $A$,
it notices this conflict before $x_1$ is chosen. Hence, the literal
$x_1$ is not the next to last choice, and by the same argument,
neither is any other literal not related to the conflict.

Despite the fact that many conflicts are discovered by the
$\lookahead$ function, there are still situations where the decision
procedure exhaustively searches through assignments that are not
relevant to a set of conflicts. A simple example is given by the union
of two programs that do not share atoms. We assume that the first
program has several stable models and that the second one has none. If
we always begin by trying atoms from the first program, then we will
search through all stable models of the first program before we
discover that the joint program has no stable models.

The excessive backtracking in the preceding example is caused by the
fact that $\smodels(P,A)$ does not notice that the conflicts in one
subprogram are independent from the truth-assignments in the
other. Fortunately, it is easy to improve $\smodels$.

\newcommand{\independent}{\mathit{independent}}

Let
$\independent(P,A,x_1,x_2)$\index{$independent$@$\independent(P,A,x_1,x_2)$}
be a function whose arguments are a program $P$, a set of literals $A$,
and two literals $x_1$ and $x_2$. Let
\begin{align*}
A_1 &= \expand(P,A\cup\{x_1\}), \\
A_2 &= \expand(P,A\cup\{x_2\}), \\
\intertext{and}
A_3 &= \expand(P,A\cup\{x_1,x_2\}).
\end{align*}
We assume that $\independent$ fulfills the two conditions
\begin{description}
\item[I1] if $\independent(P,A,x_1,x_2)$ returns true, if
$\conflict(P,A_2)$ returns false, and if $\conflict(P,A_3)$ returns
true, then $\conflict(P,A_1)$ returns true, and
\item[I2] if $\independent(P,A,x_1,x_2)$ returns true and if
$A$ is a subset of $A'$, then $\independent(P,A',x_1,x_2)$ returns
true.
\end{description}
We say that $x_1$ is independent from $x_2$ if
$\independent(P,A,x_1,x_2)$ returns true.

If $x_1$ is independent from $x_2$, if $x_2$ is just before $x_1$
on the stack, and if $x_1$ gives rise to a conflict, then one can
remove both $x_1$ and $x_2$ from the stack before pushing $\nb{x_1}$
onto it without loosing any stable models. Thus, we can skip some
literals while backtracking. We call this
backjumping\index{backjumping}.

Consider the generalized $\smodels$ procedure that only accepts a
stable model for which $\stable(P,A)$ returns true. Since we can
backtrack from a stable model and since we have to guarantee the
completeness of the decision procedure, we have to distinguish between
two modes of backtracking. Specifically, backtracking from a stable
model must be chronological while backtracking from a conflict may
take advantage of backjumping.

A decision procedure that incorporates backjumping is presented in
Algorithm~\ref{alg:backjumping}. The algorithm assumes the existence
of two global variables: $\mathit{conflict}$ and $\mathit{top}$. The
variable $\mathit{conflict}$ holds the last choice leading up to a
conflict, while the variable $\mathit{top}$ keeps the level, or depth
of recursion, above which backtracking is chronological.

\begin{algorithm}
\caption{Incorporating backjumping}
\label{alg:backjumping}
\begin{algorithmic}
\item[\textbf{function} $\smodels(P,A,\mathit{level})$]
  \index{$smodels$@$\smodels(P,A)$}
 \STATE $A:=\expand(P,A)$
 \STATE $A:=\lookahead(P,A)$
 \IF{$\conflict(P,A)$}
  \STATE return false
 \ELSIF{$A$ covers $\Atoms{P}$}
  \STATE $\mathit{top} := \mathit{level}$
  \STATE return $\stable(P,A)$ \COMMENT{$\patoms{A}$ is a stable model}
 \ELSE
  \STATE $x:=\heuristic(P,A)$
  \STATE $\mathit{conflict} := x$
  \IF{$\smodels(P,A\cup \{x\},\mathit{level}+1)$}
   \STATE return true
  \ELSIF{$\mathit{level}\geq\mathit{top}$ and
         $\independent(P,A,\mathit{conflict},x)$}
   \STATE return false
  \ELSE
   \IF{$\mathit{level} < \mathit{top}$}
    \STATE $\mathit{top} := \mathit{level}$
   \ENDIF
   \STATE return $\smodels\bigl(P,A\cup\{\nb{x}\},\mathit{level}+1\bigr)$
  \ENDIF
 \ENDIF.
\end{algorithmic}
\end{algorithm}

The soundness of $\smodels$ is obviously not affected by backjumping
since backjumping only prevents the algorithm from exploring certain
parts of the search space. Completeness, on the other hand, must be
proved.

\begin{theorem}
Let $P$ be a set of rules and let $A$ be a set of literals. If
there is a stable model $S$ of $P$ agreeing with $A$ such
that $\stable(P,S')$ returns true for $S'=S\cup\Nb{\Atoms{P}-S}$, then 
$\smodels(P,B,\mathit{top})$ returns true.
\end{theorem}

\begin{proof}
We show that completeness is not lost when backjumping is introduced.
Since backtracking is chronological, and therefore exhaustive, for a
depth of recursion smaller than $\mathit{top}$, we only have to consider
the case $\mathit{level} \geq \mathit{top}$.

Assume that there is no stable model agreeing with $A\cup\{x_0\}$ and
that when $\smodels(P,A\cup\{x_0\},\mathit{level}+1)$ returns false,
\[\independent(P,A,\mathit{conflict},x_0)\] returns true. Furthermore,
assume that $S$ is a stable model agreeing with the set
$A\cup\{\nb{x_0},x_1,\dotsc,x_n\}$ and that the set covers
$\Atoms{P}$. Then, $\conflict(P,A')$ returns true for
\[A'=\expand(P,A\cup\{x_0,x_1,\dotsc,x_n\}),\] since there is no stable
model agreeing with $A'$. Take $i$ such that $\mathit{conflict} = x_i$ 
or $\mathit{conflict} = \nb{x_i}$, and let $j$ be the smallest number
for which
\[\conflict\bigl(P,\expand(P,A\cup\{x_i,x_0,x_1,\dotsc,x_j\})\bigr)\]
returns true and
$\conflict\bigl(P,\expand(P,A\cup\{x_0,x_1,\dotsc,x_j\})\bigr)$
returns false. Then,
$\conflict\bigl(P,\expand(P,A\cup\{x_i,x_1,\dotsc,x_j\})\bigr)$
returns true by I1, I2, and E1. Hence, by C2 there is no stable model
that agrees with $A\cup\{\nb{x_0},x_1,\dotsc,x_n\}$ and this
contradicts the existence of $S$. Thus, there is no stable model
agreeing with $A\cup\{\nb{x_0}\}$.

We complete the proof by induction on the size of
\[\mathit{nc}(P,A\cup\{x\}) = \Atoms{P}-\Atoms{A\cup\{x\}}.\]
If $A\cup\{x\}$ covers $\Atoms{P}$, if the function
$\smodels(P,A\cup\{x\},\mathit{level}+1)$ returns false and if
$\mathit{level}\geq \mathit{top}$,
then there is no stable model agreeing with $A\cup\{x\}$. Let
$\mathit{nc}(P,A\cup\{x\})\neq\emptyset$, let
$\smodels(P,A\cup\{x\},\mathit{level}+1)$ return false and let
$\mathit{level}\geq \mathit{top}$. Then,
$\smodels(P,A'\cup\{x'\},\mathit{level}+2)$ returns false for
$A'=\expand(P,A\cup\{x\})$ and $x'\in \mathit{nc}(P,A')$, and by
induction there is no stable model agreeing with $A'\cup\{x'\}$. If
$\independent(P,A',\mathit{conflict},x')$ returns true, then
by the above there is no stable model agreeing with
$A'\cup\{\nb{x'}\}$ either. If, on the other hand,
$\independent(P,A',\mathit{conflict},x')$ returns false, then 
$\smodels\bigl(P,A'\cup\{\nb{x'}\},\mathit{level}+2\bigr)$ returns
false and there is again by induction no stable model that agrees with
$A'\cup\{\nb{x'}\}$. Therefore, there is no stable model that agrees
with $A\cup\{x\}$.
\end{proof}

Let the undirected graph $G=(V,E)$ of a program $P$ be defined by
$V=\Atoms{P}$ and
\[E = \{\pair{a,b}\mid a,b\in r.\head\cup\patoms{r.\body}\cup
\natoms{r.\body} \text{ for some $r\in P$}\}.\]
A simple but easy to compute instance of $\independent(P,A,x_1,x_2)$ 
returns true exactly when there is no path in $G$ between the atoms
that $x_1$ and $x_2$ cover. Since $\expand$ can not propagate any
truth values via any inactive rules, we can safely remove these rules
from $G$. Similarly, any atom in $\natoms{A}$ can also be
removed. However, we must not remove any atoms in $\patoms{A}$, as
$\patmost()$ examines them during its computation. An example of what
can go wrong if the atoms in $\patoms{A}$ are removed from $G$ is
given in the next example.

\begin{example}
Let $P$ be the program
\begin{align*}
a &\from b & b &\from a \\
a &\from c & b &\from d,\nB{d} \\
\{c\} &\from & \{d\} &\from
\end{align*}
and let $A=\{a,b\}$. If we remove the literals in $\patoms{A}$ from
the graph, then $d$ and $c$ are not connected. Hence, by first
choosing $\nB{c}$ and then $\nB{d}$ we get a conflict, and when we try
$d$ we get another conflict. Since $d$ is not connected to $c$, we
backjump past $c$ thereby missing the stable models $\{a,b,c\}$ and
$\{a,b,c,d\}$.
\end{example}

\cleardoublepage
\section{Complexity}
\label{sec:complexity}

The $\smodels$ procedure solves an NP-complete problem. It returns
true if and only if a program has a stable model. By
Proposition~\ref{prop:stable}, we can test whether a set of atoms is a
stable model in polynomial time.

\begin{theorem}
  Deciding whether a set of basic, choice, cardinality, and weight
  rules has a stable model is NP-complete.
\end{theorem}

Unsurprisingly, the $\smodels$ procedure has a worst-case time
complexity that is exponential in the number of atoms of a
program. The interesting thing about exponential computations is that
one can make them faster asymptotically by trading an arbitrary amount
of polynomial work for a tiny reduction of the exponent. For instance, 
for any $k$ \[n^k2^{\frac{(k-1)n}{k}} < 2^n\] for large $n$.

We get exponential behavior from $\smodels$ if we try to solve the
pigeon-hole problem. This is the problem of trying to stuff $n$
pigeons into $k$ holes such that there is at most one pigeon per
hole. Clearly, there is no solution if $n>k$. Let the atom $p_{i,j}$
be true when pigeon number $i$ is in hole number $j$. Then, the
pigeon-hole problem is encoded by the program $P$
\begin{gather*}
\begin{split}
\{ p_{i,1},\dotsc,p_{i,k} \} &\from \\
\mathit{false} &\from 2\,\{p_{i,1},\dotsc,p_{i,k}\} \\
\mathit{false} &\from \nB{p_{i,1}},\dotsc,\nB{p_{i,k}} \\
\mathit{false} &\from 2\,\{p_{1,j},\dotsc,p_{n,j}\}
\end{split} \\
\mathit{compute}\,\{\nB{\mathit{false}}\}
\end{gather*}
for $1\leq i \leq n$ and $1\leq j \leq k$.

Let $A$ be a set of literals having $\nB{\mathit{false}}$ as a
member. Since the program does not contain any positive loops,
$\expand(P,A) = \atleast{P,A}$. We will examine the
behavior of $\atleast{P,A}$. Notice that if
$p_{i,j}\in A$, then
\[\nB{p_{1,j}},\dotsc,\nB{p_{i-1,j}},\nB{p_{i+1,j}},\dotsc,\nB{p_{n,j}}\in
\atleast{P,A}\]
and
\[\nB{p_{i,1}},\dotsc,\nB{p_{i,j-1}},\nB{p_{i,j+1}},\dotsc,\nB{p_{i,k}}\in
\atleast{P,A}.\]
Furthermore, if
\[\nB{p_{i,1}},\dotsc,\nB{p_{i,j-1}},\nB{p_{i,j+1}},\dotsc,\nB{p_{i,k}}\in 
A,\]
then $p_{i,j}\in \atleast{P,A}$, and there is no other way of deducing 
additional literals in $\atleast{P,A}$.

Denote by $P_k$ the encoding of the pigeon-hole problem when
$n = k+1 > 3$. Then, $\smodels(P_k,\emptyset)$ will call itself
recursively at least $2^{k-2}$ times. By inspection, we
see that $\smodels(P_k,\{p_{i,j}\})$ and
$\smodels(P_k,\{\nB{p_{i,j}}\})$ are called for some $i$ and $j$ when
$k=3$. For the inductive case notice that if
$A\cap\{p_{1,k},\dotsc,p_{n,k}\} = \emptyset$, then
$\atleast{P_k,A\cup\{p_{n,k}\}}$ is equal to $\atleast{P_{k-1},A}$
modulo all atoms that handle pigeon $n$ and hole $k$. Furthermore,
under the same assumption $\lookahead(P_k,A\cup\{\nB{p_{n,k}}\})$ does
not find any conflicts if $\lookahead(P_{k-1},A)$ does not. In
addition, one can take any atom $p_{i,j}$ instead of $p_{n,k}$ as we
can freely rename atoms. Since
$\atleast{P_k,A} \subseteq \atleast{P_{k-1},A}$
for any $A$ for which $A=\atleast{P_{k-1},A}$, it immediately follows
that if $\smodels(P_{k-1},\emptyset)$ calls itself recursively at
least $2^{(k-1)-2}$ times, then $\smodels(P_k,\emptyset)$ calls itself
recursively at least $2^{k-2}$ times.

Since $\smodels(P,A)$ solves an NP-complete problem it can also solve
coNP-complete problems. In fact, since it can go through all stable
models of a logic program, it can solve $\Delta_2^p$-complete problems 
such as deciding the lexicographically largest stable model.

\begin{proposition}
  Let $L=\{P\mid P$ has a stable model and the lexicographically
  largest one contains the least significant atom$\}$. Then, $L$ is
  $\Delta_2^p$-complete.
\end{proposition}

\begin{proof}
  Given $P$ we can check whether $P\in L$ using the oracle
  \[O=\{\pair{P,A}\mid \text{there exist a stable model of $P$ that
    agrees with $A$}\}\] a linear number of time. Namely, the language
  $L$ is decided by the function
  \begin{algorithmic}
  \item[\textbf{function} $\mathit{isinL}(P)$]
    \STATE $A=\emptyset$
    \FOR{each atom $a$ in $\Atoms{P}$ in order, most significant
      first}
     \IF{there exists a stable model of $P$ that agrees with
       $A\cup\{a\}$}
      \STATE $A:=A\cup \{a\}$
     \ELSE
      \STATE $A:=A\cup \{\nB{a}\}$
     \ENDIF
    \ENDFOR
    \IF{the least significant atom is in $A$}
     \STATE return true
    \ELSE
     \STATE return false
    \ENDIF.
  \end{algorithmic}
  Hence, $L\in\Delta_2^p$. Deciding the least significant atom of the
  lexicographically largest satisfying assignment of a Boolean formula is
  $\Delta_2^p$-complete~\cite{Kre88}. Thus, $L$ is also
  $\Delta_2^p$-complete.
\end{proof}

\subsection{Function Problem Complexity}

\newcommand{\FPNP}{\mathrm{FP}^\mathrm{NP}}
\newcommand{\Opt}{\textsc{Opt}}
\newcommand{\maxsat}{\textsc{Max-weight Sat}}

As noted in~\cite{Papadimitriou} any optimization problem, whose
decision version is in NP, is in $\FPNP$. Hence, finding an optimal
stable model of a logic program is in $\FPNP$. In fact, finding an
optimal stable model is $\FPNP$-complete.

\begin{theorem}
  Let $P$ be a program containing at least one minimize statement. Let 
  $\leq_P$ order the stable models of $P$ according to the minimize
  statements in $P$. Then, the problem \Opt: given a logic
  program $P$, find a stable model $S$ of $P$ such that for any other
  stable model $S'$ of $P$, $S\leq_P S'$, is $\FPNP$-complete.
\end{theorem}

\begin{proof}
  We begin by showing that the problem is in $\FPNP$. Assume without
  loss of generality that all weights are positive. If $m$ is the
  minimize statement
  \[\mathit{minimize}\,\{a_1 = w_{a_1},\dotsc,a_n = w_{a_n},
  \nB{b_1} = w_{b_1},\dotsc,\nB{b_m} = w_{b_m}\},\]
  then define
  \[w(m)=\sum_{i=1}^n w_{a_i}+\sum_{i=1}^m w_{b_i}.\]
  Furthermore, if $S$ is a set of atoms, then define
  \[w(S,m)=\sum_{a_i\in S} w_{a_i}+\sum_{b_i\not\in S} w_{b_i}.\]
  Since the oracle
  \begin{multline*}
    O =\{\pair{P,S,k}\mid \text{there exists a stable model $S$ of
      $P$} \\
    \text{that agrees with $A$ such that $w(S,m)\leq k$}\}
  \end{multline*}
  is in NP, the procedure $\mathit{findOptimal}(P)$ in
  Algorithm~\ref{alg:findoptimal} shows that the problem of finding an
  optimal stable model of a logic program is in $\FPNP$. The procedure 
  begins with a binary search for the weight of the optimal
  model. It then constructs the model using a linear number of oracle
  calls.

  We prove that the problem is $\FPNP$-hard by noticing that the
  problem \maxsat{} can be reduced to \Opt. If
  we are given a set of clauses, each with an integer weight, then
  \maxsat{} is the problem of finding the truth
  assignment that satisfies a set of clauses with the greatest total
  weight. The problem \maxsat{} is
  $\FPNP$-complete~\cite{Papadimitriou}.

  For each clause $c=a_1\lor\dotsb\lor a_n\lor\neg
  b_1\lor\dotsb\lor\neg b_m$ with weight $w_c$, create the rule
  \[c\from \nB{a_1},\dotsc,\nB{a_n},b_1,\dotsc,b_m.\]
  For each atom $a$ that appears in a clause create the rule
  $\{a\}\from$. Finally, add the maximize statement
  \[\mathit{maximize}\,\{\nB{c}=w_c,\dotsc\}.\]
  The maximal stable model of this program is the truth assignment of
  greatest total weight.
\end{proof}

We note that if a logic program contains one minimize statement, then
we can determine the weight of the lightest stable model using a
logarithmic number of oracle calls.

\begin{algorithm}
  \caption{Proof that finding an optimal stable model is in {\protect $\FPNP$}}
  \label{alg:findoptimal}
  \begin{algorithmic}
  \item[\textbf{function} $\mathit{findOptimal}(P)$]
    \STATE $A=\emptyset$
    \IF{there is no stable model of $P$}
     \STATE return false
    \ENDIF
    \FOR{each minimize statement $m$ in $P$ in order, most significant
      first}
     \STATE $u:=w(m)$
     \STATE $l:=0$
     \WHILE{$l<u$}
      \IF{there exists a stable model $S$ of $P$ that agrees with
       $A$ such that $w(S,m)\leq\lfloor(u+l)/2\rfloor$}
       \STATE $u:=\lfloor(u+l)/2\rfloor$
      \ELSE
       \STATE $l:=\lfloor(u+l)/2\rfloor+1$
      \ENDIF
     \ENDWHILE
     \FOR{each literal $x$ in $m$}
      \IF{there exists a stable model $S$ of $P$ that agrees with
        $A\cup\{x\}$ such that $w(S,m)=l$}
       \STATE $A:=A\cup \{x\}$
      \ELSE
       \STATE $A:=A\cup \{\nb{x}\}$
      \ENDIF
     \ENDFOR
    \ENDFOR 
     \FOR{each atom $a$ in $\Atoms{P}-\Atoms{A}$}
      \IF{there exists a stable model of $P$ that agrees with
        $A\cup\{a\}$}
       \STATE $A:=A\cup \{a\}$
      \ELSE
       \STATE $A:=A\cup \{\nB{a}\}$
      \ENDIF
     \ENDFOR
    \STATE return $\patoms{A}$.
  \end{algorithmic}
\end{algorithm}

\cleardoublepage
\section{Comparison with other Algorithms}
\label{sec:comparison}

We compare the $\smodels$ algorithm with the Davis-Putnam procedure
and with some algorithms for computing stable models of logic
programs. We begin with the Davis-Putnam procedure.

\subsection{The Davis-Putnam Procedure}

The Davis-Putnam (-Logemann-Loveland) procedure~\cite{DLL62} for
determining the satisfiability of propositional formulas in
conjunctive normal form has several similarities to the $\smodels$
algorithm. We compare the two methods and highlight their
differences.

The main difference between the methods comes from the different
semantics of the underlying problem. The stable model semantics
requires that a model is grounded. Hence, the program \[a\from a\]
has only one stable model, the empty set, while the corresponding
propositional formula \[\neg a\lor a\] has two models: $\{a\}$ and the 
empty set. All other dissimilarities are the result of this
fundamental disparity.

The Davis-Putnam procedure can prune its search space in three ways:
by the subsumption of clauses, by using the pure literal rule, and by
unit propagation.

If all truth-assignments that satisfy a clause $c$ also satisfy
another clause $c'$, then $c$ subsumes $c'$. For instance, $a\lor b$
subsumes $a\lor b\lor \neg c$. Subsumed clauses can therefore be
removed from a formula without changing the set of models of the
formula. The $\smodels$ algorithm has presently no corresponding way
of pruning the search space. However, testing for subsumed clauses is
expensive and state-of-the-art satisfiability checkers only perform
subsumption in a preprocessing step.

The pure literal rule removes any clauses containing a literal whose
complement do not appear in any clause. This corresponds to setting the
literal to true in a truth-assignment. For example, as the literal
$\neg a$ does not appear in the formula
\[(a\lor b)\land (a\lor \neg c),\]
the pure literal rule makes $a$ true and removes both clauses yielding
the model $\{a\}$. Hence, the pure literal rule can discard models,
and in this case it discards the models $\{b\}$, $\{a,b\}$, and
$\{a,b,c\}$. Since $\smodels$ does not discard models, it does not
make use of any similar rule. Apart from this, it is not clear what
form a rule that does not preserve all models would have under the
stable model semantics. The presence of maximize and minimize
statements further complicates matters.

Unit propagation consists of unit resolution and unit subsumption.
Unit resolution removes all literals that are false from every clause
and unit subsumption removes all clauses that contain a literal that
is true. Since a unit is a clause of length one, it defines the truth
value of a literal and this is the reason for the name. In $\smodels$
unit propagation coincides with forward propagation and with backward
propagation of rules with false heads. Making heads without active
rules false and propagating rules with true heads backwards have no
correspondence in the Davis-Putnam procedure. Similarly, the upper
closure is unique to $\smodels$.

Lookahead is in part used by modern satisfiability checkers. They
typically only employ lookahead on a small subset of all possible
atoms and they do not avoid testing literals that follow from
lookahead tests of other literals. Their heuristics take advantage
of the lookahead that they do perform, but not in the search space
minimizing form of $\smodels$.

\subsection{Stable Model Algorithms}

We make a comparison of the $\smodels$ algorithm and some other
algorithms for computing the stable model semantics. In order to
facilitate the comparison we must first examine the well-founded
semantics~\cite{VGRS91}. We then examine the branch and bound
algorithm of~\cite{Subrahmanian95}, the SLG system of~\cite{CW96}, the
mixed integer programming methods of~\cite{BNNS94}, the modified
Davis-Putnam method of~\cite{Dimopoulos96}, and the \texttt{dlv}
system of~\cite{WNG:lpnmr99}. Finally, we end with a comparison with 
$\smodels$.

\subsubsection*{The Well-founded Semantics}

Since the definition of the well-founded
semantics~\cite{VGRS91}\index{well-founded semantics} is somewhat
complicated, we make use of an equivalent definition that suits our
needs better. Let
\[\Gamma_P(A) = \lfp{\bigcup_{r\in P}f_r(A,\cdot)}\]
for any normal logic program $P$ and any set of atoms $A$. Then, the
operator $\Gamma_P(A)$ is anti-monotonic and we can define the
well-founded semantics with the help of the least and greatest fixed
points ($\lfp{\cdot}$ and $\gfp{\cdot}$) of the monotonic operator
\[\Gamma_P^2(A) = \Gamma_P\bigl(\Gamma_P(A)\bigr).\]
Notice that \[\Gamma_P(A) = \atmost{P,A}.\]
According to~\cite{BS93}, the well-founded model $W$ of a program $P$
is defined as follows:
\begin{align*}
\patoms{W} &= \Lfp{\Gamma^2_P} \\
\natoms{W} &= \Atoms{P}-\Gfp{\Gamma^2_P}.
\end{align*}

\begin{proposition}
  If $W$ is the well-founded model of the normal logic program $P$,
  then $W\subseteq\expand(P,\emptyset)$.
\end{proposition}

\begin{proof}
  Let $B=\expand(P,\emptyset)$. Then,
  \[\Atoms{P}-\natoms{B} = \atmost{P,B} \subseteq \atmost{P,\patoms{B}}.\]
  Consequently,
  \begin{align*}
    \Gamma^2_P(\patoms{B})
    &= \Gamma_P\bigl(\atmost{P,\patoms{B}}\bigr) \\
    &\subseteq \Gamma_P\bigl(\atmost{P,B}\bigr) \\
    &= \Gamma_P\bigl(\Atoms{P}-\natoms{B}\bigr).
  \end{align*}
  In addition, $\atleast{P,B}=B$ implies
  \[\mathit{min}_r(B) = \negthickspace
  \bigcap_{\substack{\patoms{B}\subseteq C \\ \natoms{B}\cap
      C=\emptyset}} \negthickspace f_r(C,C) \subseteq\patoms{B}.\]
  As $\natoms{B}\cap C=\emptyset$ implies
  $C\subseteq\Atoms{P}-\natoms{B}$ (we can assume
  $C\subseteq\Atoms{P}$),
  \begin{align*}
    f_r\bigl(\Atoms{P}-\natoms{B},C\bigr)&\subseteq f_r(C,C), \\
    \intertext{and as $\patoms{B}\subseteq C$,}
    f_r\bigl(\Atoms{P}-\natoms{B},\patoms{B}\bigr)&\subseteq f_r(C,C).
  \end{align*}
  It follows that
  \[f_r\bigl(\Atoms{P}-\natoms{B},\patoms{B}\bigr)\subseteq\mathit{min}_r(B)
  \subseteq \patoms{B}.\]
  Hence,
  \[\bigcup_{r\in P}
  f_r\bigl(\Atoms{P}-\natoms{B},\patoms{B}\bigr)\subseteq\patoms{B}\]
  and therefore
  \[\lfp{\bigcup_{r\in P}
    f_r\bigl(\Atoms{P}-\natoms{B},\cdot\bigr)}\subseteq\patoms{B}.\]
  Thus,
  \begin{gather*}
    \Gamma_P\bigl(\Atoms{P}-\natoms{B}\bigr) =
    \atmost{P,\Atoms{P}-\natoms{B}}\subseteq\patoms{B}, \\
    \Gamma^2_P(\patoms{B})\subseteq\patoms{B},
    \intertext{and}
    \Lfp{\Gamma^2_P}\subseteq\patoms{B}.
  \end{gather*}
  Let $A=\Lfp{\Gamma^2_P}$. Then,
  \[\Gamma^2_P\bigl(\Gamma_P(A)\bigr)=
  \Gamma_P\bigl(\Gamma^2_P(A)\bigr)=\Gamma_P(A).\]
  Hence, $\Gamma_P(A)$ is a fixed point of $\Gamma^2_P$ and
  \[\Gamma_P(A)\subseteq\Gfp{\Gamma^2_P}.\]
  Now,
  \[\Atoms{P}-\Gfp{\Gamma^2_P}\subseteq\Atoms{P}-\Gamma_P(A)
  \subseteq\Atoms{P}-\Gamma_P(\patoms{B})\subseteq\natoms{B}\]
  as $\Atoms{P}-\natoms{B}\subseteq \Gamma_P(\patoms{B})$
  and we have proved that the well-founded model of $P$ is a subset of 
  $B$.
\end{proof}

In fact, the well-founded model $W = \expand(P,\emptyset)$~\cite{S95}.
Notice that since $\expand(P,A)$ can return a conflicting set of
literals if $A\neq\emptyset$, it provides a stronger pruning technique 
than the well-founded semantics. For example, consider the program
\[P = \{a\from b,\ b\from\nB{c},\ c\from\nB{d},\ d\from\nB{b}\}\]
and make the assumption that we are looking for a stable model
containing the atom $a$. A typical reduction of $P$ with respect to
$\{a\}$ would not change $P$, and an application of the well-founded
semantics produces the empty set, but not the fact that there are no
stable models containing $a$. The function call $\expand(P,\{a\})$ on the
other hand returns a set containing both $a$ and $\nB{a}$, thereby
explicitly showing that there are no models that include $a$.

\subsubsection*{Branch and Bound}

The branch and bound algorithm of~\cite{Subrahmanian95} computes the
stable models of a ground logic program in two stages. In the first
stage the logic program is simplified while the algorithm computes the
well-founded semantics of the program. In the second stage, the branch
and bound stage, all stable models of the logic program are
constructed.

Starting with the simplified program the algorithm computes all stable
models by generating smaller and smaller programs from the programs it
has already generated. Two new programs are generated from one program
by assuming that an atom, whose truth value is unknown according to
the well-founded semantics, belongs or does not belong to the stable
model that is being constructed. The search space is pruned by
disregarding every newly created program that is inconsistent or whose
partial stable model is a superset of a stable model that has already
been found.

Although the algorithm prunes the search space quite a bit, it must
keep all constructed stable models as well as all partially
constructed stable models in memory. This indicates that the algorithm
will necessarily perform badly if the number of stable or partially
constructed models is large.

\subsubsection*{The SLG System}

The SLG system~\cite{CW96} supports goal-oriented query evaluation of
logic programs under the well-founded semantics. It simplifies a logic
program during the query evaluation and produces a residual program.
If all negated literals in the residual program are ground, then the
system can compute stable models using an assume-and-reduce
algorithm.

The assume-and-reduce algorithm constructs the stable models of a
logic program in a fashion similar to the branch and bound
algorithm. However, the assume-and-reduce algorithm differs in that it
constructs one model at a time, finding all models by backtracking,
and in that it does not enlarge a partially constructed stable model
using the well-founded semantics. Instead it derives truth values by
repeatedly reducing the program. An atom that appears in the head of a 
rule with an empty body is assumed to be in the stable model and an
atom that does not appear in any head is assumed to not be in the
stable model. In addition, the algorithm can with the help of backward
propagation in some specialized circumstances derive whether an atom
belongs to the stable model or not. This slightly improves the pruning
technique.

\subsubsection*{The Mixed Integer Programming Approach}

The mixed integer programming methods of~\cite{BNNS94} computes the
stable models of a logic program by translating the program into an
integer linear program, which is then used to compute all subset
minimal models of the logic program. The models are subsequently
tested by another integer linear program and models that are not
stable are removed. As the number of minimal models can be very large
compared to the number of stable models and as the minimal models must
be stored, we conclude that this approach is very inefficient.

Since a logic program can be encoded as a satisfiability
problem~\cite{BD:ai96} and since a satisfiability problem can easily
be encoded as an integer linear program, it is possible to compute the
stable models of a logic program using only one integer linear program.
However, it would hardly be reasonable to actually utilize such a complex
encoding.

\subsubsection*{The Modified Davis-Putnam Method}

In the modified Davis-Putnam method of~\cite{Dimopoulos96}, a logic
program is translated into a set of clauses in such a way that the
propagation rules of the Davis-Putnam procedure can deduce as much as
the $\atleast{P,A}$ function of $\smodels$. In particular, the
translation needs a literal for every atom and rule in the logic
program.

In addition, the Davis-Putnam procedure is modified such that it only
branches on literals that correspond to rules of the original
program. Furthermore, the branch points are chosen such that any model 
that the procedure finds is a stable model.

The DeReS system~\cite{CMMT99}, which implements default logic, is
another system that branches on rules instead of on atoms. It does not 
prune its search space much.

\subsubsection*{The \texttt{dlv} system}

The \texttt{dlv} system of~\cite{WNG:lpnmr99} is a knowledge
representation system that uses disjunctive logic. When it searches for 
stable models of normal logic programs, it prunes the search space
using the propagation rules of the $\atleast{P,A}$ function. It is
therefore the system that is most similar to $\smodels$.

\subsubsection*{Comparison}

We compare the branch and bound algorithm, the SLG system, the mixed
integer programming approach, the modified Davis-Putnam method, and
the \texttt{dlv} system with the $\smodels$ algorithm. The
performance of the branch and bound algorithm deteriorates, due to the
amount of memory needed, when the number of stable models is
large. Both the branch and bound algorithm and the SLG system
prune the search space less effectively than $\smodels$. The main
difference between $\smodels$ without lookahead and the branch and
bound algorithm and the SLG system is that $\smodels$ does not
differentiate between assumed and derived literals. Hence, $\smodels$
automatically avoids exploring the parts of the search space that the
branch and bound algorithm avoids by storing all partially constructed
stable models and that the SLG system does not avoid.

The mixed integer programming approach computes all minimal models of
a logic program and then tests if these models are stable. This
corresponds to a very weak pruning of the search space. For instance,
the set of rules \[\{a_1\from \nB{b_1},\dotsc,a_n\from\nB{b_n}\}\]
has one stable model but $2^n$ minimal models.

The modified Davis-Putnam method and the \texttt{dlv} system do not
prune the search space using the upper closure. Hence, they also
prune less than $\smodels$. It would be quite easy to integrate the
upper closure computation into the \texttt{dlv} system as it is
similar to $\smodels$. Including the upper closure computation into
the modified Davis-Putnam method would probably be a lot harder.

We conclude that even without lookahead the $\smodels$ algorithm
prunes the search space significantly more than the branch and bound
algorithm, the mixed integer programming approach, and the SLG
system. Moreover, the $\smodels$ algorithm also prunes the search
space more than the modified Davis-Putnam method and the \texttt{dlv}
system. As the pruning in the $\smodels$ algorithm can be efficiently
implemented, $\smodels$ will compute stable models faster than the
other systems if it is given a program that requires some search.

\cleardoublepage
\section{Experiments}
\label{sec:experiments}

In order to demonstrate the performance of $\smodels$, we test
an implementation, $\smodels$ version 2.23~\cite{smodels2}, on some
combinatorially hard problems. Since there are no competitive systems
for computing stable models of logic programs, we compare $\smodels$
with some propositional satisfiability checkers. The one stable model
system that can approach the same magnitude of performance as
$\smodels$ is \texttt{dlv}~\cite{dlv}, and we also compare against it.
The intention of the tests is to assess $\smodels$ in relation to
other general purpose systems, not to compare it with a different
special purpose algorithm for each problem.

Each test instance is randomly shuffled and run ten times.  We shuffle
the test instances since a particular ordering might help the
algorithms to avoid backtracking, thereby giving a skewed picture of
their behavior. That is, we try to lessen the impact of lucky choices.
The durations of the tests are given in seconds and they represent the
time to find a solution or the time to decide that there are no
solutions. They include the time it takes to read the input and write
the result. The number of choice points, also known as branch points,
describes how many times the algorithms use their heuristics to decide
which atom to test next.

All tests were run under Linux 2.2.12 on 450 MHz Pentium III
computers with 256 MB of memory.

\subsection{3-SAT}
\label{subsec:3sat}

\newcommand{\satz}{\mathit{satz}}
\newcommand{\SATO}{\mathit{SATO}}
\newcommand{\ntab}{\mathit{ntab}}
\newcommand{\dlv}{\texttt{dlv}}

We compare $\smodels$ with three propositional satisfiability
checkers: tableau or $\ntab$~\cite{CA96}, $\SATO$
3.2~\cite{Zhang:CADE97}, and $\satz$~\cite{Li:CP97}, and with \dlv.
The test domain is random 3-SAT, i.e., randomly generated
propositional formulas in conjunctive normal form whose clauses
contain exactly three literals. The problems are chosen such that the
clause to atom ratio is $4.258+58.26a^{-5/3}$, where $a$ is the number
of atoms, since this particular ratio determines a region of hard
satisfiability problems~\cite{CA96}.

The three satisfiability checkers are all variants of the Davis-Putnam
procedure. $\SATO$ strengthens the procedure by adding clauses to the
problem during the search. Every time $\SATO$ arrives at a
contradiction it stores the negation of the choices that led to the
contradiction in a new clause. If the length of the clause is less
than 20, then it is added to the set of clauses. This approach runs
into problems if the number of added clauses grows too big.

Tableau or $\ntab$ and $\satz$ both perform lookahead on a subset of
all available atoms, but $\satz$ uses a more sophisticated heuristic,
which is also used to measure the hardness of the problem during the
search. If the problem is hard, then $\satz$ performs lookahead on all
atoms. The program $\satz$ also does some preprocessing on the
satisfiability problems during which it adds clauses to the problem.

We translate a 3-SAT problem into a logic program as follows. The
truth-assignments of the atoms $a_1,\dotsc,a_n$ of the problem are
encoded by a choice rule
\[\{a_1,\dotsc,a_n\} \from\]
and each clause
\[a_1\lor\dotsb\lor a_n\lor\neg b_1\lor\dotsb\lor\neg b_m\]
is translated into a rule
\[\mathit{false} \from \nB{a_1},\dotsc,\nB{a_n},b_1,\dotsc,b_m.\]
Since the clauses have to be satisfied, we deny the inclusion of the
atom $\mathit{false}$ in the stable models using the statement
\[\mathit{compute}\,\{\nB{\mathit{false}}\}.\]
When we test \dlv, we use the rules $a_i\from\nB{\bar{a}_i}$ and
$\bar{a}_i\from\nB{a_i}$, for $i=1,\dotsc,n$, instead of the choice
rule.

\begin{figure}
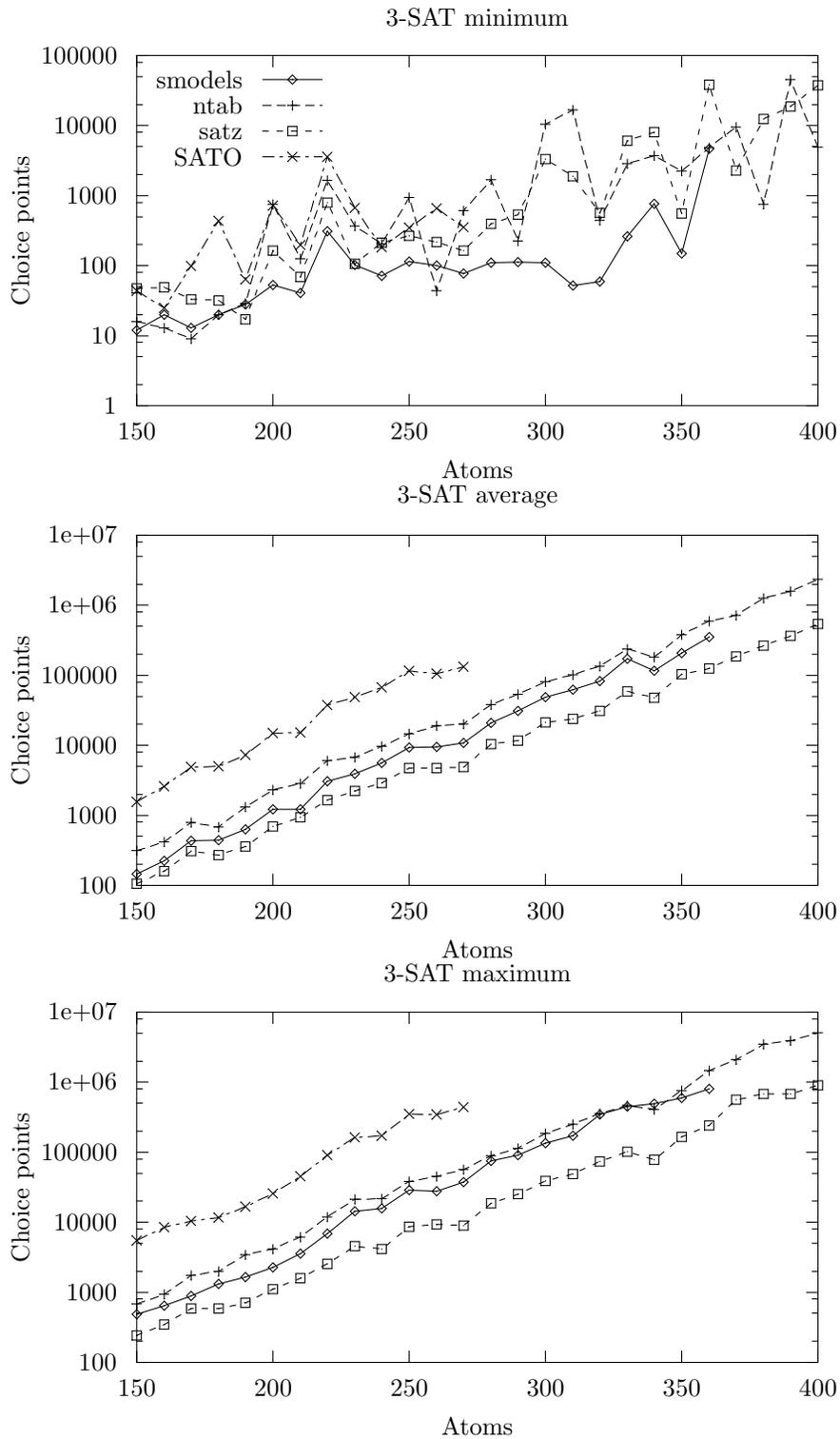

\centering
\includegraphics{satminchoice.1}
\includegraphics{satavechoice.1}
\includegraphics{satmaxchoice.1}
\caption{3-SAT, number of choice points}
\label{fig:satchoice}
\end{figure}

\begin{figure}
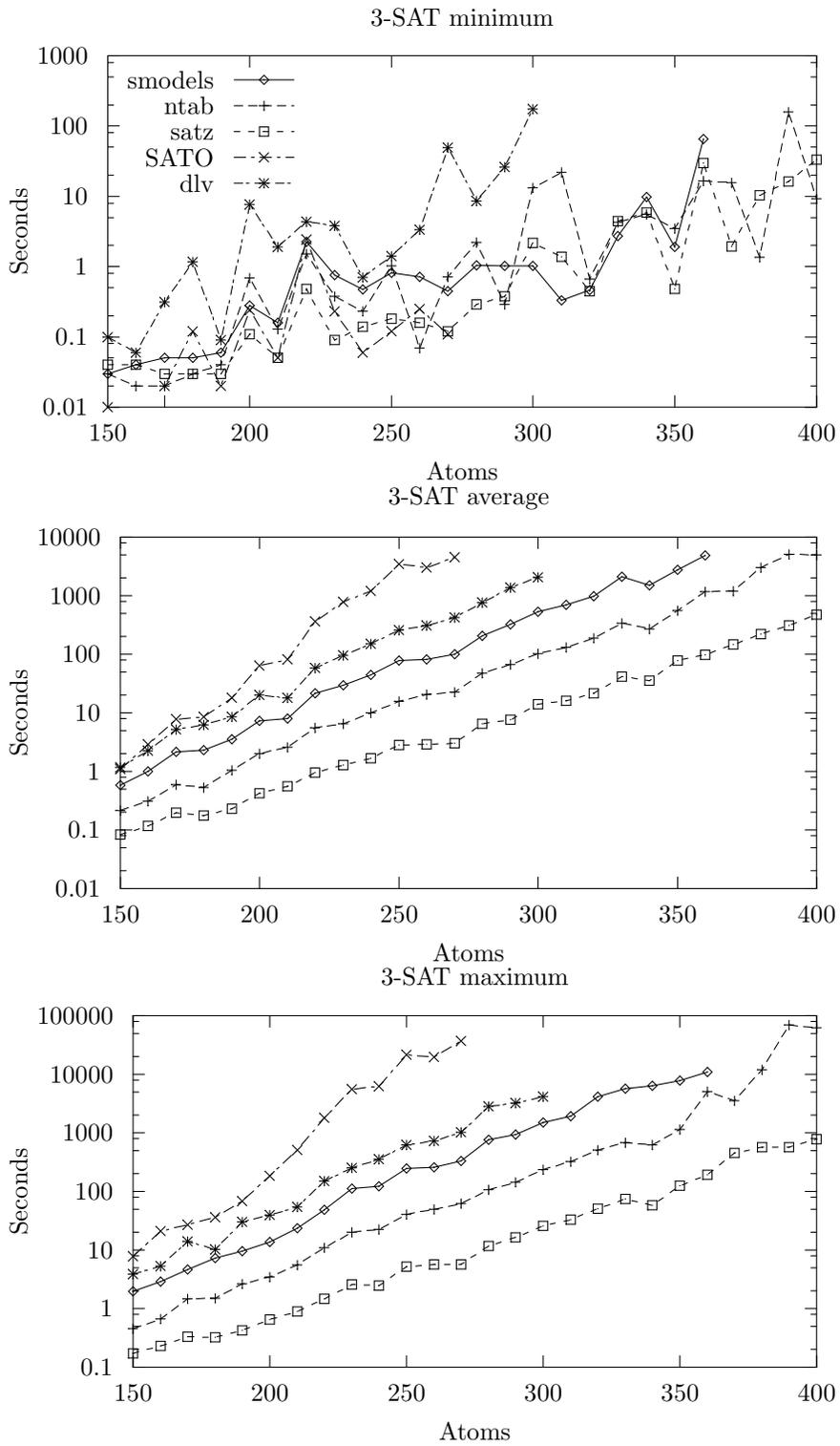

\centering
\includegraphics{satminsec.1}
\includegraphics{satavesec.1}
\includegraphics{satmaxsec.1}
\caption{3-SAT, duration in seconds}
\label{fig:satsec}
\end{figure}

We test the implementations on problems having from 150 to 400
atoms. For each problem size we generate ten satisfiability problems
using a program developed by Bart Selman~\cite{Selman:makewff}.
Every problem is randomly shuffled and tested ten times. The test
results are shown in
Figures~\ref{fig:satchoice}--\ref{fig:satsec}. The number of choice
points is not available for \dlv.

The implementation of $\smodels$ prunes the search space more than
$\SATO$ and $\ntab$, but less than $\satz$. In addition, $\SATO$
prunes the search space least of all. This indicates, as one would
expect, that doing lookahead substantially reduces the search space.
Since $\smodels$ is not as good at pruning the search space as
$\satz$, it seems that the heuristic of $\satz$ is better than that of
$\smodels$.

There is, however, a difference between how $\smodels$ and $\satz$
propagate truth values. The $\satz$ procedure makes use of the pure
literal rule. A literal is pure if its complement does not appear in
any clauses. Hence, if a set of clauses is satisfiable, then one can
remove all clauses containing pure literals and the remaining set of
clauses will still stay satisfiable. Since $\smodels$ is designed such
that it can compute all stable models of a program, it does not take
advantage of this reduction.

The heuristic of $\satz$ is similar to that of $\smodels$. The heuristic
of $\smodels$ measures how the set of undefined literals changes when
we fix the truth value of an atom, while the heuristic of $\satz$
measures how the set of clauses changes when we fix the truth value of 
an atom. Let $p$ and $n$ be a measure of the change when the truth
value of an atom is set to respectively true or false. Then,
$\smodels$ maximizes $\min(n,p)$ and $\max(n,p)$, where $\min(n,p)$ is 
more significant, and $\satz$ maximizes
\[1024np+n+p.\]
We refer to these formulas as the cost functions of $\smodels$ and
$\satz$. In order to decide whether it is the heuristic of $\satz$
that is the reason for the better pruning, we have tested the
following variants of $\smodels$ and $\satz$ on the 3-SAT problems:
$\satz$ without the preprocessing step, $\satz$ without the pure
literal rule, $\satz$ with the cost function of $\smodels$ measuring
literals and measuring clauses, $\smodels$ with the pure literal rule,
$\smodels$ with the cost function of $\satz$, and $\smodels$ with both
the pure literal rule and the cost function of $\satz$. The result is
that the variants prune the search space as much or slightly less than
the original versions. We therefore conclude that the heuristic of
$\satz$ is better than that of $\smodels$ on hard 3-SAT problems.

\begin{figure}
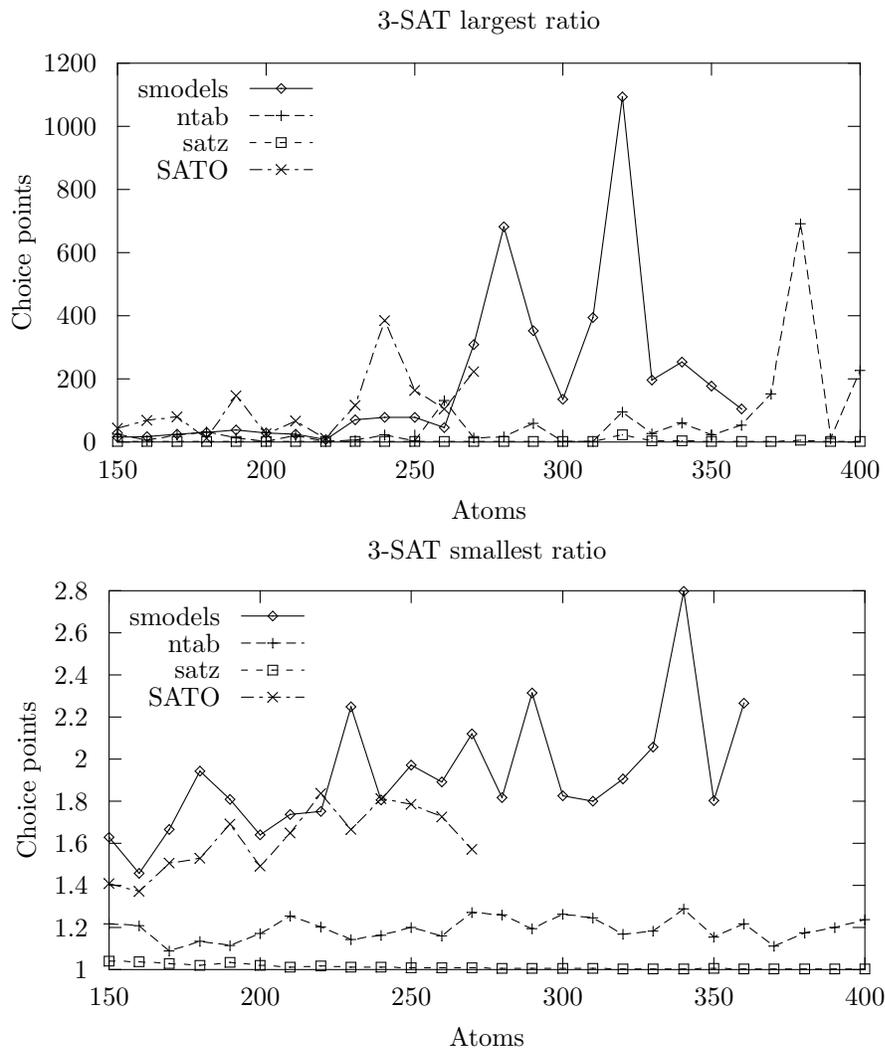

\centering
\includegraphics{ratiomaxsat.1}
\medskip

\includegraphics{ratiominsat.1}
\caption{3-SAT, choice points max/min ratio}
\label{fig:satratio}
\end{figure}

We will use the fact that we shuffled and tested each satisfiability
problem ten times, to measure how tightly the heuristics define which
literal to try next. If a heuristic is lax, then it is possible to
refine it such that it breaks the ties between the literals in a
better way. For each satisfiability problem we compute the ratio
between the largest and smallest number of choice points needed to
solve a randomly shuffled instance of the problem. The results are
shown in Figure~\ref{fig:satratio}.

Clearly, $\satz$ has a very tight and $\smodels$ a very lax heuristic.

\subsection{Pigeon-hole Problems}
\label{subsec:pigeon}

\begin{figure}
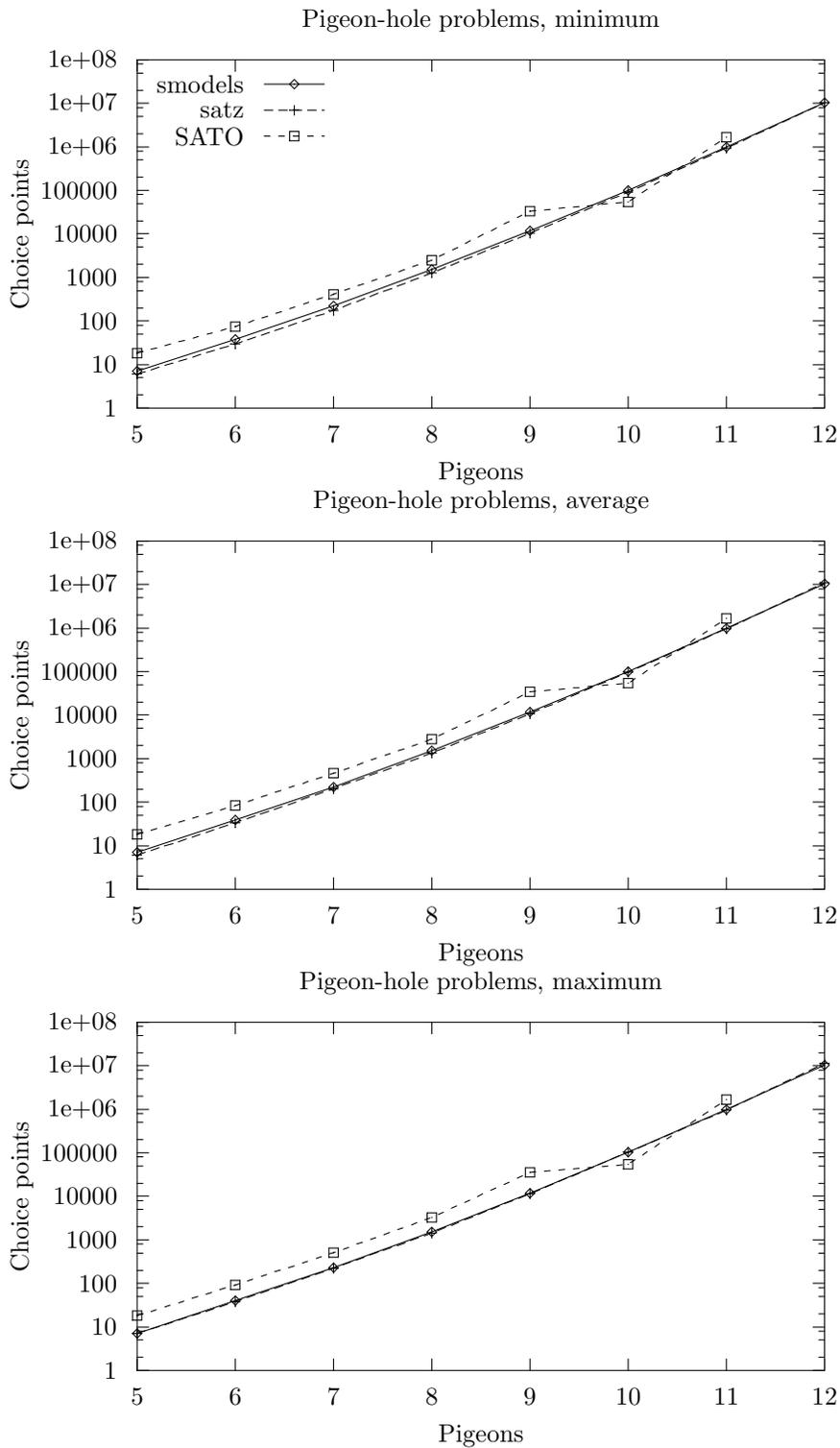

\centering
\includegraphics{holeminchoice.1}
\includegraphics{holeavechoice.1}
\includegraphics{holemaxchoice.1}
\caption{Pigeon-hole problem, number of choice points}
\label{fig:holechoice}
\end{figure}

\begin{figure}
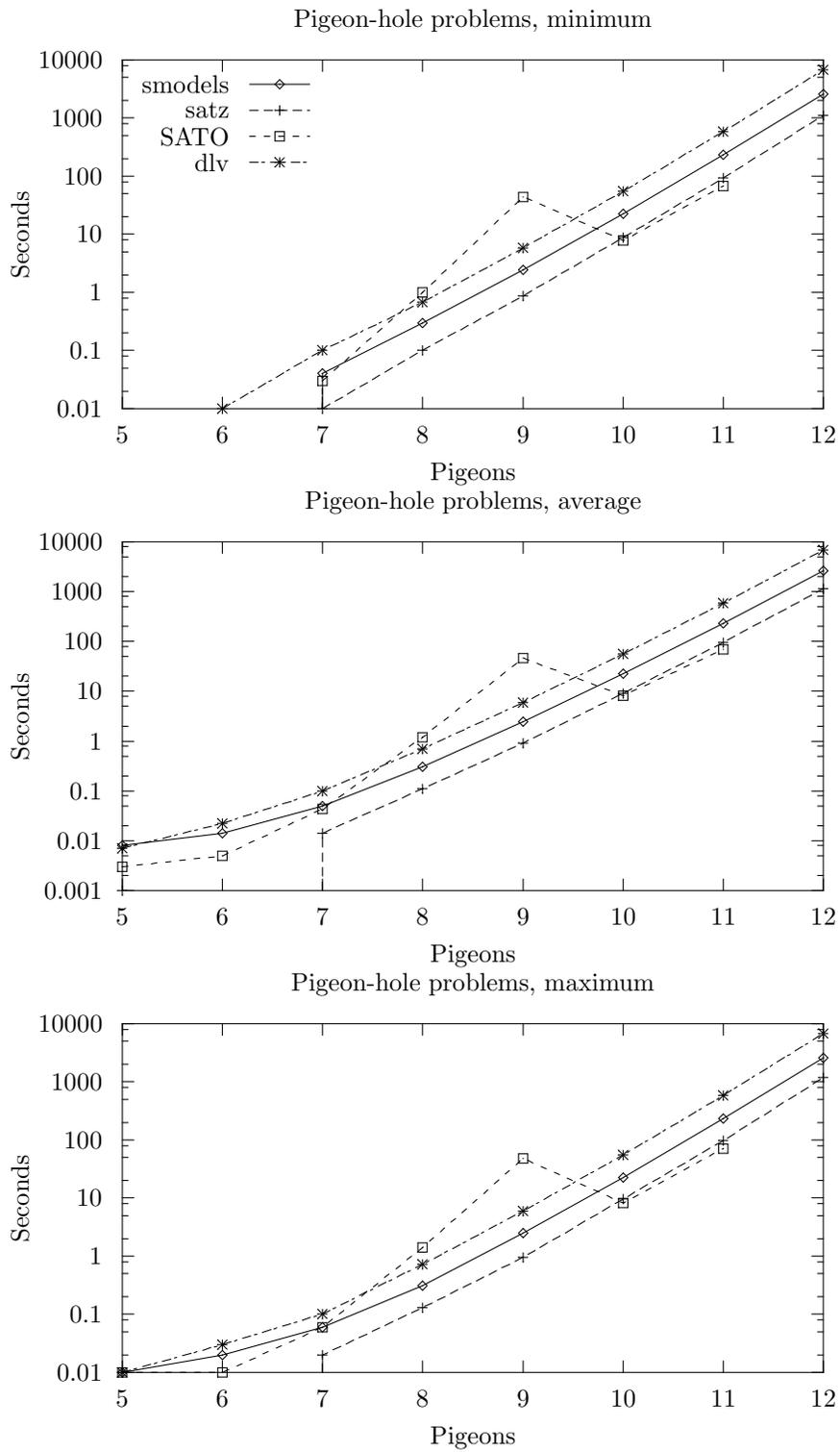

\centering
\includegraphics{holeminsec.1}
\includegraphics{holeavesec.1}
\includegraphics{holemaxsec.1}
\caption{Pigeon-hole problem, duration in seconds}
\label{fig:holesec}
\end{figure}

We test $\smodels$, $\satz$, and $\SATO$ on some pigeon-hole problems,
where the number of pigeons is one more than the number of holes. The
$\SATO$ program generates the satisfiability problems for itself and
$\satz$. Since the generated problems are not 3-SAT problems and since
$\ntab$ does not handle clauses of length greater than three, we do
not test $\ntab$. We encode the pigeon-hole problems for $\smodels$ as
in Section~\ref{sec:complexity}. When we test \dlv, we replace a
cardinality rule of the form
\begin{align*}
\mathit{false} &\from 2\,\{p_1,\dotsc,p_n\} \\
\intertext{with the basic rules}
\mathit{false} &\from p_{i_1},p_{i_2},p_{i_3}, & 1\leq
i_1<i_2<i_3\leq n.
\end{align*}
Similar translations are used in the following tests and will not be
mentioned. The results are shown in
Figures~\ref{fig:holechoice}--\ref{fig:holesec}.

We notice that $\smodels$ and $\satz$ explore very similar sized
search spaces. If we examine the durations of the tests, we see that
the overhead of $\smodels$ is smaller than it was on the 3-SAT
problems. This is explained by the compact encoding of the pigeon-hole 
problem. The program $\SATO$ experiences a strange improvement when
there are ten pigeons.

\subsection{Hamiltonian Cycles}
\label{subsec:hamiltonian}

A Hamiltonian cycle is a cycle in a graph such that the cycle visits
each node precisely once. If $G=(V,E)$ is an undirected graph, then we
encode the Hamiltonian cycles of $G$ in a program as follows. For each
edge $\{v,w\}\in E$ we make an atom $e_{v,w}$. Since the graph is
undirected, we do not distinguish between $e_{v,w}$ and $e_{w,v}$. The
idea of the encoding is that if $e_{v,w}$ is in a stable model, then
$\{v,w\}$ participates in a Hamiltonian cycle.

We notice that if $v$ is a vertex of a cycle, then there are exactly
two edges incident to $v$ that are in the cycle. Hence, we create the
three rules
\begin{align*}
  \{e_{v,w_1},\dotsc,e_{v,w_n}\} &\from &&\{v,w_i\}\in E \\
  \mathit{false} &\from 3\,\{e_{v,w_1},\dotsc,e_{v,w_n}\}
  &&\{v,w_i\}\in E \\
  \mathit{false} &\from n-1\,\{\nB{e_{v,w_1}},\dotsc,\nB{e_{v,w_n}}\}
  &&\{v,w_i\}\in E
\end{align*}
for every vertex $v$ of $G$. We must now avoid stable models that
contain more than one cycle. First we pick an arbitrary vertex $v_1\in
V$ and then we create the rule
\begin{align*}
  v_1 &\from \\
\intertext{and the rules}
  w &\from v, e_{v,w} &&\text{$\{v,w\}\in E$ and $w\neq v_1$}.
\end{align*}
It follows that a vertex $v$ is in a stable model if and only if $v$
and $v_1$ are in the same cycle. Hence, we can force the stable models
to contain exactly one cycle by including the compute statements
\begin{gather*}
  \mathit{compute}\,\{v_1,\dotsc,v_n\} \qquad v_1,\dotsc,v_n\in V\\
  \intertext{and}
  \mathit{compute}\,\{\nB{\mathit{false}}\}.
\end{gather*}

We translate the Hamiltonian cycle problem into clauses following
Papadimitriou~\cite{Papadimitriou}. The translation encodes a total
order on the vertices of the graph such that there is an edge between
any pair of vertices that are adjacent in the order. Given the
vertices $v_1,\dotsc,v_n$, we use the atom $v_{i,j}$ to encode that
vertex $v_i$ is in position $j$ using the clauses:
\begin{align*}
  &v_{i,1}\lor\dotsb\lor v_{i,n} &&
  &&\text{$v_i$ is in some position,} \\
  &\neg v_{i,j}\lor\neg v_{i,k} && j\neq k
  &&\text{$v_i$ is in at most one position,} \\
  &v_{1,j}\lor\dotsb\lor v_{n,j} &&
  &&\text{some vertex is in position $j$, and} \\
  &\neg v_{i,k}\lor\neg v_{j,k} && i\neq j
  &&\text{$v_i$ and $v_j$ are not in the same position.} \\
\intertext{Then we deny orders that do not correspond to Hamiltonian cycles:}
  &\neg v_{i,k}\lor\neg v_{j,k+1\bmod n} && \{v_i,v_j\}\not\in E.
\end{align*} 
Hence, if the graph has $n$ vertices, then we need $n^2$ atoms and on
the order of $n^3$ clauses.

\newcommand{\lequiv}{\leftrightarrow}

One can also create a more complex Hamiltonian cycle translation that
mirrors the logic programming translation. For a graph $G$ and for each
edge $\{v,w\}\in E$, we make an atom $e_{v,w}$. As before, we do not
distinguish between the atoms $e_{v,w}$ and $e_{w,v}$. For a vertex
$v\in V$, let the incident edges be $\{v,w_1\},\dotsc,\{v,w_n\}$. We
force the inclusion of at least two incident edges by creating the
clauses
\begin{align*}
  &e_{v,w_{i_1}}\lor\dotsb\lor e_{v,w_{i_k}}
  && 1\leq i_1<i_2<\dotsb <i_k\leq n,\ k=n-1 \\
  \intertext{and we deny the inclusion of more than two edges by
  creating the clauses}
  &\neg e_{v,w_{i_1}}\lor\neg e_{v,w_{i_2}}\lor\neg e_{v,w_{i_3}}
  && 1\leq i_1<i_2<i_3\leq n
\end{align*}
We avoid multiple cycles by picking a vertex and demanding that there
is a path to every other vertex. For every vertex $v$, we create an
atom $v^k$, for $k=0,\dotsc,n$, that denotes that $v$ is reachable
through a path of length $k$. We keep the final conjunctive normal
form encoding small by letting an auxiliary atom $t_{v,w}^k$ denote
that $w$ is reachable in $k$ steps and that $v$ is one step away from
$w$. Thus, the formulas
\begin{align*}
  v^{k+1}&\lequiv t_{v,w_1}^k\lor\dotsb\lor t_{v,w_n}^k
  && k=0,1,\dotsc,n-1, \\
  t_{v,w_i}^k&\lequiv e_{v,w_i}\land w_i^k
  && \text{$i=1,\dotsc,n$, and} \\
  v^1&\lor\dotsb\lor v^n
\end{align*}
ensure that there is a path from some vertex to $v$. Finally,
we pick a vertex $v$ that begins all paths by creating the clauses
\begin{align*}
  &v^0 \\
  \intertext{and}
  &\neg w^0 && \text{$w\neq v$ and $w\in V$.}
\end{align*}

\begin{figure}
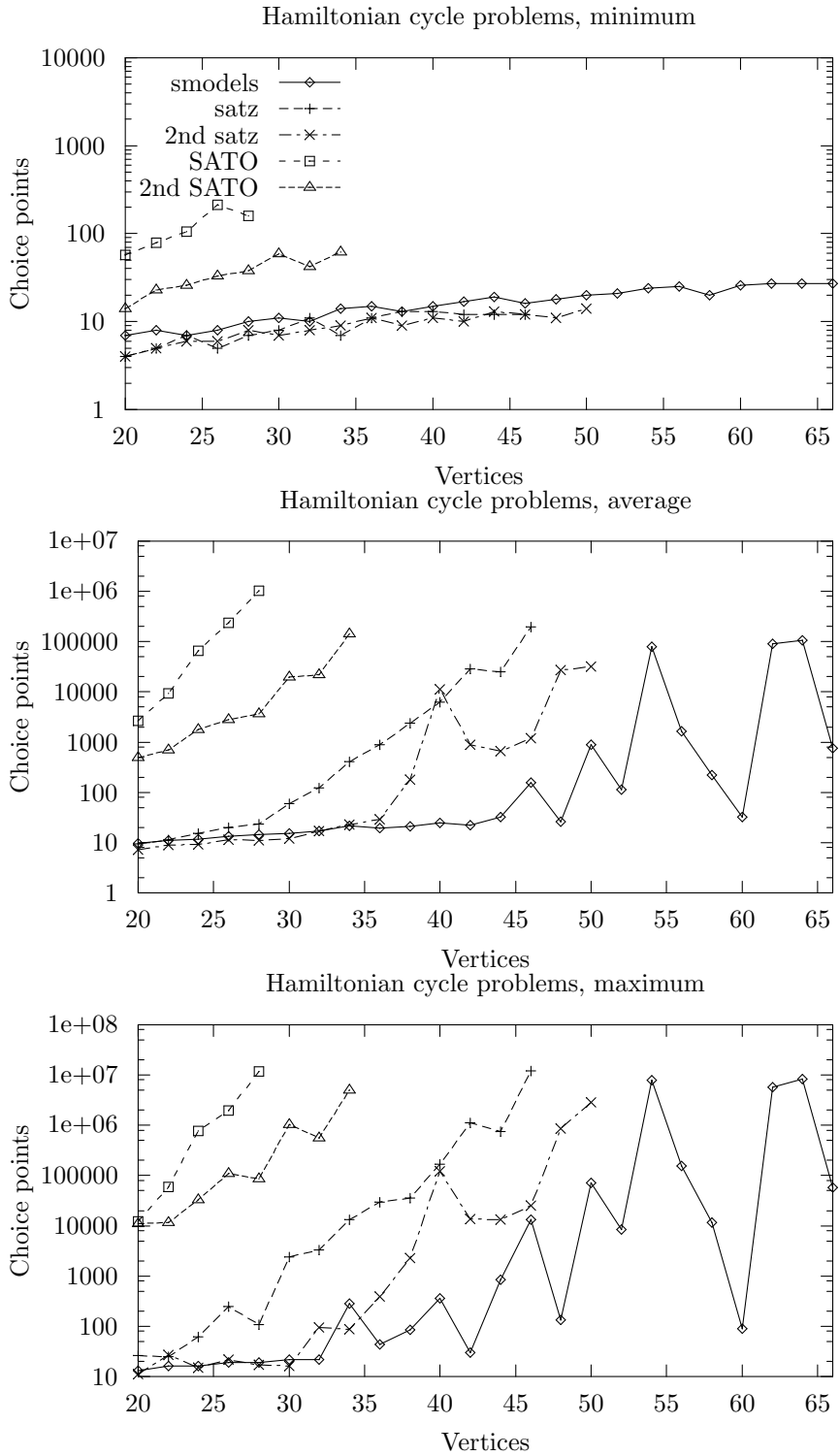

\centering
\includegraphics{hamminchoice.1}
\includegraphics{hamavechoice.1}
\includegraphics{hammaxchoice.1}
\caption{Hamiltonian cycle problem, number of choice points}
\label{fig:hamchoice}
\end{figure}

\begin{figure}
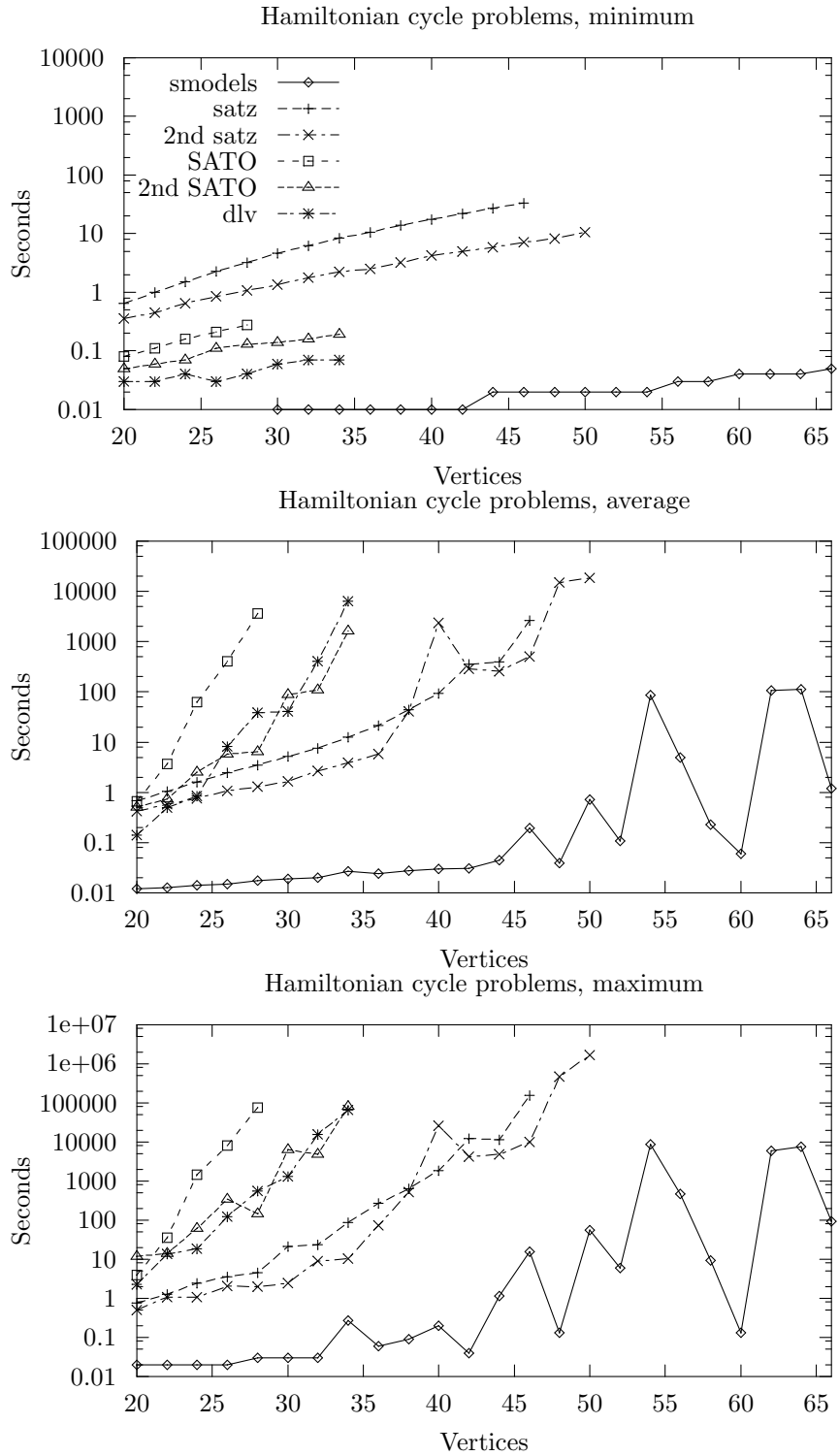

\centering
\includegraphics{hamminsec.1}
\includegraphics{hamavesec.1}
\includegraphics{hammaxsec.1}
\caption{Hamiltonian cycle problem, duration in seconds}
\label{fig:hamsec}
\end{figure}

We solve the Hamiltonian cycle problem on a special type of planar
graphs created by the $\mathit{plane}$ function found in the Stanford
GraphBase~\cite{Knuth:SGB}. We generate ten graphs for each problem
size and all these graphs have Hamiltonian cycles. The results are
displayed in Figures~\ref{fig:hamchoice}--\ref{fig:hamsec}. We observe 
that the second satisfiability encoding is better for $\SATO$ and
mostly better for $\satz$. Apparently, the heuristic of $\satz$ does
not work well on these problems.

It is precisely problems of this type that provides us with an
incentive to use a stable model semantics solver instead of a
satisfiability checker. Any problem that one can easily encode as a
satisfiability problem we can just as easily encode as a logic
program. The converse does not hold.

Consider a mapping $T$ from logic programs to sets of propositional
clauses. We say that $T$ is modular if for every program
partitioned into two disjoint parts $P_1$ and $P_2$, the program
$P_1\cup P_2$ has a stable model if and only if $T(P_1)\cup T(P_2)$ is 
satisfiable.

\begin{proposition}[Niemel\"{a}~\cite{Niemela98:cnmr}]
\label{prop:nonmodular}
There is no modular mapping from the class of logic programs to the
sets of clauses.
\end{proposition}

\begin{proof}
Consider the program $P=\{p\from\nB{p}\}$ and assume that $T$ is a
modular mapping. Then, $T(P)$ is unsatisfiable as $P$ has no stable
models. It follows that also $T(P)\cup T(\{p\from\})$ is
unsatisfiable. But this implies that $P\cup\{p\from\}$ has no stable
models, which is clearly not the case. Hence, $T$ is not modular.
\end{proof}

Even if there is no local translation of logic programs into
satisfiability problems, there are more complex ones~\cite{BD:ai96}.

\begin{example}
  The translation from a logic program to a satisfiability problem has
  two parts. The first part ensures that the rules of the logic
  program are satisfied and the second part ensures that an atom can
  not justify its own inclusion in a model. It follows that a model of
  the translation is grounded and that it corresponds to a stable
  model of the original logic program.

  Each atom in the logic program has a corresponding atom in the
  satisfiability problem. In addition, there are new atoms that assign
  indices to these atoms. By requiring that an atom can only be
  justified by atoms of lower index, we avoid circular justifications.

  For example, let $P$ be the program
  \begin{align*}
    p &\from q \\
    q &\from p.
  \end{align*}
  We translate $P$ into a set of propositional clauses as
  follows. First, we include the clauses
  \[\neg q\lor p \quad\text{and}\quad \neg p\lor q,\]
  so that if $q$ is true then also $p$ is true and if $p$ is true
  then also $q$ is true.

  Then, we introduce the new atoms $p_1$, $p_2$, $q_1$, and $q_2$
  whose truth-values decide what indices $p$ and $q$ have. The
  formulas $p_1\lor p_2$, $\neg p_1\lor\neg p_2$, $q_1\lor q_2$, and
  $\neg q_1\lor\neg q_2$ assign precisely one index to $p$ and one to
  $q$. Finally, the two formulas
  \[\neg p\lor(q\land q_1\land p_2) \quad\text{and}\quad
    \neg q\lor(p\land p_1\land q_2)\]
  guarantee that $p$ follows from $q$ only if the index of $q$ is
  lower than that of $p$ and that $q$ follows from $p$ only if the
  index of $p$ is lower than that of $q$.

  One problem remains. If $p$ is false, then either $p_1$ or $p_2$ is
  true, and we get two models instead of one. Therefore, we set the
  index of $p$ to 1 when it is false. The same must be done for
  $q$. Hence, we include the formulas
  \[p\lor p_1 \quad\text{and}\quad q\lor q_1.\]

  This satisfiability problem has only one model, $\{p_1,q_1\}$, which
  corresponds to the empty stable model of $P$.

  Take the program $P=\{p\from\nB{p}\}$ from the proof of
  Proposition~\ref{prop:nonmodular}. We can translate $P$ into a
  satisfiability problem without using any indices as the atom $p$ can
  not justify its own inclusion in a stable model. The translation
  consists of two clauses. The clause $p\lor p$ guarantees that the
  only rule in $P$ is satisfied and the clause $\neg p\lor \neg p$
  guarantees that if $p$ is true, then it is justified by the only rule.
  The two clauses have no models and the program has no stable models.

  If we add the rule $p\from$ to $P$, then we have to change the
  translation. The clause $p\lor p$ remains unchanged and we add
  the clause $p$ to keep $p\from$ satisfied. The clause
  $\neg p\lor\neg p$ is replaced by $\neg p\lor\neg p\lor\top$,
  where $\top$ is a tautology. The new translation has $\{p\}$ as its
  only model.
\end{example}

It is now clear that the more involved stable model semantics incurs a
computational overhead. But in exchange for the overhead we gain a
more powerful language. Hence, problems that can be more compactly
represented by logic programs can still be more quickly solved with
$\smodels$ than with a satisfiability checker.

\subsection{Error-correcting Codes}
\label{subsec:codes}

We will search for sets of binary words of length $n$ such
that the Hamming distance between any two words is at least $d$.
The size of the largest of these sets is denoted by $A(n,d)$. For
example, $A(5,3) = 4$ and any $5$-bit one-error-correcting code
contains at most 4 words. One such code is
$\{00000,00111,11001,11110\}=\{0,7,25,30\}$. Finding codes becomes
very quickly very hard. For instance, it was only recently proved that 
$A(10,3)=72$~\cite{OBK99}.

We construct a program whose stable models are the maximal codes with
Hamming distance $d$. If $j_1,\dotsc,j_k$ are the words whose distance
to $i$ is positive and less than $d$, then we create a rule
\[w_i \from \nB{w_{j_1}},\dotsc,\nB{w_{j_k}}\]
for every $i=0,\dotsc,2^n$. Since a code remains a code even if we
swap the zeroes for ones or permute the positions of the bits in every 
word in the same way, we can restrict ourselves to codes that include
the zero word and a word whose $d$ lowest bits are set. Therefore, we
create the rules
\begin{align*}
  w_0 &\from \\
  \mathit{false} &\from \nB{w_{j_1}},\dotsc,\nB{w_{j_{2^{n-d}}}},
\end{align*}
where $j_k=2^d(k+1)-1$, and the compute statement
\[\mathit{compute}\,\{\nB{\mathit{false}}\}.\]
Since we want to search for the largest stable model, we include the
maximize statement
\[\mathit{maximize}\,\{w_0,\dotsc,w_{2^n}\}.\]

\begin{figure}
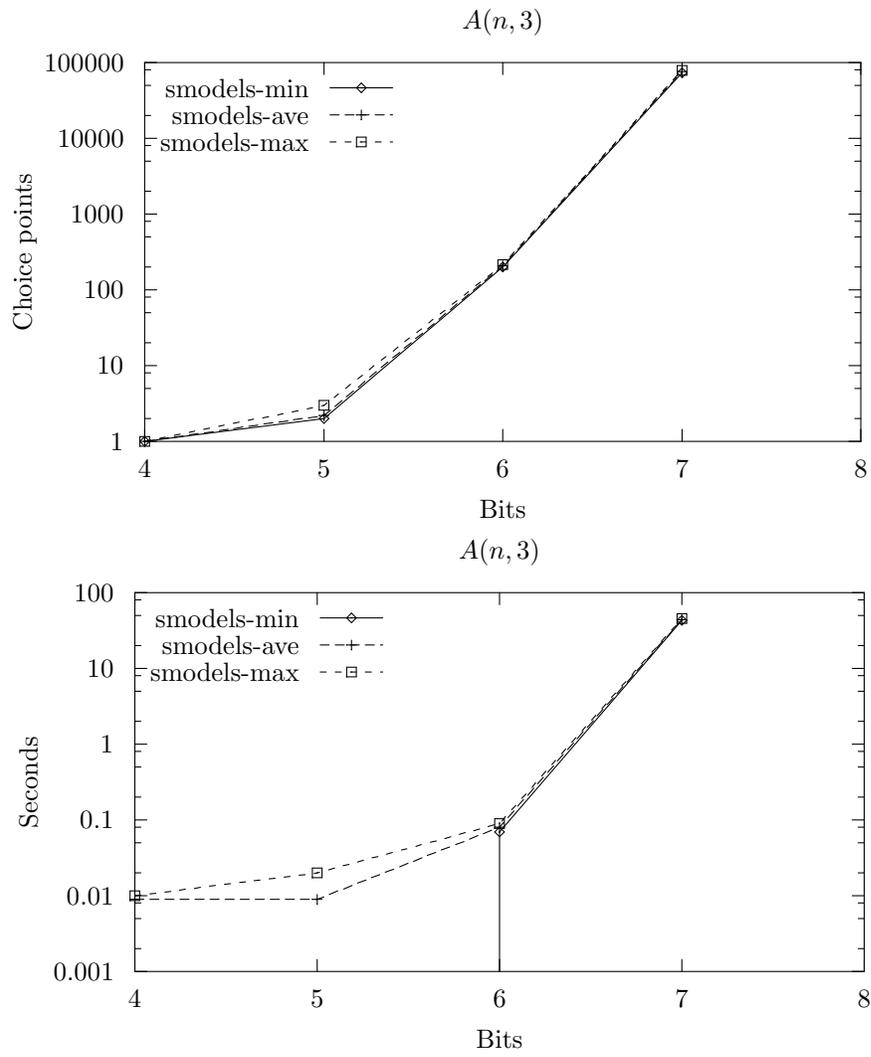

\centering
\includegraphics{codechoice.1}
\medskip

\includegraphics{codesec.1}
\caption{Maximal error-correcting codes}
\label{fig:code}
\end{figure}

The results are shown in Figure~\ref{fig:code}.

\subsection{Bin-packing}
\label{subsec:binpacking}

In order to test the weight rules we try $\smodels$ on some
bin-packing problems. The object is to pack a number of items of
varying sizes into a number of bins of equal size. We represent
the fact that item $i$ is in bin $j$ by the atom $b_{i,j}$, and we
assume that the size of item $i$ is given by the positive integer
$w_i$. If we have $n$ items and $m$ bins, then we can distribute the
items among the bins using the rules
\begin{align*}
  \{b_{i,1},\dotsc,b_{i,m}\} &\from \\
  \mathit{false} &\from 2\,\{b_{i,1},\dotsc,b_{i,m}\} \\
  \mathit{false} &\from \nB{b_{i,1}},\dotsc,\nB{b_{i,m}} & i=1,\dotsc,n.
\end{align*}
Let the bins have size $b$. We prevent the bins from containing
too many items with the help of the rules
\begin{align*}
  \mathit{false} &\from \{b_{1,j}=w_1,\dotsc,b_{n,j}=w_n\}\geq b+1 &
  j=1,\dotsc,m.
\end{align*}
Finally, we include the rule
\[\mathit{compute}\,\{\nB{\mathit{false}}\}\]
as we do not want $\mathit{false}$ to be in any stable model.

For our tests we choose the sizes of the items uniformly from the
integer interval $1,\dotsc,\mathit{max}$ for some number
$\mathit{max}$. A bit of experimentation shows that most bin-packing
problems are easy. However, $\smodels$ have problems when there are
many items and the expected total size of all the items equals the
available space in the bins. In hindsight, this is not surprising as
it includes the case when the items almost fit.

\begin{figure}
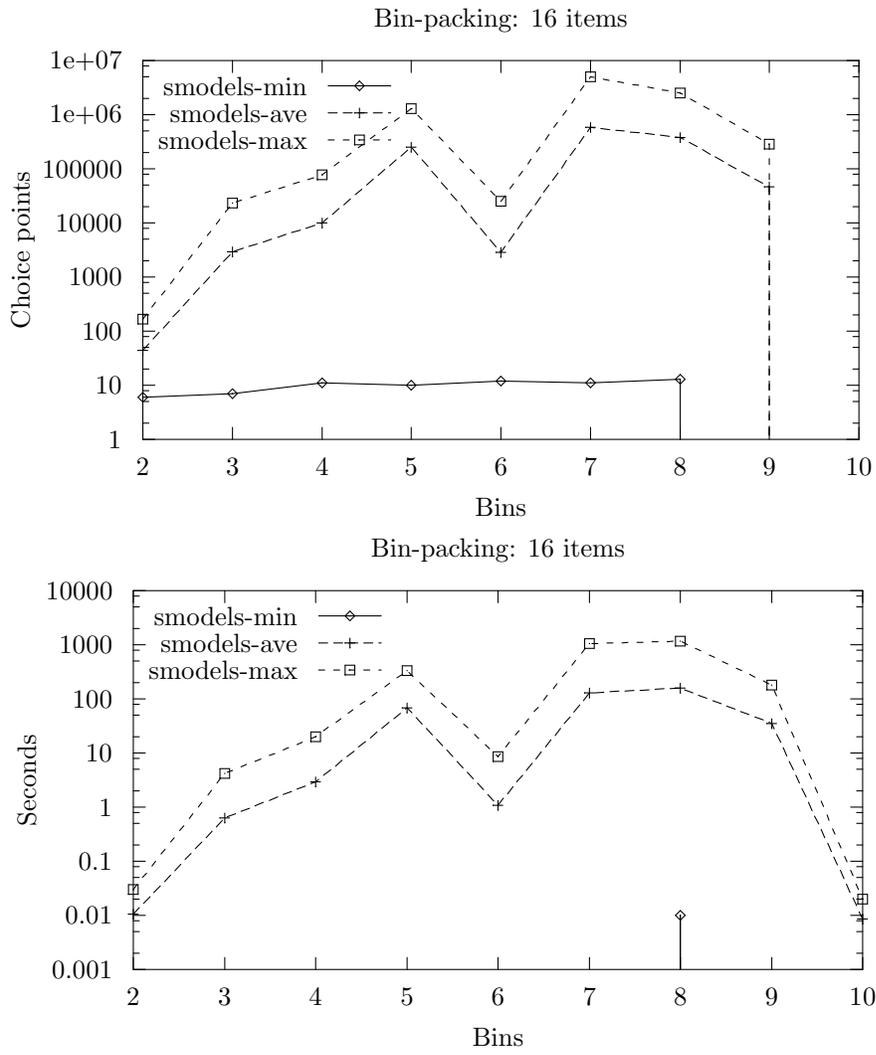

\centering
\includegraphics{binchoice.1}
\medskip

\includegraphics{binsec.1}
\caption{Bin-packing problems}
\label{fig:bin}
\end{figure}

We solve bin-packing problems with 16 items, bins of size 16, and with
the maximum size of the items set to twice the number of bins. The
results are shown in Figure~\ref{fig:bin}. As a comparison, a mixed
integer linear programming system such as
$\mathit{lp\_solve}$~\cite{lpsolve} solves each of these problems in
about 0.01 seconds.

\cleardoublepage
\section{Conclusions}
\label{sec:conclusions}

We have explored an algorithm that solves the stable model semantics
for logic programs. We have shown that it is comparatively easy to
extend the semantics and the algorithm to handle new types of rules. We
have also shown that one can easily change the algorithm to search for
specific stable models such as the lexicographically smallest or
largest one.

We have compared the algorithm with three good satisfiability solvers
on random satisfiability problems, pigeon-hole problems, and
Hamiltonian cycle problems. The best satisfiability solver goes
through a smaller search space than our algorithm when testing the
random SAT problems. We have attributed the difference to the
different heuristics of the procedures, and we have found indications
that our heuristic can be refined such that it works better. The
algorithm and the solvers behave similarly on the pigeon-hole
problems, while our algorithm is significantly better than the solvers
when it comes to the Hamiltonian cycle problems. Since the Hamiltonian
cycle problem contains more structure, it requires both a good
heuristic and full lookahead before it can be solved satisfactorily.

To conclude, the stable model semantics has a computational
overhead. But the overhead provides us with a more powerful
language. Consequently, problems that can be more compactly
represented by logic programs can be more quickly solved with
$\smodels$ than with a satisfiability checker.

\subsection{Future Work}
\label{subsec:future}

Since testing every literal before resorting to a heuristic improves
the decision procedure, one can ask if testing every two literals or
every set of $n$ literals would improve the algorithm even further. 
The problem is, of course, the overhead. Ordinary $\lookahead$ calls
$\expand$ in the worst case a linear number of times. If we test two
literals, then we call $\expand$ a quadratic number of times, and if
we test sets of $n$ literals, then the number of $\expand$ calls is
already on the order of the number of atoms to the power of $n$.

Luckily we can approximate $n$-$\lookahead$ if we are prepared to use
a quadratic amount of memory. Assume that $a,b\not\in \expand(P,A)$
and that $b\in\lookahead(P,A\cup\{a\})$. Then, any stable model $S$
that agrees with $A$ but do not contain $b$ can not contain
$a$. Hence, if we store the fact that $\nB{b}$ implies $\nB{a}$, then
we can use this information to strengthen $\expand$. Later calls to
$\lookahead$ will then take advantage of all the relations between
the literals that previous calls have found.
 
The problem is that even a quadratic amount of memory seems to create
too much overhead. Two questions are therefore left to future research.
Can one avoid the memory overhead by making use of the structure of
the programs? Does there exist another approximation that evades this
problem?


\appendix
\cleardoublepage
\section{Monotone Functions}
\label{appendix:monotone}

Let $X$ be a set and let $f:2^X\maps 2^X$ be a function. If
$A\subseteq B$ implies $f(A)\subseteq f(B)$, then $f$ is
monotonic\index{monotonic}. We have the following version of the
Knaster-Tarski fixpoint theorem.

\begin{lemma}\label{lemma:monotone}
Let $f:2^X\maps 2^X$ be a monotonic function, and let
$A\subseteq X$. If $f(A) \subseteq A$, then $\lfp{f}\subseteq A$,
where $\lfp{f}$\index{$lfp$@$\lfp{f}$} denotes the least fixed
point\index{least fixed point} of $f$.
\end{lemma}

\begin{proof}
Define
\[S = \bigcap_{f(A)\subseteq A} A \qquad\qquad
\bigl(f(X)\subseteq X\bigr).\]
Then, $f(A)\subseteq A$ implies $S\subseteq A$, which in turn implies
$f(S)\subseteq f(A)$ by the monotonicity of $f$. Hence,
$f(S)\subseteq A$, and consequently
\[f(S) = \bigcap_{f(A)\subseteq A} f(S) \subseteq
\bigcap_{f(A)\subseteq A} A = S.\]

Now, $f(S)\subseteq S$ implies $f\bigl(f(S)\bigr)\subseteq f(S)$,
which by the definition of $S$ implies $S\subseteq f(S)$. Thus,
$S = f(S)$. Moreover, for any fixed point $A$,
\[f(A) \subseteq A \quad\text{implies}\quad S\subseteq A,\] 
and hence $\lfp{f} = S$ by definition.
\end{proof}

Similarly, $A\subseteq f(A)$ implies $A\subseteq \gfp{f}$ for the
greatest fixed point\index{$gfp$@$\gfp{f}$}\index{greatest fixed
  point}
of $f$. Notice that if $X$ is finite, then
 $\lfp{f} = f^n(\emptyset)$ for some $n\leq\abs{X}$ since
$f(\emptyset)\subseteq\lfp{f}$. Furthermore, observe that if we are 
given $k$ monotonic functions $f_1,\dotsc,f_k$, then the least fixed
point of
\[g(A) = \bigcup_{i=1}^k f_i(A)\]
is the limit of any nest
\[A_{n+1} = A_n\cup f_{i(n)}(A_n),\quad \text{$A_0=\emptyset$ and
$f_{i(n)}(A_n)\subseteq A_n \Rightarrow \forall j f_j(A_n)\subseteq A_n$.}\]
In other words, the least fixed point of $g$ can be computed by
repeated applications of $f_1,\dotsc,f_k$. Finally, note that
$a\in\lfp{g}$ if and only if there is a sequence $i(0),\dotsc,i(n)$
such that $a\in A_{n+1}$.

\cleardoublepage
\section{Auxiliary Functions for {\protect $\atleast{P,A}$}}
\label{appendix:atleast}

We present the auxiliary functions used by $\atleast{P,A}$ for
cardinality, choice, and weight rules. We begin with the choice rule.
\bigskip

\begin{algorithmic}
\item[\textbf{function} $r.\fire()$]
 \STATE $r.\literal := r.\literal - 1$.
\end{algorithmic}
\bigskip

\begin{algorithmic}
\item[\textbf{function} $r.\inactivate()$]
 \STATE $r.\inactive := r.\inactive+1$
 \IF{$r.\inactive = 1$}
  \FOR{each atom $a$ in $r.\head$}
   \STATE $a.\headof := a.\headof-1$
   \IF{$a.\headof = 0$}
    \STATE $\mathit{negq}.\mathit{push}(a)$
   \ELSIF{$a.\inpos$ and $a.\headof = 1$}
     \STATE Let $r'$ be the only active rule in $a.\headlist$
     \STATE $r'.\backchaintrue()$
   \ENDIF
  \ENDFOR
 \ENDIF.
\end{algorithmic}
\bigskip

\begin{algorithmic}
\item[\textbf{function} $r.\backchaintrue()$]
 \IF{$r.\literal > 0$}
  \FOR{every $a\in\patoms{r.\body}$}
   \STATE $\mathit{posq}.\mathit{push}(a)$
  \ENDFOR
  \FOR{every $a\in\natoms{r.\body}$}
   \STATE $\mathit{negq}.\mathit{push}(a)$
  \ENDFOR
 \ENDIF.
\end{algorithmic}
\bigskip

\begin{algorithmic}
\item[\textbf{function} $r.\backchainfalse()$]
 \STATE return.
\end{algorithmic}
\bigskip

A cardinality rule $r$
\[h \from k\,\{ a_1,\dotsc,a_n, \nB{b_1}, \dotsc ,\nB{b_m}\}\]
does not need an additional variable that keeps track of $k$ if
$r.\literal$ is initialized to $k$ and $r.\inactive$ is initialized to
$k-(n+m)$.
\newpage

\begin{algorithmic}
\item[\textbf{function} $r.\fire()$]
 \STATE $r.\literal := r.\literal - 1$
 \IF{$r.\literal = 0$}
  \STATE $\mathit{posq}.\mathit{push}(r.\head)$
 \ELSIF{$r.\head.\inneg$}
  \STATE $r.\backchainfalse()$
 \ENDIF.
\end{algorithmic}
\bigskip

\begin{algorithmic}
\item[\textbf{function} $r.\inactivate()$]
 \STATE $r.\inactive := r.\inactive+1$
 \IF{$r.\inactive = 1$}
  \STATE $a := r.\head$
  \STATE $a.\headof := a.\headof-1$
  \IF{$a.\headof = 0$}
   \STATE $\mathit{negq}.\mathit{push}(a)$
  \ELSIF{$a.\inpos$ and $a.\headof = 1$}
    \STATE Let $r'$ be the only active rule in $a.\headlist$
    \STATE $r'.\backchaintrue()$
  \ENDIF
 \ENDIF.
\end{algorithmic}
\bigskip

\begin{algorithmic}
\item[\textbf{function} $r.\backchaintrue()$]
 \IF{$r.\literal > 0$ and $r.\inactive = 0$}
  \FOR{every $a\in\patoms{r.\body}$}
   \STATE $\mathit{posq}.\mathit{push}(a)$
  \ENDFOR
  \FOR{every $a\in\natoms{r.\body}$}
   \STATE $\mathit{negq}.\mathit{push}(a)$
  \ENDFOR
 \ENDIF.
\end{algorithmic}
\bigskip

\begin{algorithmic}
\item[\textbf{function} $r.\backchainfalse()$]
 \IF{$r.\literal = 1$ and $r.\inactive \leq 0$}
  \FOR{every $a\in\patoms{r.\body}$}
   \IF{$a.\inpos = $ false}
    \STATE $\mathit{negq}.\mathit{push}(a)$
   \ENDIF
  \ENDFOR
  \FOR{every $a\in\natoms{r.\body}$}
   \IF{$a.\inneg = $ false}
    \STATE $\mathit{posq}.\mathit{push}(a)$
   \ENDIF
  \ENDFOR
 \ENDIF.
\end{algorithmic}
\newpage

\newcommand{\weight}{\mathit{weight}}

A weight rule $r$
\[h \from \{ a_1 = w_{a_1},\dotsc,a_n = w_{a_n},
\nB{b_1} = w_{b_1},\dotsc,\nB{b_m} = w_{b_m}\} \geq w\]
will need some auxiliary variables. If $a$ is an atom in the body of
$r$, then let $a.\weight$ be the weight of the atom in $r$. For a
partial model $A$, let $r.\mathit{atleast} = w$, let
\[r.\mathit{max} = \sum_{a\in r.\patoms{\body}-\natoms{A}} a.\weight
    +\sum_{b\in r.\natoms{\body}-\patoms{A}} b.\weight\]
and let
\[r.\mathit{min} = \sum_{a\in r.\patoms{\body}\cap\patoms{A}} a.\weight
    +\sum_{b\in r.\natoms{\body}\cap\natoms{A}} b.\weight.\]
\bigskip

\begin{algorithmic}
\item[\textbf{function} $r.\fire()$]
 \STATE Let $a$ be the atom that is being given a truth value
 \STATE $r.\mathit{min} = r.\mathit{min} + a.\weight$
 \IF{$r.\mathit{max} \geq r.\mathit{atleast}$ and
     $r.\mathit{min} - a.\weight < r.\mathit{atleast}$}
  \IF{$r.\mathit{min} \geq r.\mathit{atleast}$}
   \STATE $\mathit{posq}.\mathit{push}(r.\head)$
  \ELSIF{$r.\head.\inneg = $ true}
   \STATE $r.\backchainfalse()$
  \ENDIF
 \ENDIF.
\end{algorithmic}
\bigskip

\begin{algorithmic}
\item[\textbf{function} $r.\inactivate()$]
 \STATE Let $a$ be the atom that is being given a truth value
 \STATE $r.\mathit{max} = r.\mathit{max} - a.\weight$
 \IF{$r.\mathit{max}+a.\weight \geq r.\mathit{atleast}$ and
     $r.\mathit{min} < r.\mathit{atleast}$}
  \IF{$r.\mathit{\max} < r.\mathit{atleast}$}
   \STATE $a := r.\head$
   \STATE $a.\headof := a.\headof-1$
   \IF{$a.\headof = 0$}
    \STATE $\mathit{negq}.\mathit{push}(a)$
   \ELSIF{$a.\inpos$ and $a.\headof = 1$}
     \STATE Let $r'$ be the only active rule in $a.\headlist$
     \STATE $r'.\backchaintrue()$
   \ENDIF
  \ELSIF{$r.\head.\inpos = $ true and $r.\head.\headof = 1$}
   \STATE $r.\backchaintrue()$
  \ENDIF
 \ENDIF.
\end{algorithmic}
\bigskip

Consider the case of $\backchaintrue()$ and $a\in r.\body$. If
\[r.\mathit{max}-a.\weight < r.\mathit{atleast},\]
then $a$ must be pushed onto the queue $\mathit{posq}$. On the other
hand, if
\[r.\mathit{max}-a.\weight\geq r.\mathit{atleast},\]
then we can immediately skip all atoms whose weight is less than
$a.\weight$.

We take advantage of this observation in the following way. We
introduce two new variables: $r.\mathit{last^+}$ and
$r.\mathit{last^-}$, and we go through the body of $r$ in decreasing
weight order. When we do backward chaining in $r.\backchaintrue()$ we
start with the atom given by $r.\mathit{last^+}$ and check every atom
until we arrive at an atom $a$ for which 
$r.\mathit{max}-a.\weight\geq r.\mathit{atleast}$. Then, we stop and
update $r.\mathit{last^+}$ to $a$. When we backtrack we must restore
$r.\mathit{last^+}$ to its previous value, and this can easily be done
without needing any extra memory. If we remove the truth value of an
atom $a$ or if we remove $a$ from a queue, and if
$a.\weight\geq r.\mathit{last^+}.\weight$, then we set
$r.\mathit{last^+}=a$.
\bigskip

\begin{algorithmic}
\item[\textbf{function} $r.\backchaintrue()$]
 \IF{$r.\mathit{min} < r.\mathit{atleast}$}
  \FOR{$a\in\Atoms{r.\body}$ in order, largest weight
    first, starting with $r.\mathit{last^+}$}
   \STATE $r.\mathit{last^+} = a$
   \IF{$a.\inpos = $ false and $a.\inneg = $ false}
    \IF{$r.\mathit{max}-a.\weight < r.\mathit{atleast}$}
     \IF{$a\in\patoms{r.\body}$}
      \STATE $\mathit{posq}.\mathit{push}(a)$
     \ELSE
      \STATE $\mathit{negq}.\mathit{push}(a)$
     \ENDIF
    \ELSE
     \STATE return
    \ENDIF
   \ENDIF
  \ENDFOR
 \ENDIF.
\end{algorithmic}
\bigskip

\begin{algorithmic}
\item[\textbf{function} $r.\backchainfalse()$]
 \IF{$r.\mathit{max}\geq r.\mathit{atleast}$ and
     $r.\mathit{min} < r.\mathit{atleast}$}
  \FOR{$a\in\Atoms{r.\body}$ in order, largest weight
    first, starting with $r.\mathit{last^-}$}
   \STATE $r.\mathit{last^-} = a$
   \IF{$a.\inpos = $ false and $a.\inneg = $ false}
    \IF{$r.\mathit{min}+a.\weight \geq r.\mathit{atleast}$}
     \IF{$a\in\patoms{r.\body}$}
      \STATE $\mathit{negq}.\mathit{push}(a)$
     \ELSE
      \STATE $\mathit{posq}.\mathit{push}(a)$
     \ENDIF
    \ELSE
     \STATE return
    \ENDIF
   \ENDIF
  \ENDFOR
 \ENDIF.
\end{algorithmic}

\cleardoublepage
\section{Auxiliary Functions for {\protect $\atmost{P,A}$}}
\label{appendix:atmost}

We present the auxiliary functions used by $\atmost{P,A}$ for
cardinality, choice, and weight rules. We begin with the choice rule.
\bigskip

\begin{algorithmic}
\item[\textbf{function} $r.\propagateFalse()$]
 \STATE $r.\upper := r.\upper + 1$
 \IF{$r.\upper = 1$ and $r.\inactive = 0$}
  \FOR{every $a\in r.\head$}
   \IF{$a.\source = 0$ or $a.\source = r$}
    \STATE $a.\source := 0$
    \STATE $\mathit{queue}.\mathit{push(a)}$
   \ENDIF
  \ENDFOR
 \ENDIF.
\end{algorithmic}
\bigskip

\begin{algorithmic}
\item[\textbf{function} $r.\propagateTrue()$]
 \STATE $r.\upper := r.\upper - 1$
 \IF{$r.\upper = 0$ and $r.\inactive = 0$}
  \FOR{every $a\in r.\head$}
   \IF{$a.\source = 0$}
    \STATE $a.\source := r$
   \ENDIF
   \STATE $\mathit{queue}.\mathit{push(a)}$
  \ENDFOR
 \ENDIF.
\end{algorithmic}
\bigskip

\begin{algorithmic}
\item[\textbf{function} $r.\isUpperActive()$]
 \IF{$r.\upper = 0$ and $r.\inactive = 0$}
  \STATE return true
 \ELSE
  \STATE return false
 \ENDIF.
\end{algorithmic}
\bigskip

\newcommand{\low}{\mathit{lower}}

For a cardinality rule $r$
\[h \from k\,\{ a_1,\dotsc,a_n, \nB{b_1}, \dotsc ,\nB{b_m}\}\]
we initialize $r.\upper$ to $k-m$. We also need a counter $r.\low$
that holds the value $k-\abs{r.\natoms{\body}-\patoms{A}}$.
\newpage

\begin{algorithmic}
\item[\textbf{function} $r.\propagateFalse()$]
 \STATE $r.\upper := r.\upper + 1$
 \IF{$r.\low \geq 1$ and $r.\inactive \leq 0$ and \\
     \quad $(r.\head.\source = 0$ or $r.\head.\source = r)$}
  \STATE $r.\head.\source := 0$
  \STATE $\mathit{queue}.\mathit{push(r.\head)}$
 \ENDIF.
\end{algorithmic}
\bigskip

\begin{algorithmic}
\item[\textbf{function} $r.\propagateTrue()$]
 \STATE $r.\upper := r.\upper - 1$
 \IF{$r.\upper = 0$ and $r.\inactive \leq 0$}
  \IF{$r.\head.\source = 0$}
   \STATE $r.\head.\source := r$
  \ENDIF
  \STATE $\mathit{queue}.\mathit{push(r.\head)}$
 \ENDIF.
\end{algorithmic}
\bigskip

\begin{algorithmic}
\item[\textbf{function} $r.\isUpperActive()$]
 \IF{$r.\upper \leq 0$ and $r.\inactive \leq 0$}
  \STATE return true
 \ELSE
  \STATE return false
 \ENDIF.
\end{algorithmic}
\bigskip

We can optimize the auxiliary functions a bit by changing $r.\low$ to
keep track of the value
\[k-\abs{r.\natoms{\body}-\patoms{A}}-\abs{B-\natoms{A}},\]
where $B$ is the set of atoms in $r.\patoms{\body}$ that are not part
of a positive loop that goes through $r$.

A weight rule $r$
\[h \from \{ a_1 = w_{a_1},\dotsc,a_n = w_{a_n},
\nB{b_1} = w_{b_1},\dotsc,\nB{b_m} = w_{b_m}\} \geq w\]
also needs the two variables $r.\upper$ and $r.\low$. 
The variable $r.\upper$ is initialized to
\[\sum_{b\in r.\natoms{\body}} b.\weight\]
and $r.\low$ holds the value
\[\sum_{a\in B-\natoms{A}} a.\weight
    +\sum_{b\in r.\natoms{\body}-\patoms{A}} b.\weight,\]
where $B$ is defined as above.
\bigskip

\begin{algorithmic}
\item[\textbf{function} $r.\propagateFalse()$]
 \STATE Let $a$ be the atom that is being removed from the upper closure
 \STATE $r.\upper := r.\upper - a.\weight$
 \IF{$r.\mathit{max}\geq r.\mathit{atleast}$ and 
     $r.\upper+a.\weight\geq r.\mathit{atleast}$
     and $r.\low<r.\mathit{atleast}$ and
     $(r.\head.\source = 0$ or $r.\head.\source = r)$}
  \STATE $r.\head.\source := 0$
  \STATE $\mathit{queue}.\mathit{push(r.\head)}$
 \ENDIF.
\end{algorithmic}
\bigskip

\begin{algorithmic}
\item[\textbf{function} $r.\propagateTrue()$]
 \STATE Let $a$ be the atom that is being added to the upper closure
 \STATE $r.\upper := r.\upper + a.\weight$
 \IF{$r.\mathit{max}\geq r.\mathit{atleast}$ and
     $r.\upper-a.\weight<r.\mathit{atleast}$
     and $r.\upper\geq r.\mathit{atleast}$}
  \IF{$r.\head.\source = 0$}
   \STATE $r.\head.\source := r$
  \ENDIF
  \STATE $\mathit{queue}.\mathit{push(r.\head)}$
 \ENDIF.
\end{algorithmic}
\bigskip

\begin{algorithmic}
\item[\textbf{function} $r.\isUpperActive()$]
 \IF{$r.\upper \geq r.\mathit{atleast}$}
  \STATE return true
 \ELSE
  \STATE return false
 \ENDIF.
\end{algorithmic}

\cleardoublepage
\section{Time-Line}
\label{appendix:timeline}

A time-line for the releases of $\smodels$ that introduced new
features.
\bigskip

\begin{tabular}{|lrl|}
\hline
Version & Date & Description \\ \hline
1.0 & 28.5.1995 & First public release \\
1.1 & 7.3.1996 & Fitting semantics \\
1.2 & 28.3.1996 & Backward chaining if head is true, \\
& & optional lookahead \\
1.3 & 27.9.1996 & Backward chaining if head is false \\
1.6 & 30.6.1997 & Lookahead is on by default, \\
& & the heuristic is introduced \\
& 31.8.1997 & Work on smodels 2.0 has begun \\
1.8 & 4.9.1997 & Strongly connected components \\
& & optimization of the upper closure \\
pre-2.0-4 & 25.3.1998 & Source pointer \\
2.0 & 22.10.1998 & New rule types \\
2.6 & 27.1.1999 & Reduction of the search space \\
\hline
\end{tabular}

\cleardoublepage
\addcontentsline{toc}{section}{References}
\bibliography{p,psi,ael,ini}
\bibliographystyle{plain}

\cleardoublepage
\addcontentsline{toc}{section}{Index}
\printindex


\end{document}